\documentclass[preprint,12pt,authoryear]{elsarticle}

\usepackage{graphicx}
\usepackage{amsmath}
\usepackage{array}
\usepackage{amssymb}
\usepackage{multirow}
\usepackage{float}
\usepackage{longtable}
\usepackage{hyperref}
\usepackage{cleveref}
\usepackage{graphicx}
\usepackage{caption}
\usepackage{subcaption}

\journal{Reliability Engineering \& System Safety}
\begin{document}
\begin{frontmatter}

\title{The Reliability of Remotely Piloted Aircraft System Performance under Aeronautical Communication Uncertainties}

\affiliation[inst1]{organization={Department of Aerospace Engineering and Engineering Mechanics},
            addressline={The University of Texas at Austin}, 
            city={Austin},
            postcode={78712}, 
            state={TX},
            country={USA}}

\author[inst1]{Yutian Pang\corref{mycorrespondingauthor}}
\cortext[mycorrespondingauthor]{Corresponding author.}
\ead{yutian.pang@austin.utexas.edu}
\author[inst1]{Andrew Paul Kendall}
\author[inst1]{John-Paul Clarke}

\begin{highlights}

\item We present mathematical formulations for the key metrics of Required Communication Performance (RCP).
\item We introduce communicability, a novel metric that integrates transaction time, availability, and continuity within the RCP concept.
\item Communication signal availability and latency parameters are incorporated into the flight simulation model of remotely piloted high-performance aircraft.
\item Through Monte Carlo simulations, we derive the mission success surface envelop for the waypoint navigation task.
\item We discover the performance decay behavior under availability and latency variabilities, and highlight the flight control limitations.
\end{highlights}

\begin{abstract}
Mission-critical operations of highly maneuverable Remotely Piloted Aircraft Systems (RPAS) require reliable communication to ensure safe integration into existing airspace. Understanding system-level performance under stochastic communication conditions is essential for estimating mission success and assessing operational risks. This study quantifies the impact of communication latency and complete signal loss on the mission completion performance of a highly maneuverable RPAS. The mission is defined as a static waypoint tracking task in three-dimensional airspace. We first derive mathematical formulations for key reliability metrics within the Required Communication Performance (RCP) framework. These stochastic communication factors, including latency and availability, are then incorporated into flight control simulations to evaluate system behavior. Extensive multiprocessing Monte Carlo simulations are conducted using high-performance computing to generate mission success rate and mission completion time envelopes. Results show significant degradation in flight performance as communication latency increases or availability decreases, which directly reduces the system stability margin. To better characterize this relationship, we introduce a new reliability metric, communicability, which integrates three key RCP metrics and provides insight into the maximum tolerable latency for flight control. The proposed framework informs RPAS design by revealing trade-offs between communication capability and flight control performance. The code used in this study is publicly available at this \href{https://github.com/YutianPangASU/comm-dynamics}{repository}.

\end{abstract}

\begin{keyword}
Remotely Piloted Aircraft System, Required Communication Performance, Latency and Availability, Flight Simulations, Mission Success Envelop
\end{keyword}

\end{frontmatter}


\section{Introduction \label{sec: introduction}}
The Remotely Piloted Aircraft System (RPAS) is an end-to-end system in which an unmanned Remotely Piloted Aircraft (RPA) is flown from a Remote Pilot Station (RPS) by a Remote Pilot (RP). An RPAS can carry cameras, sensors, communication equipment, or other payloads to perform specific tasks. Although RPAS have been used for reconnaissance and information gathering of military operations since the 1950s (i.e., High-Altitude Long Endurance unmanned aerial vehicle) \citep{ma2013simulation, chaturvedi2019comparative}, their future role has expanded considerably to the integration of RPAS into existing civilian airspace due to the gradual maturity of aerospace autonomy capabilities \citep{kephart2010see}. The growing interest in RPAS stems from the versatility and operational advantages. RPAS can undertake missions in hazardous environments without placing aircrews at physical risk, thus reducing the threat to fatalities. They are also invaluable in tasks such as real-time surveillance \citep{iscold2010development, thammawichai2017optimizing}, search and rescue \citep{scherer2015autonomous}, and environmental monitoring \citep{asadzadeh2022uav}. The ability of RPAS to remain airborne for extended periods while providing constant updates and imagery makes them an increasingly important tool for both the official and the private sector \citep{abeyratne2019legal}.

Air Traffic Management (ATM) serves as the backbone of safe and efficient flight operations within the active field of research on RPAS. ATM consists of three major functions (air traffic services, airspace management, and air traffic flow management) to ensure safe and efficient travel of aircraft within the given section of airspace \citep{ICAO9869}, by integrating human, software and hardware supported by communications, navigation, surveillance (CNS) capabilities \citep{flavio_vismari_safety_2011}. Air traffic services provides services to support the daily aeronautical operations such as collision avoidance, conflict resolution, and maintain aeronautical separation minima via timely information provision \citep{wgc1}. In recent years, emerging aviation concept such as Advanced Air Mobility (AAM), is pushing the boundaries of ATM adaptation even further. Research institutes and government agencies such as the Federal Aviation Administration (FAA), National Aeronautics and Space Administration (NASA), and the European Union Aviation Safety Agency (EASA) are leading the research effort on Advanced Air Mobility (AAM) \citep{goyal2022advanced, johnson2022nasa}. AAM pushes air transportation to the new stage by integrating physical and digital infrastructures to efficiently move the people and cargo in the airspace, with multi-rotor vertical or short takeoff and landing aircraft ideally being electrical (eVTOL/STOL). Urban Air Mobility (UAM) is viewed as the subset of AAM where the operation are mostly at urban scenarios \citep{mitre5g}. The Urban Mobility Concept of Operation defines the AAM operates above the AAM corridors at 400 feet above ground level (AGL), and cruises between $1,500$ and $4,000$ AGL at the speed between 50 knots and 140 knots. Besides normal data and voice communication requirements in ATM, video streams communication is the desired function for, (a) real-time monitoring of passenger cabins; (b) command and control (C2) link of pilot video; (c) passenger entertainments \citep{uam2}. To address the increased demand for AAM/UAM operations in the urban scenario, the communication system that can works smartly becomes vital to maintain a proper workload for the controller maintaining safety and efficiency in the performance-based airspace.

However, this expanding role for RPAS comes with significant challenges to the community, particularly regarding the safe and seamless integration into the existing ATM system. One needs to ensure that RPAS can safely share the airspace without causing disruptions to manned aircraft. This requires the establishment of airworthiness standards and verification techniques that do not impose undue burdens on air traffic services. The lack of the standard regulations of integrating remotely piloted aircraft systems (RPAS) into existing airspace has been identified as a major challenge \citep{icao2017remotely, serafino2019link}. The integration of RPAS has been outlined by the Federal Aviation Administration (FAA) \citep{huerta2013integration} and Single European Sky ATM Research (SESAR) joint undertaking \citep{sesar-2035}, as well as the International Civil Aviation Organization (ICAO). The long-term objective is to ensure all RPAS are regulated as routine civil operations \citep{teutsch2022communication}. Under this big picture, the safety guarantee of RPAS system reliability remains a challenge, along with detect and avoid system verification, and human factor satisfactory. The command and control (C2) link performance on communication latency, availability, and continuity can be evaluated in real-time, while latency and availability are reported to be sufficient for communication link performance evaluation \citep{serafino2019link}. Safety-critical communication networks have been the subject of reliability modeling in other domains, including nuclear power plant instrumentation and control systems \citep{lee_reliability_2015} and fieldbus-based distributed control systems \citep{cauffriez_design_2004}, providing transferable insights for RPAS communication dependability.  

Researchers have been working on studying the reliability of safety-critical sub-systems in aviation \citep{pettit2001general, weckman2001modeling, lu2015reliable, muecklich2023safety}. Complementary efforts in the reliability community have advanced the understanding of accident causation and system safety through both organizational and technical lenses \citep{saleh_highlights_2010}, analyzed trends in aviation maintenance risk \citep{marais_analysis_2012}, and developed systematic approaches to model general aviation accidents such as high-risk occurrence chains in helicopter operations \citep{rao_high_2018} and state-based accident models \citep{rao_state-based_2020}. These lines of research also extend to the emerging field of system-level verification in either deterministic or probabilistic sense \citep{tomlin2001safety, reinhardt2013safety, hook2016certification}. Aircraft performance is one of the prime examples for verification of reliability, as they can be modeled in the simulation environment \citep{heidlauf2018verification}. However, existing works usually adopt simple approximations which assume the aircraft can be modeled with a point mass \citep{hu2020uas, guo2021safety, hu2022obstacle} or reduced number of freedom (i.e., Dubins Model) \citep{dubins1957curves, molnar2024collision}. On the other hand, communication latency and complete signal loss are critical events that impact RPAS system reliability, and drastically deteriorate flight performance, especially during autonomous operations. Additionally, the desired level of automation during certain RPAS missions is also an open-ended question to the community. For instance, three levels of control architectures are in service, (i) human-in-the-loop (HITL), where the control system requires continuous and direct human input; (ii) human-on-the-loop (HOTL), where human act as the supervisor to the machine; (iii) fully autonomous, where the RPAS determines procedures to fulfill mission requirements. RPAS are moving objects, proper autonomous functions needs to be put onboard to maintain stability as well as perform basic maneuvers. Studying the impact of C2 link issues helps answer the question of, \textit{how much autonomy and what type of control architecture do we need to be have onboard for RPAS in order to complete the given task?}

One closely-related work was actually done on a within-visual-range (WVR) air-to-air (A2A) dogfight RPAS virtual reality simulator \citep{thirtyacre2021effects, thirtyacre2022remotely}. The author employed repeated experiments with experienced pilots to test the combat success score, with a set of sparse latency values ranging from 0 up to 1.25 seconds, during high- and low-speed engagement scenarios. The author discovered that when latency cannot be eliminated, conducting high-speed beyond-line-of-sight (BLOS) engagements may lessen its negative impact, which provides insights for WVR combat in real world. This study makes notably contribution by considering the communication latency in the dynamic dogfight task. However, this study has not considered the impact of complete C2 link loss, while only six latency values are considered in the simulator. Another closely related work studies the impact of communication latency and signal loss on separation assurance of RPAS for autonomous cargo operations \citep{bulusu2022impact}. They remarkably consider communication availability by introduce message drop probability (MDP) to approximate the C2 link drop behavior. A sparse set of five values of latency values, and six values of MDP are selected in their simulation. However, only smaller MDP values are selected, and the complete C2 link loss situation is not considered in their study.

To the best of the authors' knowledge, no existing work has examined the effects of complete communication loss and communication latency simultaneously on the performance of an RPAS within a full six-degree-of-freedom (6DOF) simulation, let alone considered these challenges for high-performance systems (i.e., the F-16 plant). Furthermore, there are no comparable studies that determine the mission success and mission completion time envelopes through extensive Monte Carlo simulations while varying extremely dense latency and availability parameters. Therefore, we highlight the contributions of our work here,
\begin{itemize}
    \item We examine the concept of RCP and provide the complete mathematical formulation of key metrics on availability and continuity, based on state transition probabilities. The communication availability is modeled as a continuous-time Markov chain (CTMC), which is defined as a continuous-time Markov chain with the \textit{on} and \textit{off} states, with exponential waiting times for each state.
    \item We propose a new RCP metric, \textit{communicability}, to represent the possibility that a system can support the transmission of messages with a certain duration at any random given time, with a communication latency value.
    \item We incorporate the communication latency and C2 link loss into the flight simulation of a remotely piloted F-16. The communication latency impacts the time steps to apply the remote control inputs. 
    \item Given the waypoint following task, we generate the mission success rate envelope and mission completion time envelope for successful simulations by conducting extensive Monte Carlo simulations with multiprocessing on a high-performance computer. Mission success is defined as the aircraft reaching the slant range of each waypoint in order. We discover the performance decay behavior by providing a complete study of flight performance under availability and latency variabilities.
    \item We further generate the figure showing the mission success rate versus communicability under different latency intervals, to help understanding the maximum toleratable latency values for the given waypoint following task. 
    \item We notice an intriguing phenomenon where the remotely piloted F-16 enters an infinite looping behavior while tracking a waypoint during certain latency values with full availability, and slightly decrease or increase the latency can alleviate the issue. 
\end{itemize}

In the remainder of this paper, we provide a literature review of related works in \Cref{sec: literature review}. Then we look into the concept of RCP, and provide the mathematical formulations of key RCP metrics in \Cref{sec: rcp}. We briefly introduce the F-16 flight dynamics and control used in the simulation in \cref{sec: simulation}. \Cref{sec: experiments} provides the necessary details on our experiments, with extra discussions in \Cref{subsec: discussion}. \Cref{sec: conclusion} concludes the paper and explain possible future research directions.

\section{Literature Review \label{sec: literature review}}
\begin{figure}[hbt!]
\centering
\includegraphics[width=0.95\textwidth]{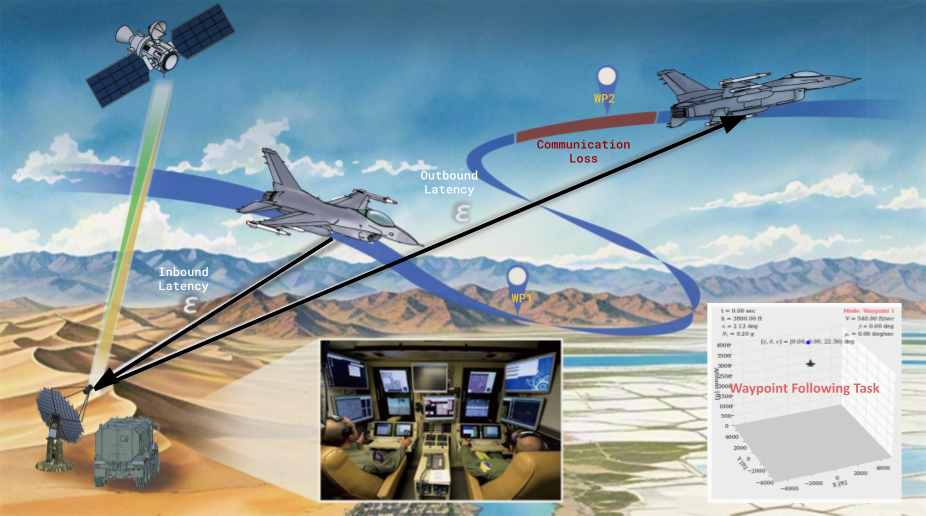}
\caption{Flight simulation case study in this research. The task has two waypoints: WP1 and WP2. The communication link between the remote piloted station  (RPS) and the remotely piloted aircraft (RPA) encounters inbound/outbound latency $\varepsilon$ and complete signal loss duration (i.e., availability $P_A$) during flight, leading to degraded performance on mission completion. }
\label{fig: carton-f16}
\end{figure}

\subsection{Reliability of Aeronautical Communication}
The special committee on Future Air Navigation System (FANS) of ICAO has identified the enhancement of communications, navigation, surveillance and air traffic management (CNS/ATM) as one of the first objectives \citep{wgc1}. C2 link relies on signal propagation to transmit either higher-level operational instructions or immediate flight control parameters for Air-to-Ground (A2G) or Air-to-Air (A2A) communications. Signal strength estimation is one of the most critical part of signal propagation \citep{alsamhi2018predictive} to ensure the receiver gaining sufficient level of power. Signal propagation can be affected by three main phenomena: reflection, diffraction, and scattering. Reflection occurs when a signal bounces off large, smooth surfaces, causing multipath fading due to multiple signal paths arriving at different times. Diffraction involves the bending of signals around sharp edges, allowing them to reach shadowed areas not in direct line of sight but causing attenuation in the process. Scattering happens when signals encounter small or irregular objects, dispersing energy in multiple directions and reducing signal strength, yet sometimes assisting in overcoming obstacles by reaching otherwise blocked areas.

Specifically, the term \textit{communication} stands for accurate information transfer between the transmitter and receiver. Voice and data communications are considered within the scope of current ATM services. Voice communication for A2G systems within the continental U.S. airspace are two-way between the pilots and the ground stations, achieved by Remote Communication Facilities (RCFs) maintained by FAA. RCFs use Very High Frequency (VHF) and Ultra High Frequency (UHF) antenna transmitters and receivers mounted on buildings or towers. VHF operates from 118 MHZ to 136.975 MHZ are for civil aviation use, while UHF channel takes effect from 225 MHZ to 400 MHZ of military usage \citep{markus2020existing}. High Frequency (HF) aeronautical voice communication also serves A/G communications for a much long range. It is available over international airspace (or continental U.S. only in the case of emergency), which covers frequency from 3 MHZ to 30 MHZ. Iridium and Inmarsat provide complete or approximate world-wide satellite voice communication (SATVOICE) service in Lower Earth Orbit (LEO) or Geostationary Earth Orbit (GEO). The reliability of satellite communication systems has been analyzed using Markov Bayesian Networks to model subsystem interactions and optimize redundancy designs in the space environment \citep{chen_reliability_2024}. On the other side, data communication refers to textual data providing routes information to the pilots, which also referred as data link. Similar to voice communication, there is HFDL/VDL refers to data link using HF/VHF, as well as Iridium and Inmarsat providing satellite-based data communication (SATCOM) services primarily used in oceanic and remote airspace \citep{PARKSATVOICE}. 

The concept, \textit{reliability}, stands for the probability that the system functioning correctly under certain operation conditions over a specific period of time \citep{villemeur1992reliability, ahmad2017reliability}. The intellectual roots and evolution of reliability engineering as a discipline have been well documented \citep{saleh_highlights_2006}, and the methods for system reliability optimization have advanced considerably over several decades \citep{coit_evolution_2019}. Understanding the financial implications of reliability is also important, as the net present value framework illustrates how system reliability directly impacts revenue generation capability and investment decisions \citep{saleh_reliability_2006}. The reliability of aviation communication system has been investigated by either the RCP concept by ICAO for manned commercial aviation, or the Required Link Performance (RLP) by the Joint Authorities for Rulemaking of Unmanned Systems (JARUS). RCP and RLP have identical key metrics to quantify the reliability of communication systems as, transaction time, availability, continuity, and integrity. Similar metrics has also been used in measuring reliability of satellite-based Differential Global Positioning System (DGPS) for maritime navigation \citep{moore2002maritime} and other navigation application domains. The detailed explanations and mathematical derivations is given in \Cref{sec: rcp}, along with our newly proposed metric, \textit{communicability}, to consider the first three key metrics together. For consistency of terminologies, we use RCP instead of RLP throughout this paper.

\subsection{Communication Latency}
Although RPAS has been deployed in the United States (US) for decades \citep{gertler2012us}, the latency in C2 link remains an inherent limitation of RPAS operations \citep{trsek2007last}. The emerging automation capabilities of RPAS urgently require a thorough understanding of communication latency to flight performance. As shown in \Cref{fig: carton-f16}, the mission we are considering in this work are typically BLOS operations, where the C2 links travel thousands of miles through terrestrial networks and satellites, from the RPS to RPA (i.e., inbound latency). In return, the RPAS sensor measurements will transmit back to the RPS through the same path (i.e., outbound latency). The latency impacts the measurements of the RPAS true states and what displayed to the RPS, as well as the RP's control input and RPA receiving the command. As pointed out in the Observation-Orientation-Decision-Action (OODA) loop theory, the capability to operate a faster decision-making process than the opponent will lead to better chance of winning \citep{brehmer2005dynamic}. Latency is viewed as delays in each step of the OODA loop. Understanding latency impact, especially in highly dynamic environment (i.e., RPAS), is critical and will greatly increase the chance of mission success.

Remotely operated systems, which enable remote operations in environments ranging from remote surgeries to the surfaces of distant planets (i.e., Moon/Mars exploration), are critically affected by latencies in communication links, which reduce the operators' ability to efficiently control and monitor the robot’s actions \citep{avgousti2018medical, garcia2016enabling, burns2019science, taylor2016mars, vozar2014driver}. Latency originates from various sources, including network delays, processing inefficiencies, and sensory input limitations, and can necessitate suboptimal control strategies (i.e.,, move-and-wait open-loop), thereby hindering mission success \citep{lupisella2018low, vozar2018development}. Latency exceeding certain thresholds diminishes situational awareness and adaptability, particularly when it is variable rather than constant, or when it delays feedback to the user rather than the robot \citep{garcia2016enabling, luck2006investigation}. This challenge is significantly magnified over interplanetary distances, where communication delays of up to 20 minutes between Earth and Mars demand advance programming and careful planning of rover operations to mitigate risks and ensure mission objectives are met \citep{burns2019science, taylor2016mars}. As space exploration and other remote applications advance, efforts to reduce or compensate for latency—through low-latency rovers, improved bandwidth, and enhanced levels of autonomy—are seen as essential for enabling human-like precision and creativity, while retaining robotic endurance and safety in hostile or inaccessible environments \citep{botta2017networking, lupisella2018low}. The \textit{move-and-wait} approach is usually employed while communication latency prevents continuous, responsive control, but significantly slow task completion \citep{sheridan1963remote}. Moreover, \textit{move-and-wait} is not possible for RPAS scenarios. \cite{sollenberger2003effect} investigates the impact of ground communication system delays to the controller workload, while replacing the aging radio device with newer models. The HITL simulation study concludes that maximum tolerable ground system delay is 350 ms without impacting the performance on the air traffic controllers. 

Possible solutions to fight latencies are improving robot design, employing predictive modeling, and advancing communication technologies. In related transportation domains, model-based verification methods have been developed to address parameter uncertainty in train control systems where communication delays and packet-loss factors directly affect safety \citep{cheng_model-based_2016}, and formal verification using Colored Petri Nets has been applied to validate safety communication protocols \citep{lijie_verification_2012}. For instance, the typical approach involves inserting a predicted compensatory delay into the feedback loop of the control system, allowing the operator to experience instantaneous synthetic feedback, which is particularly effective when latency is stable and predictable \citep{zhao2015stability, wang2017analysis}. Emerging high-frequency communication technologies (i.e., 5G), promise both high data throughput and ultra-low latency, enabling more responsive and precise teleoperations \citep{voigtlander20175g}. Additionally, control augmentation strategies such as trajectory prediction \citep{pang2021data, pang2022bayesian} can alleviate latency impacts. However, unpredictable external disturbances can make these predictive models unreliable \citep{chen2007human}. Long-term solutions point toward autonomous operations, where vehicles sense, decide, and maneuver without human interference \citep{byrnes2014nightfall, mayer2015new}. A hybrid approach that combines the aforementioned solution strategies may offer promising avenues for mitigating latency, with further necessary improvements. 

\subsection{Communication Loss}
Many factors can cause complete communication signal loss for RPAS, which requires accurate measurement of sensor input (i.e., radar altimeters) for awareness of situations within the area of interest. From an organizational perspective, communication loss events can be understood within the broader framework of risk dynamics, where operational pressures may erode safety margins over time \citep{marais_conceptualizing_2008}. Communication redundancy is a primary design countermeasure, though the value of such redundancy is contingent on the prevalence of common-cause failures that can simultaneously affect multiple communication channels \citep{hoepfer_value_2009}. The sensor data can be inaccurate, delayed, corrupted, or spoofed. Cyber-physical attacks toward communication infrastructures and links is the major source of complete communication signal loss \citep{dave2022cyber, ukwandu2022cyber}. Based on the awareness of the attacker blocking communication channels, we classify them into intentional and unintentional attacks. Intentional attacks toward aviation communication systems include jamming, spoofing, and tampering. Signal jamming refers to deliberately disrupting radio signals using devices that broadcast noise or create interference on the same frequency as aviation communication channels. Spoofing injects false signals to mimic legitimate communication channels to disrupt ATC and pilots. Tempering, also known as man-in-the-middle attack, refers to unauthorized access to the system or even damage physical communication infrastructures. Distributed-Denial-of-Service (DDoS) is viewed as a special form of spoofing by massively injecting false signals into the communication system. Unintentional attacks include the electromagnetic interference (EMI) caused by onboard equipment or passenger electronics, cross-modulation caused the mixing of signals from different frequencies. Space weather events is one example of unintentional attacks towards the aviation communication systems. Space weather events disrupt ionosphere reflections and create blockage of shortwave radio transmissions \citep{xue2024optimizing}. The formation of space weather events is highly unpredictable, as it is related to the moving of particles inside of the Earth's magnetic field towards the poles caused by solar activities \citep{tsurutani2005october, tsurutani2009brief, tsurutani2022space}. North Polar Flights traversing the area north of $82^o$N primarily rely on HF to maintain connection with ground stations, or SATCOM for most remote areas. However, both HF and SATCOM are heavily impacted by space weather events in the area north of $82^o$N, and even inaccessible for HF bands below 30MHz \citep{xue2024optimizing}. In these scenarios, flights are required to divert or land near the North Pole due to the lack of timely and accurate space weather event prediction. 

Few research works considered the issue of communication system loss. A closely related work examines how C2 link latency and communication reliability affect the separation assurance capability of RPAS operating within the national airspace system \citep{bulusu2022impact}. Using a generic arrival pattern scenario with three merging flows, the study models latency as the time delay from when air traffic control identifies a required maneuver to when the RPA initiates it, and models reliability as the probability that a control message will fail to be received. The paper discovers that when response times exceed 30 seconds or when message drop probability surpasses $0.2$, the chance of loss of safe separation increased. The severity of these impacts varies by RPA aircraft type, highlighting the influence of performance characteristics on safety outcomes. The paper provides detailed insights into how these methods could be applied in future studies as automation in airspace operations continues to grow. Another closely-related work explores the impact of C2 latency on RPA performance during WVR A2A simulated combat in virtual reality environments \citep{thirtyacre2021effects, thirtyacre2022remotely}. The research focuses on how latency affects decision-making efficiency using Boyd's OODA Loop, comparing performance under high-speed and low-speed engagements. The author used a repeated-measures design where fighter pilots simulated one-on-one combat with discrete latency levels, and revealed that latency above 0.25 seconds begins to impact performance, with delays over 1.25 seconds leading to severe degradation in control and combat scores. Pilots in high-speed engagements benefited from consistent turn geometry and sustained higher acceleration loads, which partially offset the latency effects. This study remarkably considers latency effect in a dynamic environment, but fail to include the C2 loss situations.

The existing literature examining the effects of latency and command-and-control (C2) link disruptions on fast-moving vehicles requiring precise, robust control remains limited. While some studies consider latency, they often neglect the impact of C2 link loss altogether or rely solely on probabilistic models, rather than rigorously testing the flight control system with systematically varied inputs. Moreover, current research tends to focus on a narrow set of manually chosen latency values, rather than conducting comprehensive assessments across a dense range of latencies. Although previous work has observed general performance degradation as latency increases, the specific patterns and magnitude of this degradation have not been thoroughly investigated. All of the above discussions motivates this investigation on communication system reliability of RPAS.

\section{Required Communication Performance (RCP) \label{sec: rcp}}
In this section, we will first introduce the conceptual background of RCP within the Performance-Based Airspace (PBA) and the four major parameters defined by ICAO \Cref{subsec: rcp-concept}. Then, we provide the mathematical formulation of RCP parameters in \Cref{subsec: rcp-formulation}. Lastly, we propose the new metric to represent communication system status, communicability, representing transaction time, availability, and continuity all in one.

\subsection{The RCP Concept \label{subsec: rcp-concept}}

Performance Based Communication \citep{ICAO9869} provides the foundations and specifications to achieve and maintain the \textbf{Required Communication Performance (RCP)}. RCP concept is used to specify the required communication performance requirements to support specific functions, operations, and procedures within the performance-based airspace (PBA). While the RCP was proposed by ICAO, JARUS proposed the \textit{Required C2 Link Performance} (RLP) concept for RPAS, to address the emerging need on RPAS communication standardization and provide better terminology to RPAS C2 supporting systems. RLP adopts the same parameters as RCP, except that the performance of C2 link is evaluated on the entire system level \citep{jarus4required}. For simplicity, we uses RCP throughout the remaining part of the paper. RCP is based on a set of parameters as defined below,
\begin{itemize}
    \item Transaction Time (TT): The maximum nominal time span required in which the operational communication transaction is completed before alternating to another procedure. 
    \item Availability: The probability of the communication service can be initiated when needed. 
    \item Continuity: Given the service is available, the probability that the communication can be completed within the transaction time.
    \item Integrity: The probability of transactions completed within the transaction time without detected error.
\end{itemize}

In RCP, a transaction is the process of the human uses to give clearance, order commands, and send flight information, and is completed based on subjective confidence. Moreover, the transaction time include communication steps such as the speaking time (e.g., averaged 2.6 seconds for pilot and 3.8 seconds for controllers \citep{sollenberger2003effect}) and the reaction time of the controller and the pilot (i.e., 1 second), as well as the Air-to-Ground and Ground-to-Ground signal transmission time. Notably, the European Organisation for Civil Aviation Equipment (EUROCAE) ED136 provides the acceptable maximum latency of $100ms$ from the air traffic controller to the ground stations, $2ms$ for A/G communication, and $50ms$ for radio activation \citep{acpwgf24}. In the case of multiple transactions, the TT is the time to complete the most stringent transaction. RCP type is the specific label defines the communication transaction performance standard, denoting values for transaction time, availability, continuity and integrity. Five RCP types are prescribed (RCP10, RCP60, RCP120, RCP240, RCP400) \citep{ICAO9869}. RCP10 standard requires the communication transaction is likely to be finished within 10 seconds, where 10 seconds is the transaction time. It is mostly used for enhancing the aeronautical horizontal separation violation (i.e., 5nm) intervention capability of controller. Continuity can be critical for RCP10, where the transaction is likely to be finished within 10 seconds without the time to repeat the voice messages. To meet RCP10, the transaction should be completed with a single attempt. RCP60 usually complements RCP10 with data communication (i.e., data link). RCP120 and RCP240 are used for supporting separation violation intervention capability in the extended range of 15nm radius or 30/30 (30nm lateral and 30nm longitudinal). RCP240 is the mandatory for oceanic airspace of application with certain separation standards as noted by the FAA \citep{PARKSATVOICE}. Beyond 30/30 separation, RCP400 is the standard for SATVOICE and SATCOM where communication services are provided through HF signal. A recent collaborative study between the FAA and JetBlue Airways conducted in Fall 2017 validated the onboard SATVOICE system is able to satisfy the RCP400 criteria and SATVOICE is viable for RCP400 \citep{PARKSATVOICE}. The communication equipment carried by the aircraft should at least satisfy the appropriate required RCP type to receive approval of operations. Other RCP types may also be established based on experience and usage. The development of RCP concept enables regulatory authorities to grant operational approvals for specific areas based on the capability to meet a specified RCP, rather than solely on the installation or carriage of specific communications equipment. The airspace authority prescribes the RCP for a given area of airspace or a specific procedure after determining RCP types, as well as maintaining compliance of RCP among different airspace users.

\subsection{Mathematical Formulations \label{subsec: rcp-formulation}}
We provide the mathematical formulation of Availability and Continuity in this section. Integrity refers to the trust that can be placed in a data transmission being complete and unaltered, without any undetected loss or corruption of data packets. In this study, we are not concerned with the technical aspects of data packet integrity and only focus on the three other metrics.

We start by defining $\tau_{msg}$ as the duration of the message to be transmitted in the communication system (i.e., Transaction Time). We further assume that the signal transition between \textit{on} and \textit{off} states follows the Continuous-Time Markov Chain (CTMC), with exponential waiting times for each state. Markov-based methods have been widely adopted in reliability engineering for modeling dynamic systems. For example, Markov models have been combined with event-tree/fault-tree analysis for dynamic reliability assessment \citep{bucci_construction_2008}, Petri nets with aging tokens have been used for modeling degrading and repairable systems \citep{volovoi_modeling_2004}, discrete-time Bayesian networks have been applied to systems with common cause failures \citep{guo_discrete-time_2021}, and Dynamic Object Oriented Bayesian Networks have been formalized for complex system reliability evaluation \citep{weber_complex_2006}. That is,

\begin{itemize}
    \item $X(t) = 1$: the \textit{on} state where the service is available.
    \item $X(t) = 0$: the \textit{off} state where the service is unavailable.
\end{itemize}

We model the duration of communication availability, $T_{on}$ and the duration of communication unavailability, $T_{off}$ as two independent Exponential random variables,

\begin{align}
    T_{on} \sim \mathsf{Exp}(\lambda_{off})\\
    T_{off}\sim \mathsf{Exp}(\lambda_{on})
\end{align}

\noindent where the parameters used are defined as,
\begin{itemize}
    \item $\lambda_{on}$: This rate governs the transition from the \textit{off} state ($X(t) = 0$) to the \textit{on} state ($X(t) = 1$). It is the rate at which the communication service becomes available while staying unavailable.
    \item $\lambda_{off}$: This rate governs the transition from the \textit{on} state ($X(t) = 1$) to the \textit{off} state ($X(t) = 0$). It is the rate at which the communication service becomes unavailable while staying available.
    \item $T_{on}$: This means that the time duration for which the service stays in the \textit{on} state before switching to \textit{off} is exponentially distributed with rate $\lambda_{off}$.
    \item $T_{off}$: This means that the time duration for which the service stays in the \textit{off} state before switching to \textit{on} is exponentially distributed with rate $\lambda_{on}$.
\end{itemize}

\subsubsection{Availability}
Availability is defined as the proportion of time the system is \textit{on} over the long term. Given the definition of the probability of the system states,
\begin{itemize}
    \item $P_A(t)$: The system is at $X(t)=1$ state at time $t$.
    \item $P_{\Bar{A}}(t)$: The system is at $X(t)=0$ state at time $t$.
\end{itemize}
It's clear that $P_A(t) + P_{\Bar{A}}(t) = 1$ since there are only two discrete state in such system. Further given the Exponential distributions of $T_{on}$ and $T_{off}$, we can model the transitions between system states with a continuous-time Markov chain (CTMC) with inflow rate and outflow rate as $\lambda_{on}$ and $\lambda_{off}$, and the corresponding stochastic matrix $Q$ for state transitions can be expressed as,
\begin{equation}
Q =  
    \begin{bmatrix}
    -\lambda_{on} & \lambda_{on} \\
    \lambda_{off} & -\lambda_{off} 
    \end{bmatrix}  
\end{equation}
The rate of change of $P_{A}(t)$ is defined as the difference between the inflow rate and outflow rate,
\begin{align}
    \frac{dP_{A}(t)}{dt} 
    &= \text{inflow rate} - \text{outflow rate} \\
    &= \lambda_{on} \cdot P_{\Bar{A}}(t) - \lambda_{off} \cdot P_{A}(t) \\
    &= \lambda_{on} \cdot (1-P_{A}(t)) - \lambda_{off} \cdot P_{A}(t) \\
    &= \lambda_{on} - (\lambda_{on} + \lambda_{off}) \cdot P_{A}(t)
\end{align}
The solution of the first-order differential equation has the general form as,
\begin{equation}
    P_{A}(t) = \mathcal{C} \cdot e^{-{(\lambda_{on} + \lambda_{off})}\cdot t}  + \frac{\lambda_{on}}{\lambda_{on} + \lambda_{off}}
\end{equation}
\noindent with initial condition $P_{A}(0) = 1$ (assuming the system starts at the \textit{on} state). Then the constant $\mathcal{C}$ is,
\begin{equation}
    \mathcal{C} = 1 - \frac{\lambda_{on}}{\lambda_{on} + \lambda_{off}}
\end{equation}
Finally, the availability is represented as,
\begin{equation}
\label{eq: avail}
    P_{A}(t) = (1 - \frac{\lambda_{on}}{\lambda_{on} + \lambda_{off}}) \cdot e^{-{(\lambda_{on} + \lambda_{off})}\cdot t}  + \frac{\lambda_{on}}{\lambda_{on} + \lambda_{off}}
\end{equation}
Notably, the steady state availability is given by the ratio of the expected \textit{on} duration to the total expected cycle time (which is the sum of the expected \textit{on} and \textit{off} times). It is simply the proportion of time that the system is \textit{on} over the infinite horizon ($t \rightarrow \infty$). 
\begin{align}
\label{eq: steady-solution-avail}
    P_{A}(t\rightarrow\infty) 
    &= \lim_{t\to\infty} (1 - \frac{\lambda_{on}}{\lambda_{on} + \lambda_{off}}) \cdot e^{-{(\lambda_{on} + \lambda_{off})}\cdot t}  + \frac{\lambda_{on}}{\lambda_{on} + \lambda_{off}} \\
    &= \frac{\lambda_{on}}{\lambda_{on}+\lambda_{off}} \\
    &= \frac{\mathbb{E}[T_{on}]}{\mathbb{E}[T_{on}] + \mathbb{E}[T_{off}]} \\
\end{align}

\subsubsection{Continuity}
Continuity is defined as the probability that once the service is available at time $t$ (i.e., \textit{on}), it remains available for the required duration $\tau_{msg}$. If we discretize $\tau_{msg}$ into $n$ steps $\tau_0, \tau_1, \ldots, \tau_n$, with infinitesimal time interval $\Delta_\tau$. Then we have,
\begin{equation}
    \tau_{msg} = \Sigma_{i=0}^{n-1} \Delta_\tau
\end{equation}

The probability that the system is available on any discretized timestamp is $P(\tau_i)$. By discretizing time duration, we have the continuity expressed as the joint probability at all timestamps,
\begin{align}
P_C(\tau_{msg})
&= P\bigl(X(\tau_0)=1, X(\tau_1)=1, \ldots , X(\tau_n)=1\bigr) \nonumber\\
&= P\bigl(X(\tau_0)=1\bigr) \cdot P\bigl(X(\tau_1)=1 \mid X(\tau_0)=1\bigr) \nonumber\\
&\quad \cdot P\bigl(X(\tau_2)=1 \mid X(\tau_0)=1 \cap X(\tau_1)=1\bigr) \cdots \nonumber\\
\quad \cdot P\bigl(X(\tau_n)=1 \mid &X(\tau_0)=1 \cap X(\tau_1)=1 \cap \cdots \cap X(\tau_{n-1})=1\bigr)
\end{align}

For Markov process, the joint probability can be simplified by assuming the future state only depends on the current state,
\begin{align}
P_C(\tau_{msg})
&= P\bigl(X(\tau_0)=1\bigr) \cdot P\bigl(X(\tau_1)=1 \mid X(\tau_0)=1\bigr) \cdots \nonumber\\
&\quad \cdot P\bigl(X(\tau_{n})=1 \mid X(\tau_{n-1})=1\bigr) \nonumber\\
&= P\bigl(X(\tau_0)=1\bigr) \cdot \prod_{k=1}^{n} P\bigl(X(\tau_{k})=1 \mid X(\tau_{k-1})=1\bigr) \nonumber\\
&= P_A(\tau_0) \cdot \prod_{k=1}^{n} P\bigl(X(\tau_{k})=1 \mid X(\tau_{k-1})=1\bigr)
\end{align}

The transition probability $P(X(\tau_{n})=1 | X(\tau_{n-1})=1)$ of the Markov process is the probability that the service is at the \textit{on} state at $\tau_{n}$ given the service is at the \textit{on} state at $\tau_{n-1}$. That is,
\begin{equation}
    P(X(\tau_{k})=1 | X(\tau_{k-1})=1) = e^{-\lambda_{off} \cdot \Delta_\tau}, \quad \forall k \in {1, 2, ..., n}
\end{equation}

Therefore, the probability that the system remains available for the required duration is,
\begin{align}
P_C(\tau_{msg})
&= P_A(\tau_0) \cdot \Pi_{k=1}^n P(X(\tau_{k})=1 | X(\tau_{k-1})=1)) \\
&= P_A(\tau_0) \cdot (e^{-\lambda_{off} \cdot \Delta_\tau})^{n-1} \\
&= P_A(\tau_0) \cdot e^{-\lambda_{off} \cdot \Sigma_{i=0}^{n-1} \Delta_\tau} \\
&= P_A(\tau_0) \cdot e^{-\lambda_{off} \cdot \tau_{msg}}
\end{align}

Similarly, assuming the system starts at the \textit{on} state (i.e., $P_A(\tau_0) =1$), continuity is expressed as,
\begin{equation}
    P_C(\tau_{msg}) = e^{-\lambda_{off} \cdot \tau_{msg}}
\end{equation}

Notably, this is also the probability that the service maintains the \textit{on} state for at least $\tau_{msg}$ before transitioning into the \textit{off} state as,
\begin{align}
P_C(\tau_{msg}) 
&= P(T_{on} \geq \tau_{msg}) \\
&= \int_{\tau_{msg}}^{\infty} \mathsf{Exp}(\lambda_{off}) \cdot dx \\
&= e^{-\lambda_{off} \cdot \tau_{msg}}
\end{align}


Continuity is related to the time duration $\tau_{msg}$, for which we desire the service to remain continuously available. This makes it a useful metric for assessing the reliability of service over specific time intervals, independent of when those intervals start, utilizing the memoryless feature of exponential distributions. 

\subsection{Communicability \label{subsec: communicability}}

\begin{figure}[hbt!]
\centering
\includegraphics[width=0.95\textwidth]{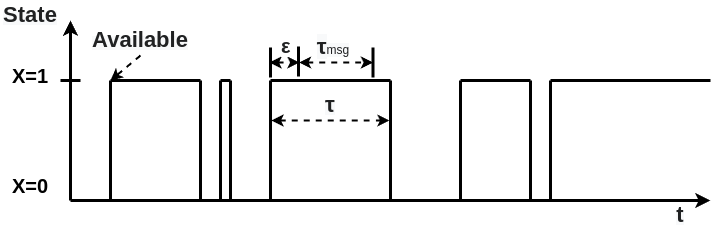}
\caption{The concept of the control mask square wave. }
\label{fig: squarewave}
\end{figure}

To further consolidate transaction time, availability, and continuity into a single metric, we propose the \textit{communicability} metric. We define $\tau$ as the time since the system most recently became available (as shown in \Cref{fig: squarewave}), following an exponential decay, and $\tau_{msg}$ as the duration of the message to be transmitted in the communication system. $\varepsilon$ is the latency parameter representing one way system latency in general. The probability density function (PDF) of $\tau$ is defined as,
\begin{equation}
    f(\tau) = (1-P_A) \cdot e^{-\frac{\tau}{k}}
\end{equation}
\noindent where the scaling factor $k$ to normalize $f(\tau)$ and $k$ is $\frac{1}{1-P_A}$ by solving,
\begin{equation}
    \int_0^\infty f(\tau) d\tau = 1
\end{equation}
The final representation of $f(\tau)$ becomes, 
\begin{equation}
    f(\tau) = (1-P_A) \cdot e^{-(1-P_A)\tau}
\end{equation}

\begin{figure}[hbt!]
\label{fig: dist_tau}
\centering
\includegraphics[width=0.5\textwidth]{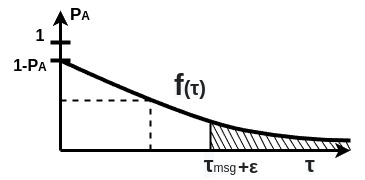}
\caption{The probability density function of $\tau$}
\end{figure}

Based on these, we propose a new metric, communicability, representing the probability that a communication system will support the transmission of messages of duration $\tau_{msg}$ with latency $\varepsilon$ that are initiated at any random time $t$. The expression for communicability is:
\begin{align}
\label{eq: P_comm}
P_{\text{comm}} 
&= P_A \cdot \bigg[ \frac{\int_{\tau_{msg}+\varepsilon}^\infty (\tau - (\tau_{msg}+\varepsilon)) \cdot f(\tau) d\tau}{\int_0^\infty \tau \cdot f(\tau) d\tau} \bigg] \\
&= P_A \cdot e^{-(1-P_A) \cdot (\tau_{msg}+\varepsilon)}
\end{align}
\noindent where $P_A$ is the availability as defined above. In the case where $\lambda_{on}+\lambda_{off}=1$, the $P_{comm}$ can be expressed as,
\begin{align}
\label{eq: P_comm}
    P_{\text{comm}} 
    &= \lambda_{on} \cdot e^{-\lambda_{off} \cdot (\tau_{msg}+\varepsilon)}
\end{align}

\section{Flight Simulator \label{sec: simulation}}
In this section, we briefly introduce the 6DOF flight simulation environment used in this research. Flight dynamics simulation has long served as a key tool for reliability and safety verification of aerospace systems, from early applications such as the Dynamic Flowgraph Methodology demonstrated on the Titan II space launch vehicle digital flight control system \citep{yau_demonstration_1995}, to recent risk-based control system designs for autonomous vehicles verified through formal methods and simulation-based testing \citep{johansen_development_2023}. We adopt the autonomous aerospace verification challenge benchmark, which was released as the basis for analyzing the detailed behavior of autonomous maneuvers for a remotely piloted high performance F-16 Falcon \citep{heidlauf2018verification}. 

The benchmark model is based on the textbook model, where the performance of F-16 has been extensively studied since the 1970s \citep{nguyen1979simulator, seto1999case, stevens2015aircraft}. The benchmark uses the linearized version of the nonlinear flight dynamics of the aircraft about a steady-state trim point (i.e., without gain scheduling), by assuming the autonomous aircraft is operating without sideslip, on a flat earth using the north-east-down frame, and with approximated aerodynamic coefficients. The benchmark model further ignores the leading edge flaps and experimental thrust vectoring control \citep{sonneveldt2006nonlinear}. The aerodynamic coefficients are reasonable approximations for the subsonic flight regime. Euler angle based equations of motion are used, however, they are limited to moderate intensity maneuvers, as they will encounter singularities when the pitch angle $\theta$ exceed $\pm \frac{\pi}{2}$, in which case quaternions should be used \citep{sonneveldt2006nonlinear}. Further explanation of the equations of motion is given in \Cref{subsec: dynamics}. 

For the flight controller, the system adopts a \textit{fly-by-wire} control scheme \citep{favre1994fly}, where the aircraft still takes the four standard inputs but the control surfaces are actuated indirectly through the higher-level controller. The higher-level controller (HLC), or Autopilot, adopts integral tracking control, to generate the reference vector as the input to the lower-level controller (LLC) (e.g., Section 5.5 of \cite{stevens2015aircraft}). The LLC decouples the longitudinal and lateral motions and uses two Linear Quadric Regulators (LQRs) with a low-pass filter. Although LQRs have certain limitations on handling high angle-of-attack scenarios and are not as flexible as Nonlinear Dynamic Inversion (NDI) \citep{yang2024flight}, they were still used in practice by General Dynamics for flight control of the F-16C Fighting Falcon. The specific task we are focusing on is a waypoint following (reaching) task in the NED frame, where the aircraft is expected to reach the current waypoint and then proceed to the next one. The control system adds a performance output to track the given reference inputs and system outputs. This dual-loop controller design uses the digital control implementation of the system-plus-integrator(compensator) augmented state equations to track a command \citep{stevens2015aircraft}. Further explanations are given in \Cref{subsec: controller}.

\subsection{Flight Dynamics of the Simulator \label{subsec: dynamics}}

\begin{figure}[hbt!]
\label{fig: er-dynamics}
\centering
\includegraphics[width=0.9\textwidth]{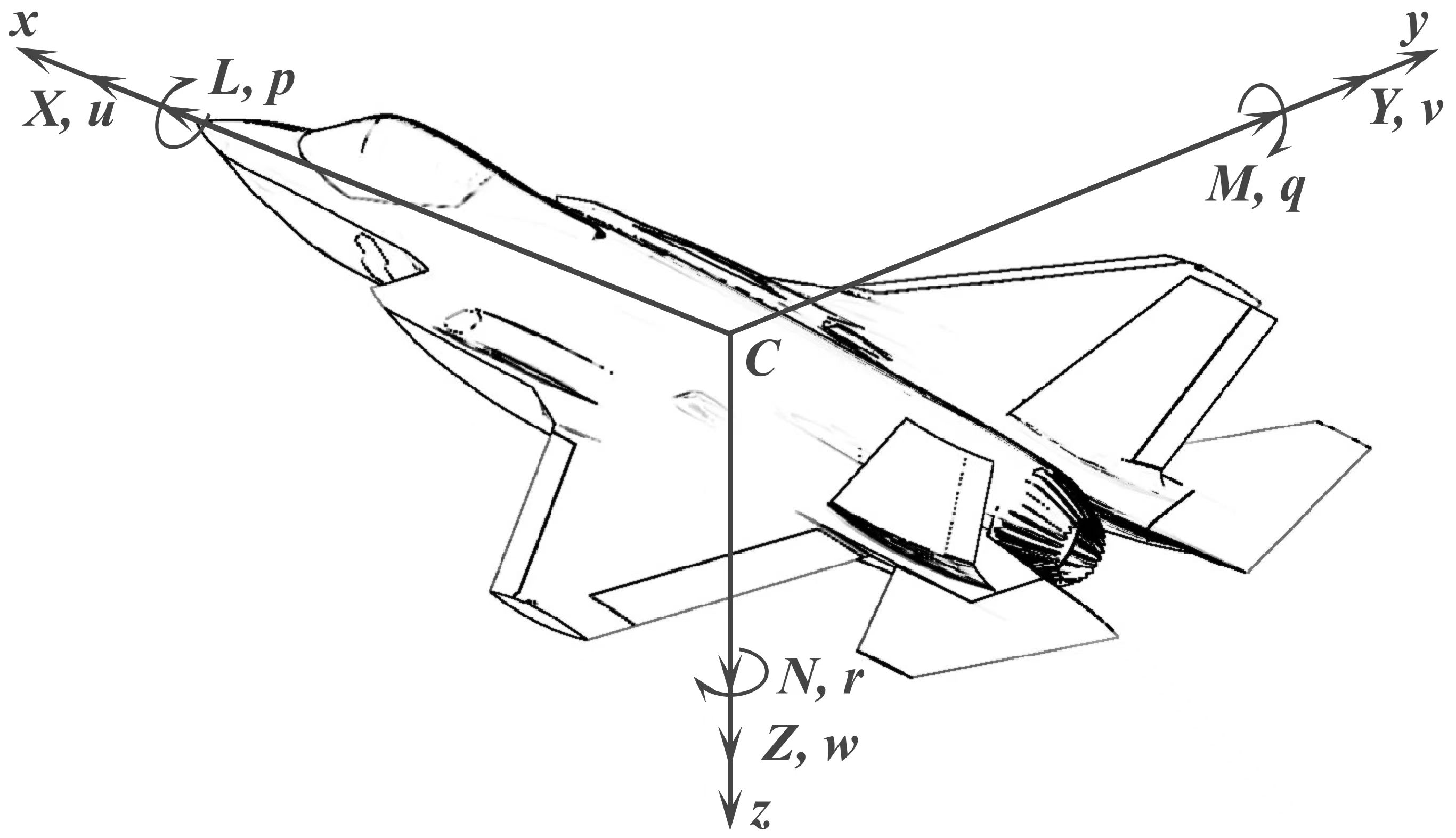}
\caption{Forces and Moments Acting on the Airplane}
\end{figure}

In the context of aircraft dynamics and control, navigation equations are the set of equations that determine the aircraft's position and orientation over time. They relate the aircraft's motion (velocity and angular rates) to changes in its geographic position (latitude, longitude, altitude) and orientation (roll, pitch, yaw angles). These equations are derived from the kinematic relationships between the aircraft's body-fixed frame (attached to the aircraft) and the inertial frame (fixed relative to the Earth). They account for the aircraft's translational and rotational motions to determine how its position and orientation change over time. The translational motion of the aircraft is governed by Newton’s second law,
\begin{equation}
    \mathbf{F} = m \mathbf{a}
\end{equation}

\noindent where $\mathbf{F}$ is the vector of forces acting on the aircraft in the body axis frame, including aerodynamic forces $\mathbf{F_d}$, gravitational forces $\mathbf{F_g}$ and engine thrust forces $\mathbf{F_{th}}$. In the Earth frame, \( \mathbf{F}_{g} \) is given by:
\begin{equation}
    \mathbf{F}_{g, earth} = \begin{bmatrix} 0 \\ 0 \\ mg \end{bmatrix}
\end{equation}

In the body frame, we need to transform the gravitational force using the aircraft's orientation (Euler angles: roll \(\phi\), pitch \(\theta\), yaw \(\psi\)). The gravitational force in body axes becomes:
\begin{equation}
    \mathbf{F}_{g} = m g \begin{bmatrix} -\sin(\theta) \\ \cos(\theta)\sin(\phi) \\ \cos(\theta)\cos(\phi) \end{bmatrix}
\end{equation}

Without considering thrust vectoring, the force vector can be represented by,
\begin{align}
    \mathbf{F} 
    &= \mathbf{F_d} + \mathbf{F_g} + \mathbf{F_{th}} \\
    &= \begin{bmatrix} X \\ Y \\ Z \end{bmatrix} + mg\begin{bmatrix} -\sin(\theta) \\ \cos(\theta)\sin(\phi) \\ \cos(\theta)\cos(\phi) \end{bmatrix} + \begin{bmatrix} T \\ 0 \\ 0 \end{bmatrix}
\end{align}

The acceleration $\mathbf{a} = \begin{bmatrix} \dot{u} & \dot{v} & \dot{w} \end{bmatrix}^T$ can be related to the velocity vector \(\mathbf{V} = \begin{bmatrix} u & v & w \end{bmatrix}^T\) (velocity components in the body frame) and the angular rates \(\mathbf{\omega} = \begin{bmatrix} p & q & r \end{bmatrix}^T\) (roll, pitch, and yaw rates) using the following relation:

\begin{equation}
    \dot{\mathbf{V}} = \mathbf{a} + \mathbf{\omega} \times \mathbf{V}
\end{equation}

Thus, the translational dynamics in matrix form are,
\begin{equation}
    \begin{bmatrix} \dot{u} \\ \dot{v} \\ \dot{w} \end{bmatrix} = \frac{1}{m} \begin{bmatrix} X+T \\ Y \\ Z \end{bmatrix} +g\begin{bmatrix} -\sin(\theta) \\ \cos(\theta)\sin(\phi) \\ \cos(\theta)\cos(\phi) \end{bmatrix} - \begin{bmatrix} 0 & -r & q \\ r & 0 & -p \\ -q & p & 0 \end{bmatrix} \begin{bmatrix}
        u \\ v \\ w
    \end{bmatrix}
\end{equation}

The Earth frame velocities can be expressed as the product of three rotation matrices that describe the transformation from the body frame to the Earth frame. These rotation matrices are based on the Euler angles: roll (\(\phi\)), pitch (\(\theta\)), and yaw (\(\psi\)).

The relationship between the body frame velocities \([u, v, w]^T\) and the Earth frame velocities \([\dot{x}_e, \dot{y}_e, \dot{h}_e]^T\) can be written as the product of three rotation matrices:
\begin{align}
\begin{bmatrix} \dot{x}_e \\ \dot{y}_e \\ \dot{z}_e \end{bmatrix} &= R_y(\psi)\, R_p(\theta)\, R_r(\phi) \begin{bmatrix} u \\ v \\ w \end{bmatrix}
\end{align}
where the product of the three rotation matrices yields the Direction Cosine Matrix (DCM):
\begin{equation}
\resizebox{.9\textwidth}{!}{$
R_y R_p R_r = \begin{bmatrix}
\cos\psi & -\sin\psi & 0 \\
\sin\psi & \cos\psi & 0 \\
0 & 0 & 1
\end{bmatrix} \begin{bmatrix}
\cos\theta & 0 & \sin\theta \\
0 & 1 & 0 \\
-\sin\theta & 0 & \cos\theta
\end{bmatrix}\begin{bmatrix}
1 & 0 & 0 \\
0 & \cos\phi & -\sin\phi \\
0 & \sin\phi & \cos\phi
\end{bmatrix}
$}
\end{equation}
\begin{equation}
= \begin{bmatrix}
\cos\theta \cos\psi & \cos\theta \sin\psi & -\sin\theta \\
\sin\phi \sin\theta \cos\psi - \cos\phi \sin\psi & \sin\phi \sin\theta \sin\psi + \cos\phi \cos\psi & \sin\phi \cos\theta \\
\cos\phi \sin\theta \cos\psi + \sin\phi \sin\psi & \cos\phi \sin\theta \sin\psi - \sin\phi \cos\psi & \cos\phi \cos\theta
\end{bmatrix}
\end{equation}
so that $[\dot{x}_e, \dot{y}_e, \dot{z}_e]^T = R_y R_p R_r\, [u, v, w]^T$.

Rotational dynamics describe how the aircraft rotates about its center of mass in response to external moments. There are two common sets of rotation dynamics equations, (a) kinematic equations relate angular velocities to the rate of change of the Euler angles; (b) moment equations bridge moments to angular accelerations. 

For kinematic equations, the body frame angular velocities (\(p\), \(q\), \(r\)) are related to the time derivatives of the Euler angles (\(\phi\), \(\theta\), \(\psi\)) via the following transformation:

\begin{equation}
\begin{bmatrix} \dot{\phi} \\ \dot{\theta} \\ \dot{\psi} \end{bmatrix} = \begin{bmatrix} 
1 & \sin{\phi} \tan{\theta} & \cos{\phi} \tan{\theta} \\
0 & \cos{\phi} & -\sin{\phi} \\
0 & \sin{\phi}/\cos{\theta} & \cos{\phi}/\cos{\theta}
\end{bmatrix} \begin{bmatrix} p \\ q \\ r \end{bmatrix}
\end{equation}

\noindent where,
\begin{itemize}
    \item \(\dot{\phi}, \dot{\theta}, \dot{\psi}\) are the time derivatives of the Euler angles (roll, pitch, and yaw rates).
    \item \(p, q, r\) are the body angular velocities about the \(x\), \(y\), and \(z\)-axes.
\end{itemize}

In rotational dynamics, the rotational motion of the aircraft is governed by Euler’s rotational equations:
\begin{equation}
\label{eq: euler-rotational-equations}
    \mathbf{M} = \mathbf{I} \dot{\mathbf{\omega}} + \mathbf{\omega} \times (\mathbf{I} \mathbf{\omega})
\end{equation}

\noindent where \(\mathbf{M} = \begin{bmatrix} L & M & N \end{bmatrix}^T\) is the vector of roll, pitch, and yaw moments, \(\mathbf{\omega} = \begin{bmatrix} p & q & r \end{bmatrix}^T\) is the angular velocity in the body frame, \(\dot{\mathbf{\omega}} = \begin{bmatrix} \dot{p} & \dot{q} & \dot{r} \end{bmatrix}^T\) is the angular rates, while $\mathbf{I} = \begin{bmatrix} I_{xx} & -I_{xy} & -I_{xz} \\ -I_{xy} & I_{yy} & -I_{yz} \\ -I_{xz} & -I_{yz} & I_{zz} \end{bmatrix}$ is the inertia of the aircraft. From \Cref{eq: euler-rotational-equations}, the equation of motion for angular velocity $\omega$ is,
\begin{equation}
    \dot{\omega} = \mathbf{I}^{-1}(\mathbf{M} - \omega \times \mathbf{I}\omega)
\end{equation}

This can be simplified as follows,
\begin{align}
    \dot{p} &= (C_1 r+C_2 p)q + C_3 L + C_4 N \\
    \dot{q} &= C_5 p r - C_6 (p^2 - r^2) + C_7 M \\
    \dot{r} &= (C_8 p + C_2 r)q  + C_4 L + C_9 N
\end{align}
\noindent where,
\begin{align}
  C_1 &= \frac{(I_{yy}-I_{zz})I_{zz} - I^2_{xz}}{\Gamma} ,
  C_2 = \frac{I_{xz}(I_{xx}-I_{yy}+I_{zz})}{\Gamma}, 
  C_3 = \frac{I_{zz}}{\Gamma} \\
  C_4 &= \frac{I_{xz}}{\Gamma},
  C_5 = \frac{I_{zz} - I_{xx}}{I_{yy}},
  C_6 = \frac{I_{xz}}{I_{yy}} \\
  C_7 &= \frac{1}{I_{yy}}, 
  C_8 = \frac{(I_{xx}-I_{yy})I_{xx} - I^2_{xz}}{\Gamma},
  C_9 = \frac{I_{xx}}{\Gamma} \\
  \Gamma &= I_{xx}I_{zz} - I^2_{xz}
\end{align}

Specifically, the aerodynamic force vector $\mathbf{F_d}$ and aerodynamic moment vector $\mathbf{M}$ are computed by,

\begin{align}
\mathbf{F_d} &= \begin{bmatrix} X \\ Y \\ Z \end{bmatrix} = \Bar{q} S 
    \begin{bmatrix}
        C_{XT}(\alpha, \beta, p, q, r, \delta_e, \delta_a, \delta_r) \\
        C_{YT}(\alpha, \beta, p, q, r, \delta_e, \delta_a, \delta_r) \\
        C_{ZT}(\alpha, \beta, p, q, r, \delta_e, \delta_a, \delta_r) 
    \end{bmatrix}\\
    \mathbf{M} &= \begin{bmatrix} L \\ M \\ N \end{bmatrix} = \Bar{q} S 
    \begin{bmatrix}
        b C_{LT}(\alpha, \beta, p, q, r, \delta_e, \delta_a, \delta_r)\\
        \Bar{c} C_{MT}(\alpha, \beta, p, q, r, \delta_e, \delta_a, \delta_r)\\
        b C_{NT}(\alpha, \beta, p, q, r, \delta_e, \delta_a, \delta_r)
    \end{bmatrix}
\end{align}


\noindent where the total aerodynamic force coefficients $C_{XT}, C_{YT}, C_{ZT}$ and total moment coefficients $C_{LT}, C_{MT}, C_{NT}$ are usually obtained from wind-tunnel tests of a scaled aircraft model. In this case, we use the aerodynamic coefficients provided in Appendix A.8 of \cite{stevens2015aircraft} with polynomial interpolations from \citep{morelli1998global}. The detailed calculation of the coefficients is provided in \cite{sonneveldt2006nonlinear}. It worth noting that leading edge flaps (e.g., $\delta_F$) are not considered as in \cite{yang2024flight}, resulting in a low-fidelity aircraft model used for demonstration and verification purposes as in \cite{stevens2015aircraft}.

These 12 equations of motion describe the aircraft’s position, velocity, and orientation in typical 6DOF flight simulation models. In this research, we have the 13th additional equation of motion that models the thrust lag for a turbojet engine. The state variable associated with this equation is the actual engine thrust \( \delta_t(t) \). The thrust lag function refers to the dynamics of how the aircraft's thrust changes in response to throttle inputs from the pilot. This is due to the physical inertia of the engine and delays in the fuel control system. The thrust produced by an engine doesn't change instantaneously when the throttle is adjusted, as it takes a time for the engine to spool up (increase thrust) and spool down (decrease thrust). This delay is often modeled as a first-order lag system of the form,

\begin{equation}
    \dot{\delta_t} = \frac{1}{\kappa} \left( \delta_t^{\text{cmd}} - \delta_t \right)
\end{equation}

\noindent where \( \delta_t^{\text{cmd}}\) is the commanded thrust (i.e., the thrust the pilot intends based on the throttle position). \( \kappa \) is the time constant of the thrust lag, representing how quickly the engine can respond to changes in throttle position. A small \( \tau \) means a fast response (minimal lag), while a larger \( \tau \) means a slower response (more lag). \( \dot{\delta_t}\) is the time derivative of the thrust, or the rate of change of thrust.

This completes the equations of motion for the aircraft position, velocity, orientation, and engine response. Higher-fidelity F-16 flight simulation models may consider leading edge flaps and coupling between longitudinal and lateral dynamics \citep{sonneveldt2006nonlinear}.

\subsection{Controller Design \label{subsec: controller}}

We are interested in a waypoint following task where the flight controller follows a reference command signal provided by an autopilot. The control system adopts the design called system-plus-integrator(compensator) with augmented states of the following form,

\begin{align}
    \dot{\mathbf{x}}(t) &= A \cdot\mathbf{x}(t) + B\cdot\mathbf{u}(t) + G\cdot\mathbf{r}(t) \\
    \mathbf{y}(t) &= C\cdot\mathbf{x}(t) + F\cdot\mathbf{r}(t) \\
    \mathbf{z}(t) &= H\cdot\mathbf{x}(t)
\end{align}

\noindent and the control outputs of the form,
\begin{equation}
\label{eq: control_low}
    \mathbf{u}(t) = -K \cdot \mathbf{y}(t)
\end{equation}

\noindent with $\mathbf{x}(t) \in \mathbb{R}^n$ the state variables, $\mathbf{u}(t) \in \mathbb{R}^m$ the control inputs, and $\mathbf{y}(t) \in \mathbb{R}^p$ the measured outputs. The optimal feedback control gain matrix $K \in \mathbb{R}^{m \times p}$ is determined by the design procedure. If the objective is to guarantee stability by driving any initial condition errors to zero, then the optimal control input $\mathbf{u}(t)$ to optimize the quadratic cost of,

\begin{equation}
\min \mathcal{J} = \frac{1}{2} \int_{0}^{\infty} \left( \mathbf{x}(t)^T Q \mathbf{x}(t) + \mathbf{u}(t)^T R \mathbf{u}(t) \right) dt
\end{equation}

\noindent is desired, where $Q$ and $R$ are penalty metrics in symmetric positive-semidefinite forms (i.e., $\mathcal{J} \geq 0$). The optimal solution can be solved with Riccati equation, or gradient-based approach when an analytical solution is infeasible \citep{choi1999lqr, ashraf2018design, priyambodo2020model}. In the context of reliability-constrained control, LQR has also been extended to incorporate interval uncertainty and nonprobabilistic time-dependent reliability constraints for satellite attitude-vibration control \citep{yang_reliability-constrained_2023}.

Moreover, this flight simulator adopts a cascading control scheme where an outer loop (Autopilot) responsible for high-level tasks (i.e, waypoint navigation), and an inner loop (LLC) ensures precise tracking of the Autopilot's reference commands $\mathbf{r}(t)$. The Autopilot computes $\mathbf{r}(t)$ based on the aircraft's current state and the desired waypoints, providing heading adjustment, altitude changes, and airspeed maintenance. The LLC uses an LQR controller with integral augmentation to compute control inputs that minimize tracking errors and ensure robustness. Lastly, elements of the feedback gain matrix $K$ coupling lateral and longitudinal states are regulated to zero, such that the coupling between certain input-output pairs are eliminated (i.e., decoupled lateral and longitudinal motion controls). This is also known as the constrained output feedback design. The detailed derivation of the compensator dynamics with augmented states can be found from Section 5.4 to Section 5.7 of \cite{stevens2015aircraft}.

\subsubsection{Higher-Level Control \label{subsubsec: hlc}}
Higher-level control is referred to as Autopilot in the benchmark. Autopilot provides \textit{pilot relief} such that the pilot does not need to pay continuous attention to controlling the aircraft, especially when the aircraft is designed slightly unstable \citep{stevens2015aircraft}. Autopilot provides automatic control systems to handle certain pilot relief functions (i.e., altitude holding) and special procedure functions (i.e., automatic landing). The Autopilot uses Proportional-Derivative (PD) controllers to compute high-level reference commands for the aircraft's heading, roll angle, altitude, and airspeed. These reference commands are then passed to the Low-Level Controller (LLC), which handles the precise tracking of these commands using an LQR with integral action. Autopilot has four control objectives, (i) heading control; (ii) roll angle control; (iii) altitude control; (iv) airspeed control.

For our waypoint following task, we have the waypoint autopilot, which uses geometrically derived guidance based on the kinematic relationships \citep{lee2010integrated, jensen2011waypoint, he2019minimum} fr RPAS navigation. Given the current waypoint $\text{WP} = (x_{wp}, y_{wp}, z_{wp})$, and the current aircraft states $(x_e, y_e, z_e)$, all in the NED frame, we have the distance to the waypoint as,
\begin{align}
    \Delta_x = x_{wp} - x_e, \quad \Delta_y = y_{wp} - y_e, \quad \Delta_z = z_{wp} - z_e.
\end{align}

The desired heading $\psi^{\text{cmd}}$ to the waypoint $WP$ is computed by,
\begin{equation}
    \psi^{\text{cmd}} = \frac{\pi}{2} - \tan^{-1}\left(\frac{\Delta_y}{\Delta_x}\right)
\end{equation}
\noindent where the heading $\psi=0$ is aligned along the x-axis.

\textbf{Heading control} uses variables $\psi, \psi^{\text{cmd}}, r$ where $\psi$ and $\psi^{\text{cmd}}$ are the aircraft current and reference yaw angles respectively, and $\phi^{\text{cmd}}$ is the reference roll angle provided by heading control. In heading control, $e_\psi = \psi^{\text{cmd}} - \psi$ computes the heading error between the desired heading to the waypoint and the current heading where the PD controller is applied to the heading error to calculate the desired roll angle $\phi^{\text{cmd}}$ that steers the aircraft towards the waypoint as,
\begin{equation}
    \phi^{\text{cmd}} = k_{\psi_p} e_\psi - k_{\psi_d} r = k_{\psi_p} (\psi^{\text{cmd}} - \psi) - k_{\psi_d} r
\end{equation}

\noindent where $k_{\psi_p}$ is the proportional gain for heading error, and $k_{\psi_d}$ is the derivative gain for yaw rate damping. The derivative term uses the yaw rate $r$ to provide damping, preventing overshoot and improving stability. The desired roll angle is limited to ensure the aircraft does not exceed safe bank angles. 

\textbf{Roll angle control} considers $\phi, \phi^{\text{cmd}}, p$ by calculating the error between the current and desired roll angle $e_\phi = \phi^{\text{cmd}} - \phi$ with PD control,
\begin{equation}
    p_s^{\text{cmd}} = k_{\phi_p} e_\phi - k_{\phi_d} p = k_{\phi_p} (\phi^{\text{cmd}} - \phi) - k_{\phi_d} p
\end{equation}
\noindent where $k_{\phi_p}$ is the proportional gain for roll angle error, and $k_{\phi_d}$ is the derivative gain for roll rate damping. This PD control calculates the desired roll rate in the stability axis $p_s^{\text{cmd}}$ needed to achieve the desired roll angle, as the reference for controlling aileron deflection. 

\textbf{Altitude control} calculates the desired normal load factor $N^{\text{cmd}}_z$ needed to achieve the desired altitude, with a PD control on the altitude error $e_{z_e}$ and rate of climb/descent $\dot{z_e}$ with,
\begin{equation}
    N^{\text{cmd}}_z = k_{z_p} e_{z_e} - k_{z_d} \dot{z_e} = k_{z_p} ({z_e}_{\text{cmd}} - z_e) - k_{z_d} \dot{z_e}
\end{equation}
\noindent where $e_{z_e} = {z_e}_{\text{cmd}} - z_e$ is the altitude error and $\dot{z_e} = V_t \sin(\gamma)$ is the projection of airspeed $V_t$ onto the flight path angle $\gamma$. Similarily, proportional gain is applied to the altitude error, and a derivative gain is added to $\dot{z_e}$.

\textbf{Airspeed control} is the proportional control (P-Control) for the throttle command,
\begin{equation}
    \delta_t^{\text{cmd}} = k_{V_t} e_{V_t} = k_{V_t} (V_t^{\text{cmd}} - V_t)
\end{equation}
which controls the airspeed error $e_{V_t}$ between the desired airspeed ($V_t^{\text{cmd}}$) and the current airspeed ($V_t$). $\delta_t^{\text{cmd}}$ directly adjusts the engine thrust output. 

Autopilot outputs the reference vector $\mathbf{r} = [N_z^{\text{cmd}}, p^{\text{cmd}}, N_{yr}^{\text{cmd}}, \delta_t^{\text{cmd}}]$ to the lower level controller.  The Autopilot sets $N_{yr}^{\text{cmd}} = 0$ in standard maneuvers by default to avoid introducing sideslip or yawing motion, while the lower level control can adjust the rudder deflection $\delta_r$ to maintain coordinated flight.

\subsubsection{Lower-Level Control}
LLC handles the precise tracking of reference commands by computing the necessary control surface deflections and throttle settings, to ensure that the aircraft's response closely follows the reference commands. LLC operates by a Linear Quadratic Regulator (LQR) with integral actions for integral tracking control. LQR with integral tracking control can be achieved by adding integral state errors, resulting in the augmented states. Furthermore, in digital control implementations, the control of lateral and longitudinal motions is separated into two LQR controllers. 



For longitudinal motion, the short-period approximation to the aircraft dynamics is linearized around the nominal flight condition, as referenced in Table 3.6-3 of \citep{stevens2015aircraft}. The states of interests are $\mathbf{x}_{\text{long}} = [\alpha, q, \int e_{N_z} dt]^T$, and $\mathbf{y}_{\text{long}} = [\alpha, q, N_z]^T$, where $\int e_{N_z} dt$ is the integral of the normal acceleration error. $e_{N_z} = N^{ref}_z - N_z$ is the difference between the reference normal acceleration from the Autopilot and the measured normal acceleration. The control input for longitudinal motion is the elevator deflection in radians $\mathbf{u}_{\text{long}} = [\delta_e]$. 

For lateral motion, a separated LQR is introduced with the state variables $\mathbf{x}_{\text{lat}} = [\beta, p, r, \int e_\psi dt, \int e_{N_{yr}} dt]^T$, and $\mathbf{y}_{\text{lat}} = [\beta, p, r, \psi, N_{yr}]^T$ with control inputs $\mathbf{u}_{\text{lat}} = [\delta_a, \delta_r]^T$. Similar to the longitudinal controller, the lateral controller uses state feedback combined with error integrals to ensure accurate tracking and disturbance rejection. The control inputs are computed as a linear combination of the current states and the integral of the error states. 

The LQR controller ensures optimal performance for the longitudinal dynamics of the F-16 aircraft, focusing on minimizing steady-state errors in normal acceleration while considering control effort. The control law for $\mathbf{u}_{\text{long}}$ and $\mathbf{u}_{\text{lat}}$ is thus given as \Cref{eq: control_low}, with corresponding control gain matrices $K_{\text{long}}$ and $K_{\text{lat}}$. 

\begin{figure}[H]
\centering
\includegraphics[width=0.8\textwidth]{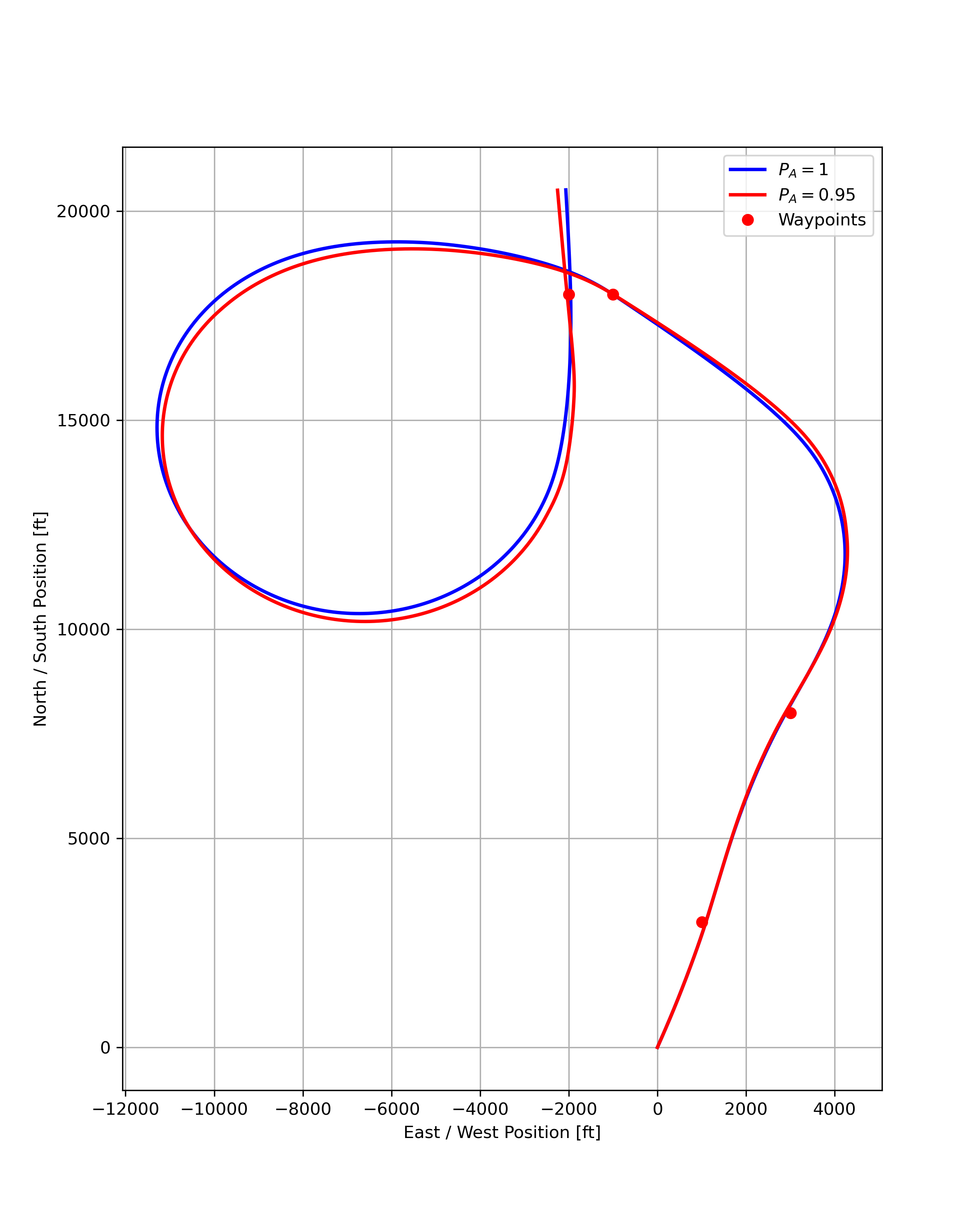}
\caption{Aircraft trajectory in the vertical plane when comparing $P_A=1$ and $P_A=0.95$ with no latency in both cases. For a given $P_A=0.95$, we first generate the signal state masks square wave for the entire simulation duration as in \Cref{fig: compare_controlstates}. Then, apply to the mask to $\mathbf{u}(t)$ to inference the state derivatives. }
\label{fig: compare_trajectory}
\end{figure}

\begin{figure}[H]
\centering
\includegraphics[width=\textwidth,height=10cm]{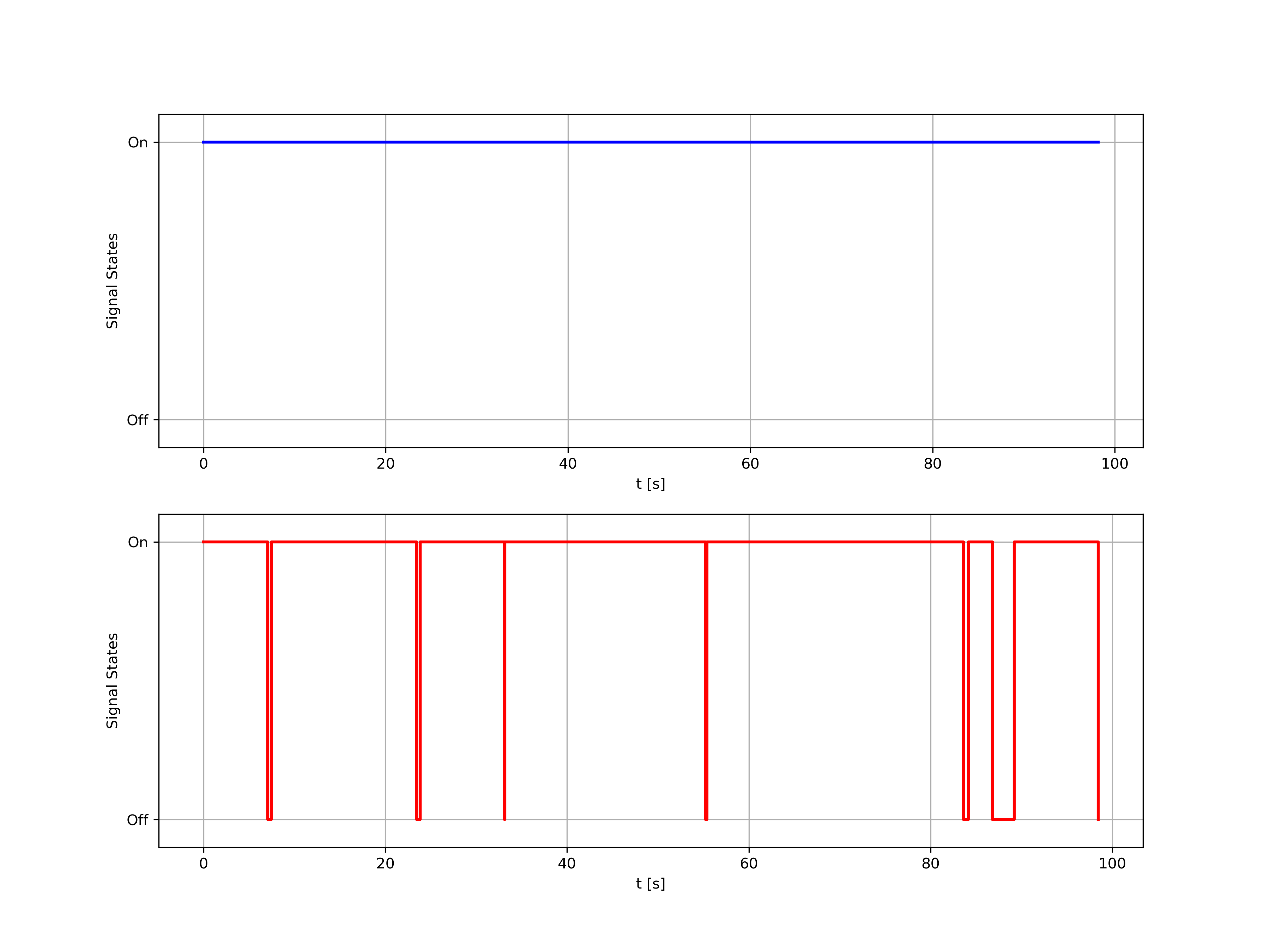}
\caption{Control signal states during simulation.}
\label{fig: compare_controlstates}
\end{figure}

\begin{figure}[H]
    \centering
    \begin{subfigure}[t]{0.9\textwidth}
        \centering
        \includegraphics[width=\textwidth,height=4cm]{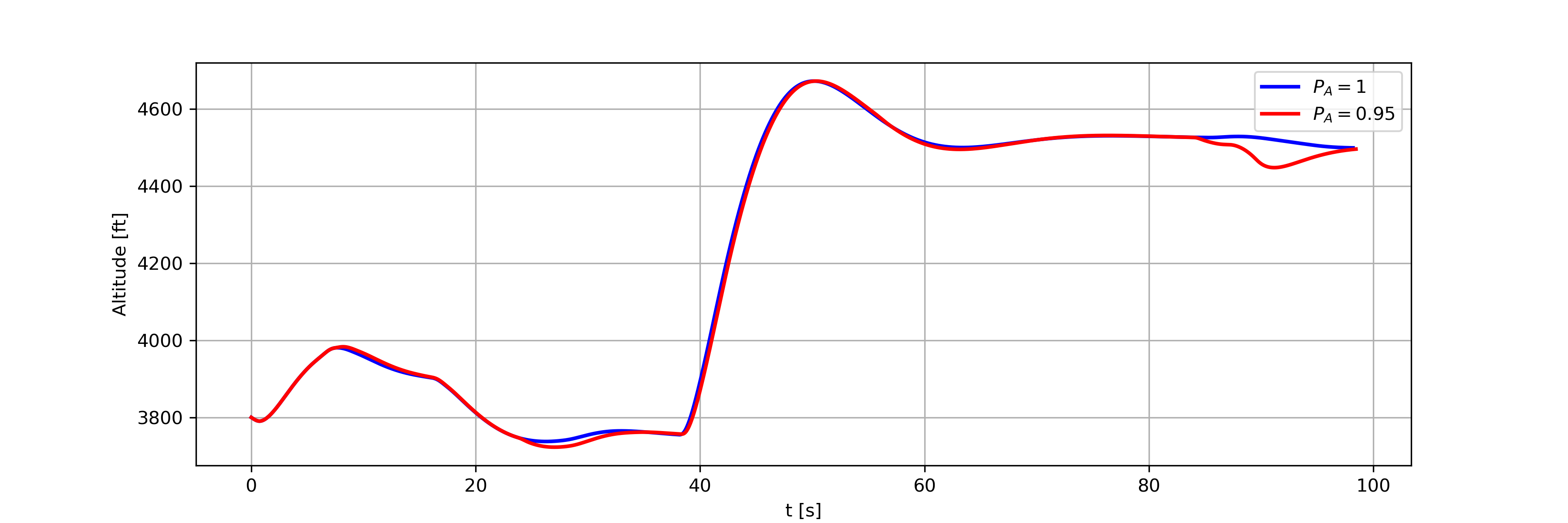}
        \caption{Altitude along trajectory.}
        \label{fig: compare_alt}
    \end{subfigure}
    ~
    \begin{subfigure}[t]{0.9\textwidth}
        \centering
        \includegraphics[width=\textwidth,height=4cm]{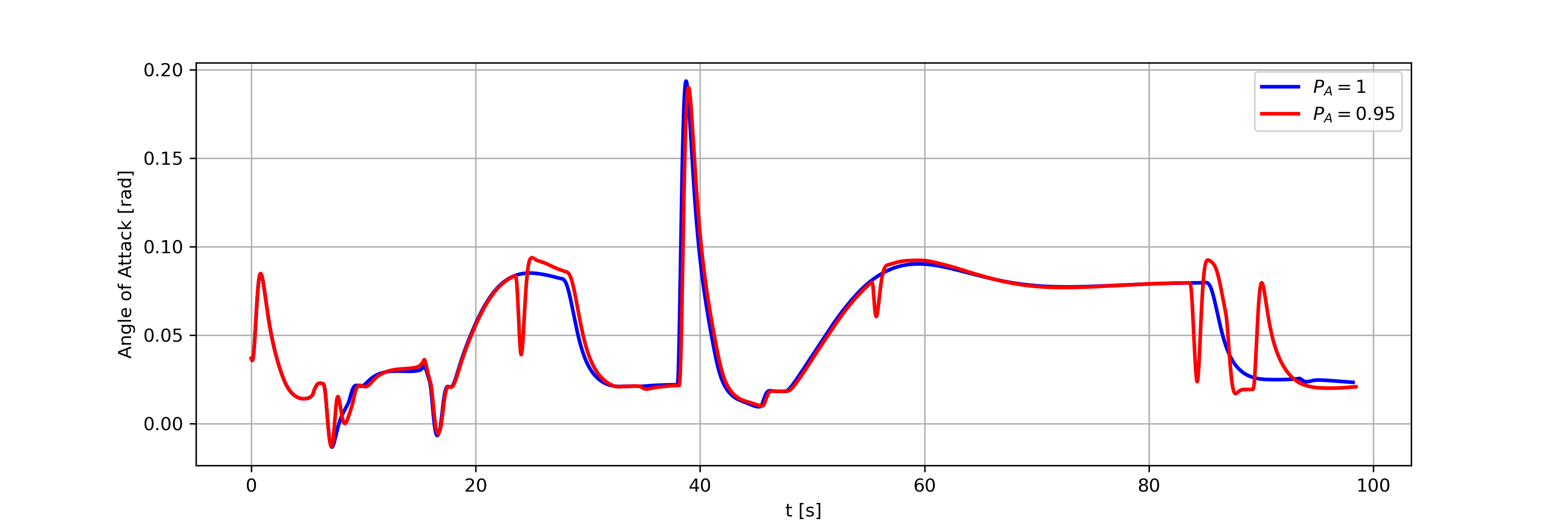}
        \caption{Angle of Attack $\alpha$ versus simulation time.}
        \label{fig: compare_aoa}
    \end{subfigure}
    ~
    \begin{subfigure}[t]{0.9\textwidth}
        \centering
        \includegraphics[width=\textwidth,height=4cm]{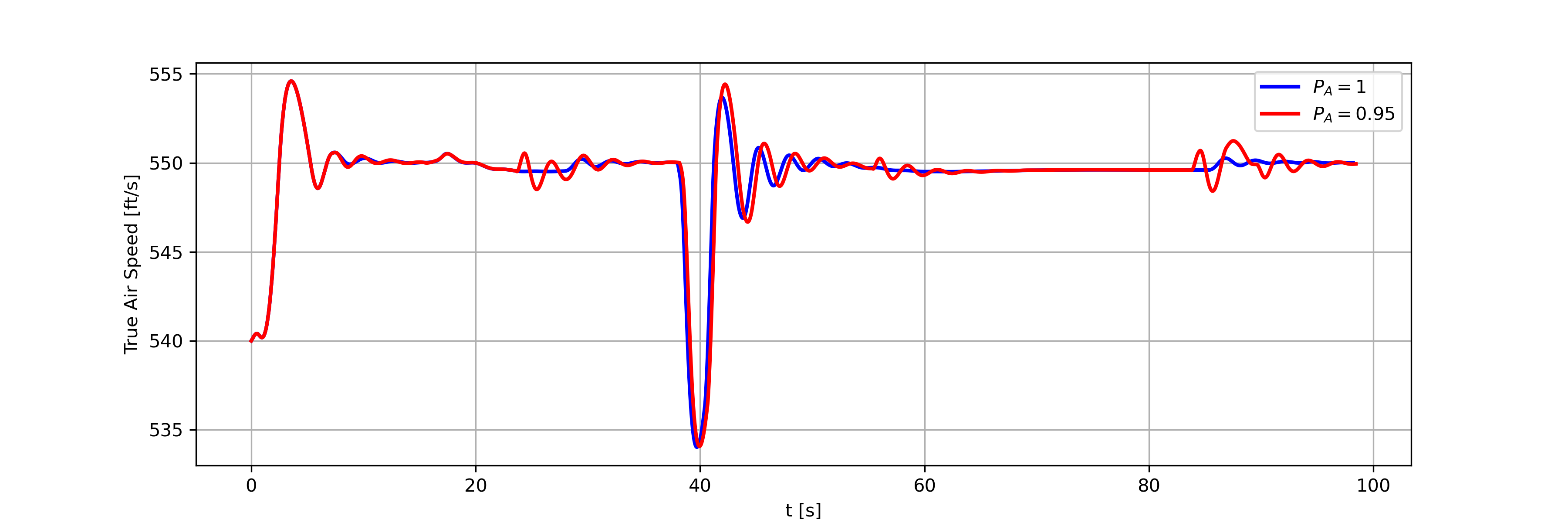}
        \caption{Airspeed $V_t$ along trajectory.}
        \label{fig: compare_vt}
    \end{subfigure}
    \caption{Altitude, angle of attack, and airspeed under different communication availabilities during the waypoint reaching simulation. }
    \label{fig: compare-alt-aoa-vt}
\end{figure}

\begin{figure}[H]
    \centering
    \begin{subfigure}[t]{0.9\textwidth}
        \centering
        \includegraphics[width=\textwidth,height=4cm]{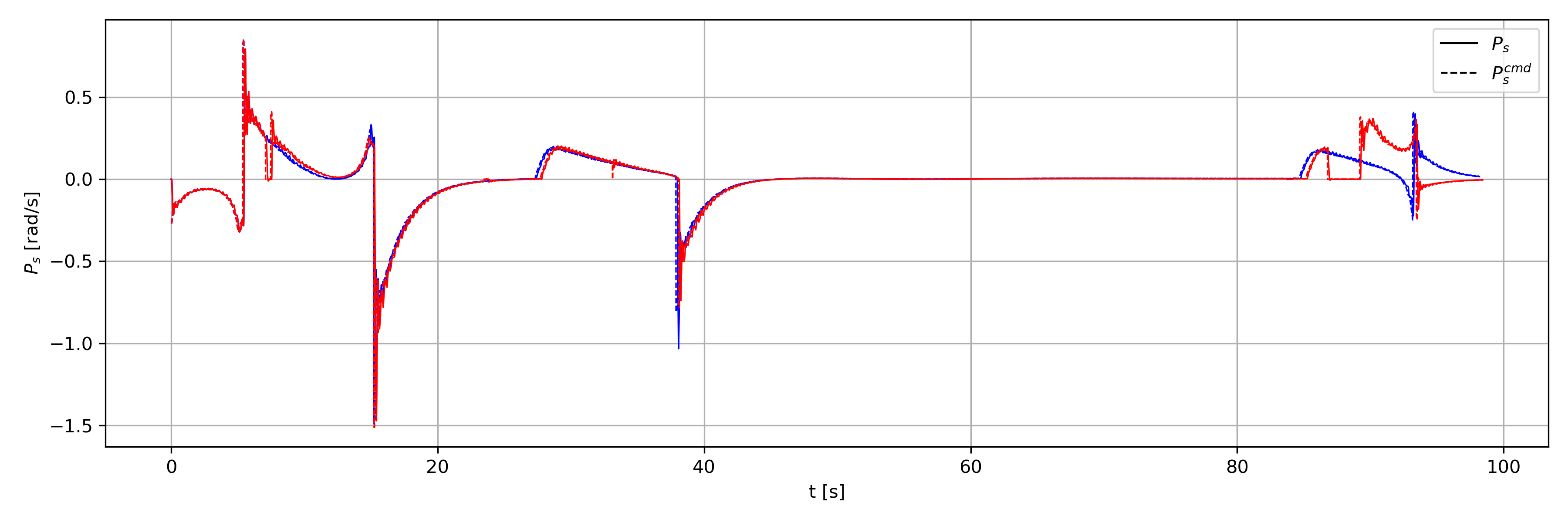}
        \caption{Stability roll rate $p_s$ along trajectory .}
        \label{fig: compare_ps}
    \end{subfigure}
    ~
    \begin{subfigure}[t]{0.9\textwidth}
        \centering
        \includegraphics[width=\textwidth,height=4cm]{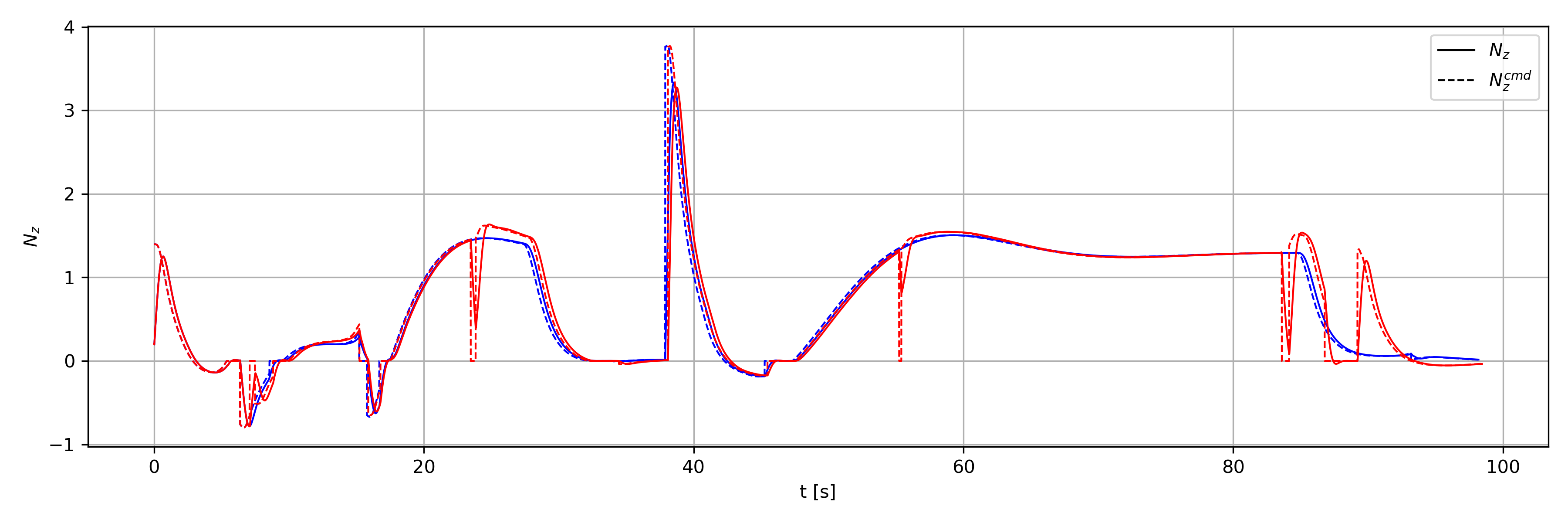}
        \caption{Normalized load factor $N_z$ along trajectory .}
        \label{fig: compare_nz}
    \end{subfigure}
    ~
    \begin{subfigure}[t]{0.9\textwidth}
        \centering
        \includegraphics[width=\textwidth,height=4cm]{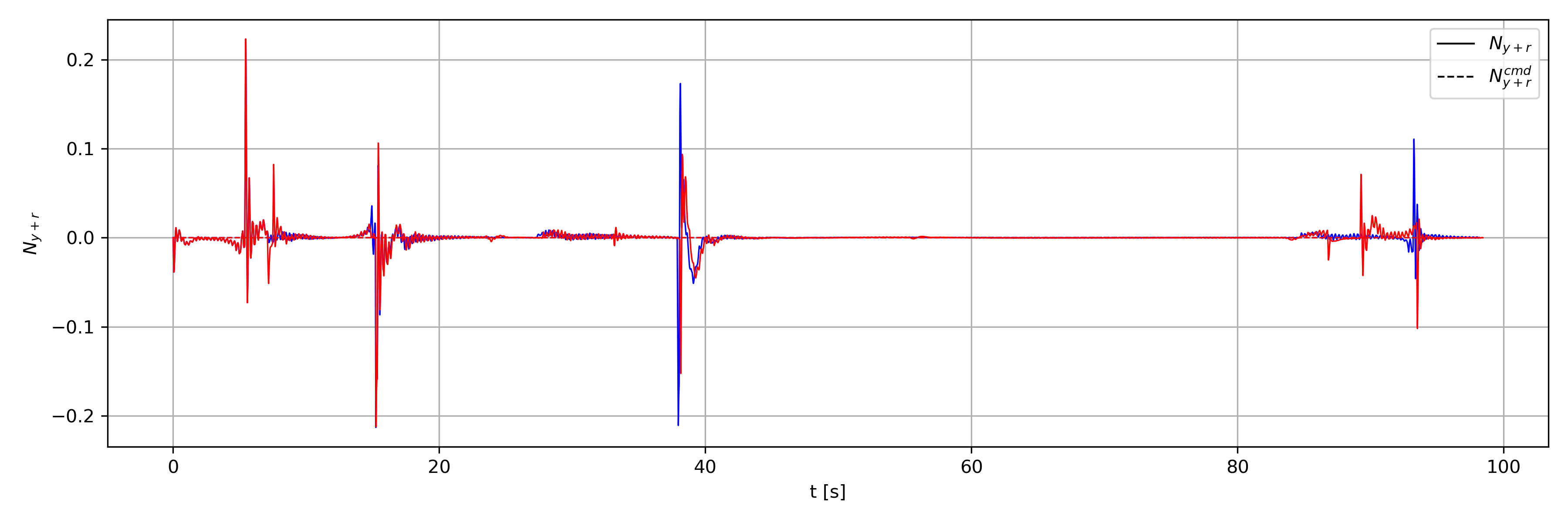}
        \caption{Sum of the side acceleration and yaw rate $N_{y}+r$ along trajectory.}
        \label{fig: compare_nyr}
    \end{subfigure}
    \caption{Higher-level control variables under different communication availabilities during the waypoint reaching simulation. }
    \label{fig: compare-hlc}
\end{figure}

\begin{figure}[H]
    \centering
    \begin{subfigure}[t]{0.9\textwidth}
        \centering
        \includegraphics[width=\textwidth,height=4cm]{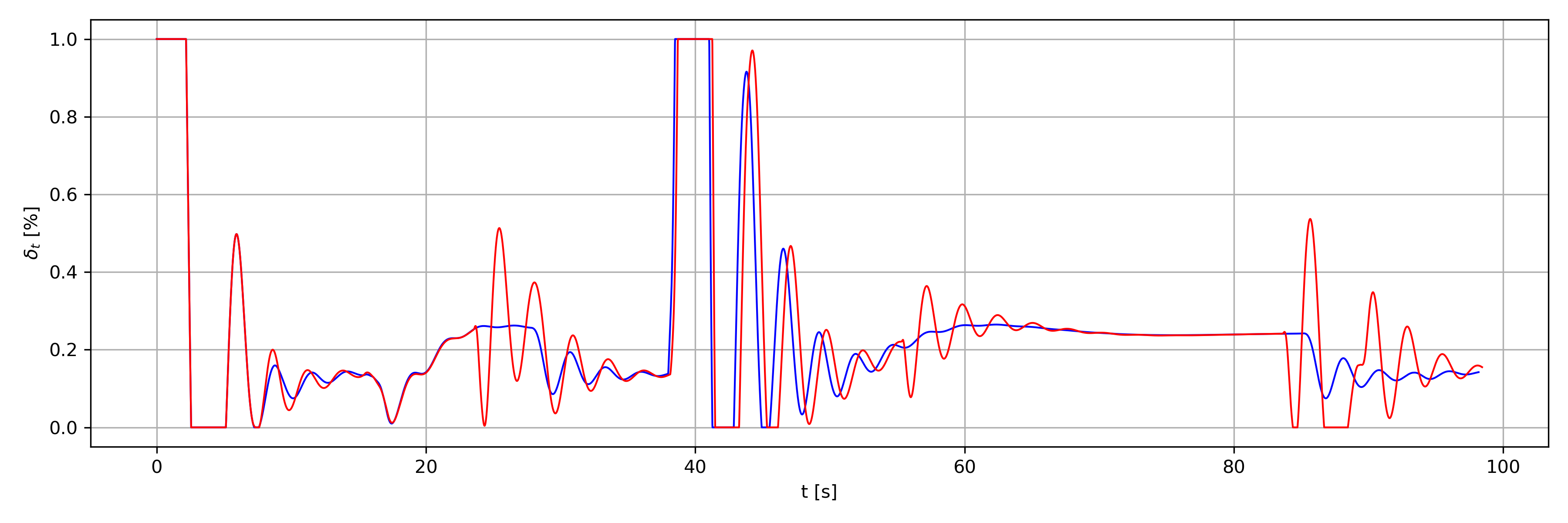}
        \caption{Throttle power command $\delta_t$ along trajectory.}
        \label{fig: compare_dt}
    \end{subfigure}
    ~
    \begin{subfigure}[t]{0.9\textwidth}
        \centering
        \includegraphics[width=\textwidth,height=4cm]{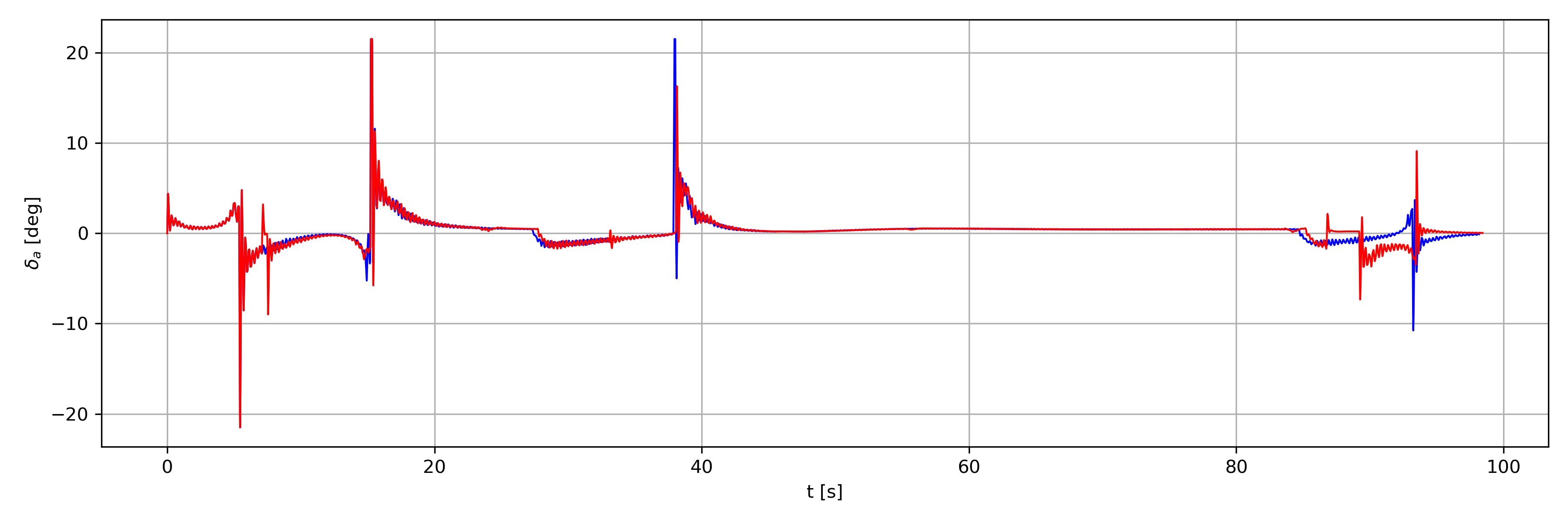}
        \caption{The aileron surface deflection $\delta_a$ along trajectory.}
        \label{fig: compare_da}
    \end{subfigure}
    ~
    \begin{subfigure}[t]{0.9\textwidth}
        \centering
        \includegraphics[width=\textwidth,height=4cm]{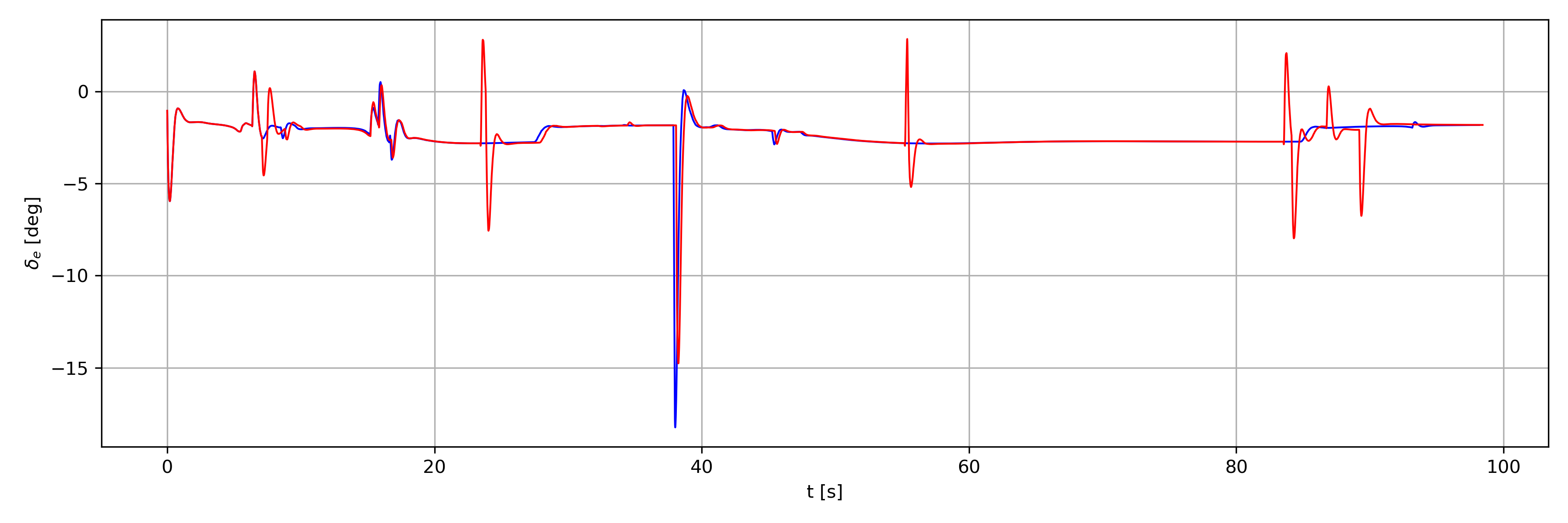}
        \caption{The elevator surface deflection $\delta_e$ along trajectory.}
        \label{fig: compare_de}
    \end{subfigure}
    ~
    \begin{subfigure}[t]{0.9\textwidth}
        \centering
        \includegraphics[width=\textwidth,height=4cm]{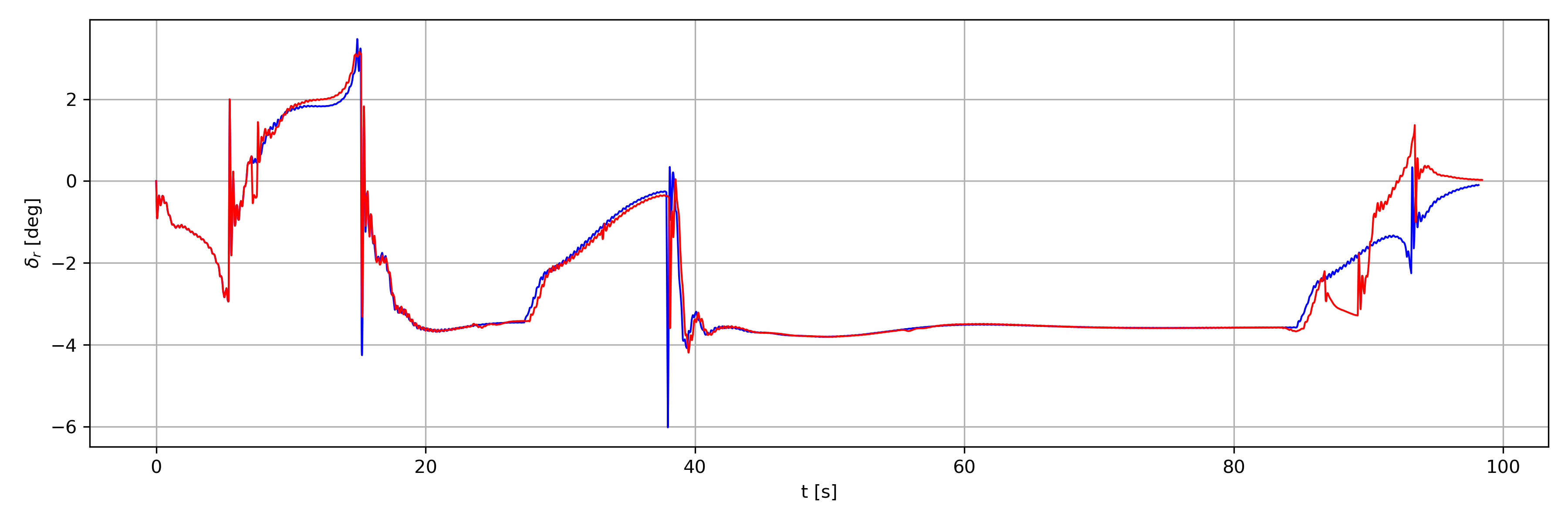}
        \caption{The rudder surface deflection $\delta_a$ along trajectory.}
        \label{fig: compare_dr}
    \end{subfigure}
    \caption{Lower-level control variables under different communication availabilities during the waypoint reaching simulation. }
    \label{fig: compare-llc}
\end{figure}

Following the formulation in \Cref{subsec: rcp-formulation}, we simulate communication C2 link loss with CTMC between the \textit{on} and \textit{off} states \Cref{fig: compare_controlstates}. The overhead aircraft trajectory from two simulation runs with $P_A=1$ and $P_A=0.95$ is also shown \Cref{fig: compare_trajectory}, along with a comparison of higher-level control inputs (\Cref{fig: compare-hlc}) and lower-level control inputs (\Cref{fig: compare-llc}). The mission is defined as a waypoint following task, where the RPAS is commanded to reach a minimum slant range to each waypoint before proceeding to the next. The RPAS starts from the origin, while the airspace of interest is $[-12,000, 4,000]$ feet on the east/west direction, and $[0, 20,000]$ feet to the north/south direction. The altitude range is around $[3,800, 4,600]$ feet with Vt between $[535, 555]$ feet/second (\Cref{fig: compare-alt-aoa-vt}). We choose to regulate first three higher-level reference values $\mathbf{r} = [N_z^{\text{cmd}}, p^{\text{cmd}}, N_{yr}^{\text{cmd}}, \delta_t^{\text{cmd}}]$ to $0$ once the C2 link is lost. The objective of regulating these reference values to zero is to make sure the F-16 can return to a steady straight and level flight once the RP connection is lost.

\section{Monte Carlo Simulations \label{sec: experiments}}
\begin{figure}[hbt!]
\label{fig: flowchart}
\centering
\includegraphics[width=0.95\textwidth]{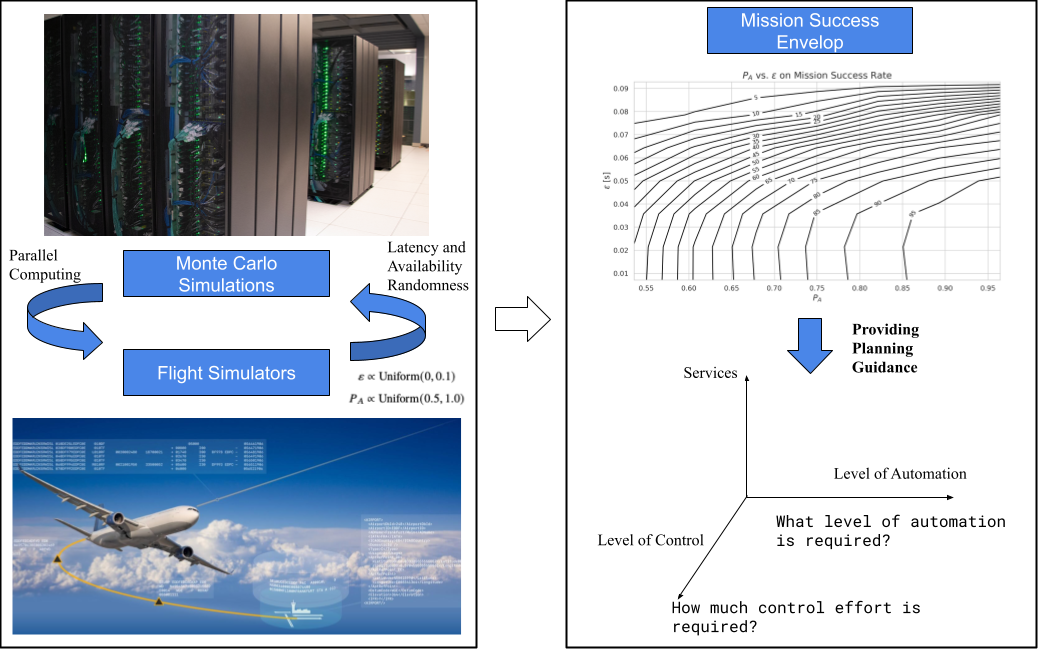}
\caption{Flowchart of the experiment. }
\end{figure}

In this section, we show the details of our Monte Carlo simulation experiments. We start by introducing the definition of the mission in this study. Then, we explain the steps to include communication loss and latency to the RPAS simulation. We further give necessary technical details on the Monte Carlo simulations to generate the communicability performance envelopes -- the key concept we intend to introduce with this study.  As a reference. \Cref{fig: flowchart} shows the flowchart of the procedures. 

\subsection{Definition of Mission Success \label{subsec: definition}}

\begin{figure}[H]
    \centering
    \begin{subfigure}[t]{0.45\textwidth}
        \centering
        \includegraphics[width=\textwidth,height=6cm]{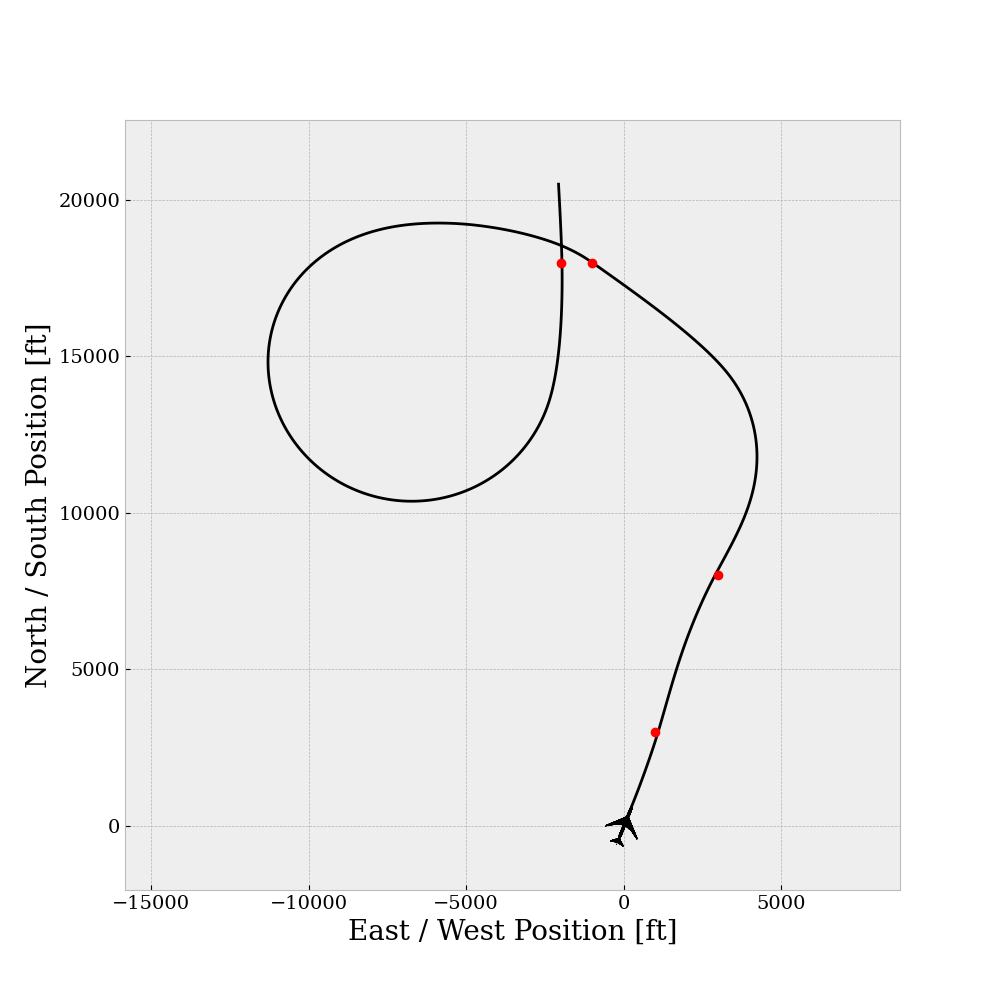}
        \caption{Scenario 1.}
        \label{fig: trajectory-wp1}
    \end{subfigure}
    \begin{subfigure}[t]{0.45\textwidth}
        \centering
        \includegraphics[width=\textwidth,height=6cm]{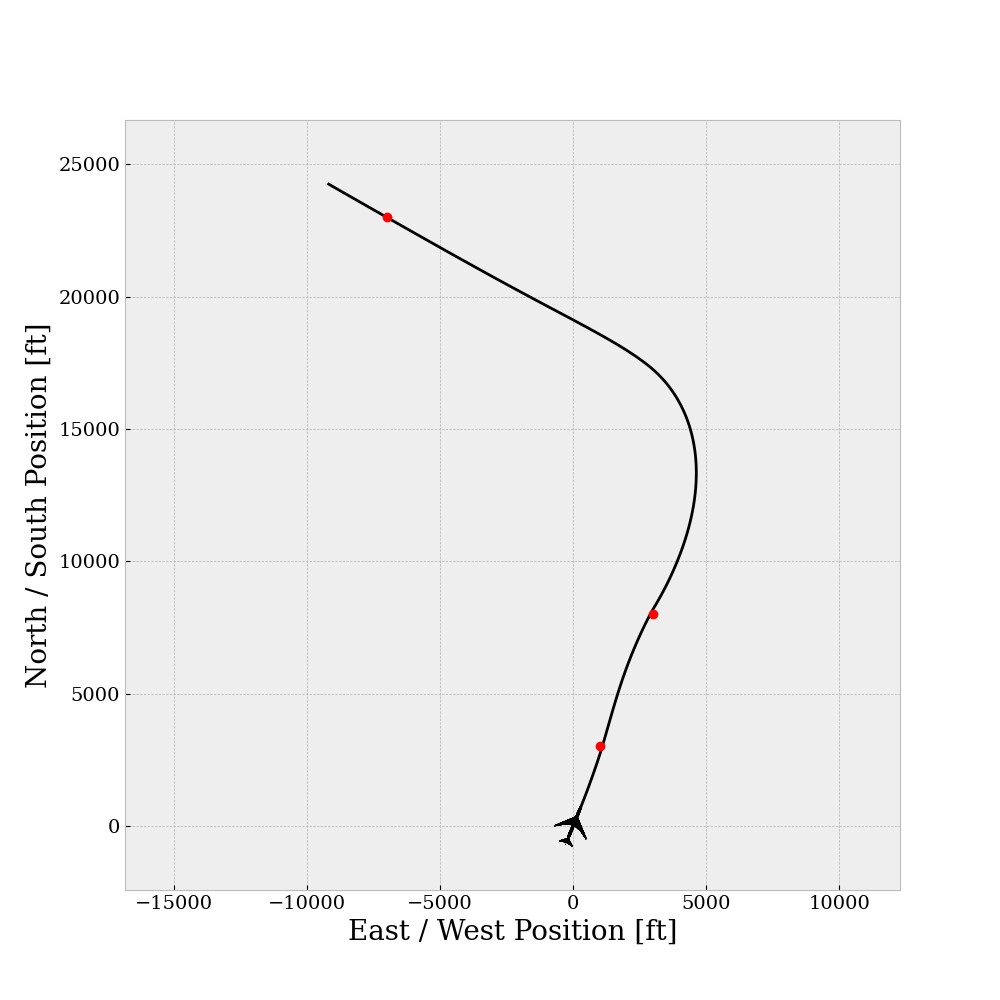}
        \caption{Scenario 2.}
        \label{fig: trajectory-wp2}
    \end{subfigure}
    \caption{Flight trajectory in the vertical plane for two scenarios used in this work. The first scenario is considered the hard case with 4 waypoints and requires a minimum radius turn of the RPAS. The second scenario is the simple case with three waypoints.}
    \label{fig: trajectory-two}
\end{figure}

The mission is defined as the waypoint following task in the 3D airspace. Following the notations and derivations used for HLC in \Cref{subsubsec: hlc}, we further compute the slant range $R$ by the $L^2$ distance between the current waypoint and the current position (e.g., $R = \sqrt{\Delta_x^2 + \Delta_y^2 + \Delta_z^2}$). The simulation checks the slant range to each current waypoint, and switches to the next waypoint only if the condition $R < R_{\text{threshold}}$ is met. Once all waypoints are reached, the simulation changes the mode to \textit{Done}, and sets $\psi^{\text{cmd}}$ to $0$ in HLC to maintain a level flight.  Mission completion is defined as the RPAS reaches all given waypoint in the current scenario with a \textit{Done} status from the simulator. Specifically, we select two waypoint scenarios to represent a hard case and a simple case as in \Cref{fig: trajectory-two}. \Cref{fig: trajectory-wp1} is considered the hard case with 4 waypoints and requires a minimum radius turn of RPAS, while \Cref{fig: trajectory-wp2} is the simple scenario.

\subsection{Mission Success Envelope \label{subsec: envelop}}
\begin{table}[h!]
\centering
\resizebox{\textwidth}{!}{%
\begin{tabular}{c|cc|cc|c}
\hline
\multirow{2}{*}{} & \multicolumn{2}{c|}{$P_A$}                   & \multicolumn{2}{c|}{$\varepsilon$} & \multirow{2}{*}{Number of Simulations} \\ \cline{2-5}
                  & \multicolumn{1}{c|}{Distribution}    & Step & \multicolumn{1}{c|}{Distribution}      & Step   &                                        \\ \hline
Scenario 1        & \multicolumn{1}{c|}{Uniform(0.5, 1)} & 1e-3 & \multicolumn{1}{c|}{Uniform(0, 0.1)}   & 1e-3   & 1,124,413                              \\ \hline
Scenario 2        & \multicolumn{1}{c|}{Uniform(0.5, 1)} & 1e-3 & \multicolumn{1}{c|}{Uniform(0, 0.1)}   & 1e-3   & 1,109,958                              \\ \hline
\end{tabular}%
}
\caption{Monte Carlo Simulation random parameters for two waypoints following scenarios. For each simulation run, we random sample $P_A$ from $\text{Uniform}(0.5, 1)$, and $\varepsilon$ from $\text{Uniform}(0, 0.1)$. A total of over one million sample results are collected for each scenario.}
\label{tab: mc-parameters}
\end{table}

We assume the inbound and outbound latency are identical, thus the total delay is assumed to be $2\varepsilon$. For the given simulation time step $t$, the control vector $\mathbf{u}(t)$ is,
\begin{equation}
    \mathbf{u}(t) = X(t) \cdot \mathbf{u}(t - 2\varepsilon)
\end{equation}
\noindent where $X(t)$ is the signal state at time $t$. There are two source of uncertainty in each simulation run, the latency $\varepsilon$, and the signal states $X(t)$ obtained from CTMC with transition probability $P_A$ between the \textit{on} and the \textit{off} states. From \Cref{subsec: rcp-formulation}, the signal availability is determined by the duration of the \textit{on} state and the \textit{off} states, which follows exponential distributions governed by $\lambda_{on}$ and  $\lambda_{off}$. Further assuming $\lambda_{on} + \lambda_{off} = 1$, we have $P_A = \lambda_{on}$. Thus, the random variables for Monte Carlo simulations narrow down to $P_A$ and $\varepsilon$. \Cref{tab: mc-parameters} provides the detailed selection of the random variables. We choose the dense grid interval of $0.001$ for both $P_A$ and $\varepsilon$. 

The simulation was completed on a high-performance workstation with Intel Xeon w9-3495X processor of 56 cores, 112 threads, and a maximum turbo frequency of up to 4.8 GHz. We collect slightly over 1 million Monte Carlo Simulation samples for each scenario, utilizing the multiprocessing tools \citep{aziz2021python}. We further store the Boolean indicating whether the waypoint following mission is completed, and the mission completion time if completed.

\subsubsection{Mission Success Rate \label{subsubsec: msr}}
Mission success rate is a straightforward performance indicator. It quantifies how frequently an operation meets the intended goals, thereby informing engineers or decision-makers on overall system reliability expectations, and the regions in need of improvement \citep{chamberlain1966space, jacklin2019small, zhao2020mission}. Related formulations for phased-mission systems (PMS), commonly found in aerospace applications, have been studied using Semi-Markov Processes for non-exponential component lifetimes \citep{li_reliability_2018} and Markov regenerative processes for missions subjected to random shocks \citep{li_reliability_2018-1}. Moreover, Bayesian networks have been employed to evaluate complex multi-state system reliability under common cause failures and epistemic uncertainties \citep{mi_reliability_2018}, and to assess the reliability of redundant control systems \citep{cai_using_2012}. In this work, mission success rate is calculated as the ratio between completed simulations and total number of simulations for any given parameter pair $\varepsilon$ and $P_A$. It stands for the percentage or proportion of attempts that meet the predefine success criteria given above.

\begin{figure}[hbt!]
\centering
\includegraphics[width=0.95\textwidth]{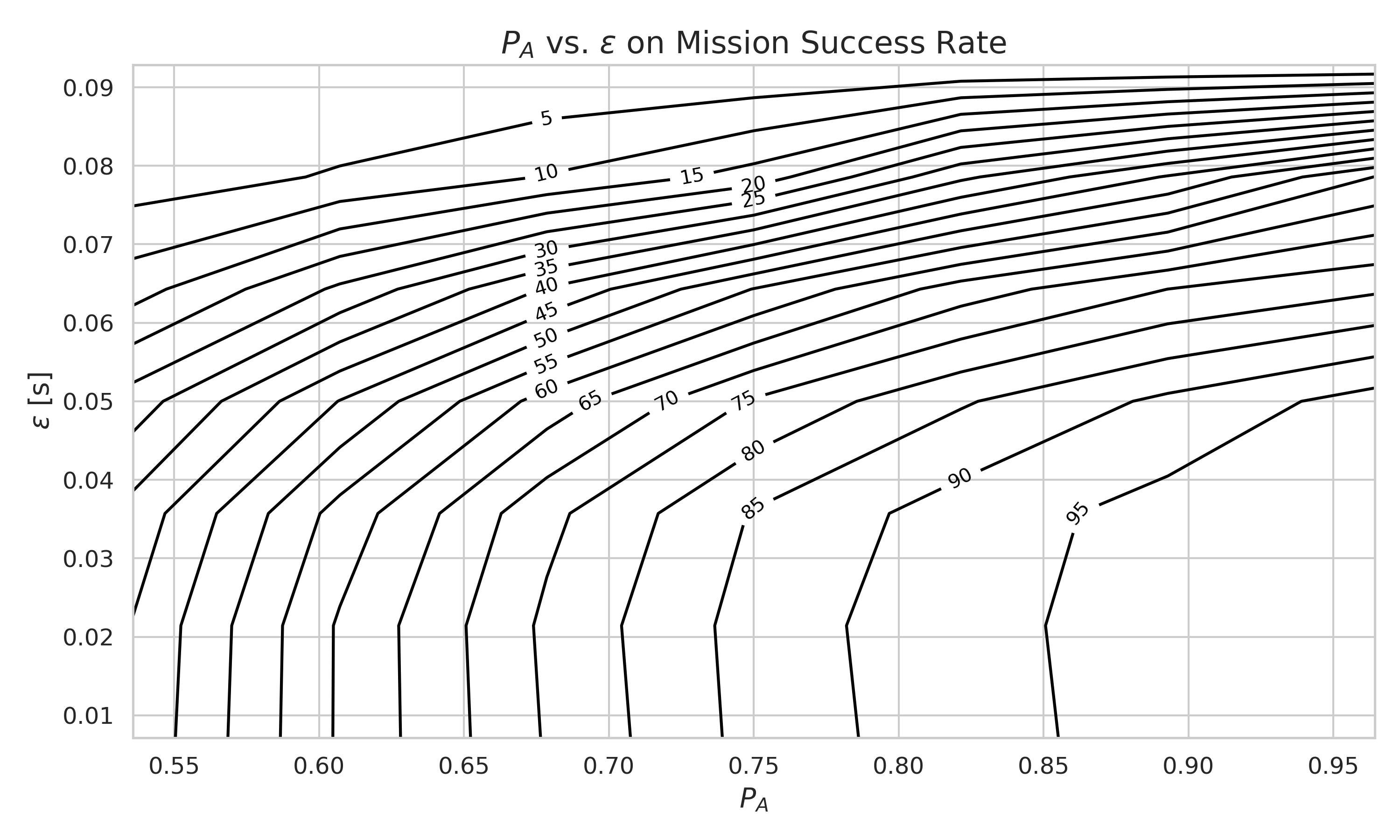}
\label{fig: mission-success-contour-lines}
\caption{An example of the figure showing the contour lines on mission success levels. Mission success rate percentage is marked on the contour lines. }
\end{figure}

As shown in \Cref{fig: mission-success-contour-lines}, the probability of mission success decreases monotonically as the random variables $\varepsilon$ and $P_A$ increase. Moreover, once these variables exceed a certain threshold, the success rate deteriorates sharply, with the rate of deterioration continuing to accelerate thereafter.

\begin{figure}[H]
    \centering
    \begin{subfigure}[t]{0.32\textwidth}
        \centering
        \includegraphics[width=\textwidth,height=4cm]{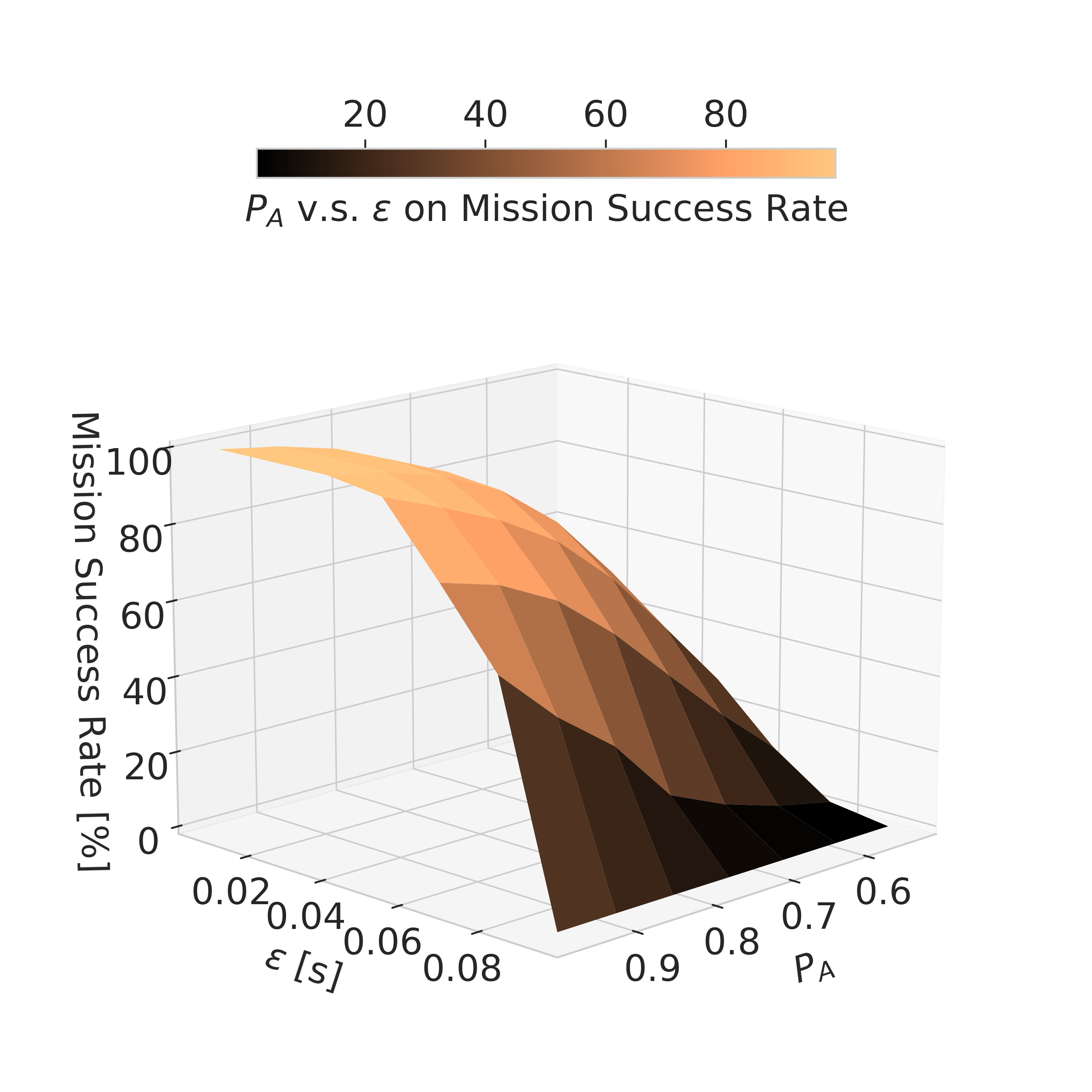}
    \end{subfigure}
    \begin{subfigure}[t]{0.32\textwidth}
        \centering
        \includegraphics[width=\textwidth,height=4cm]{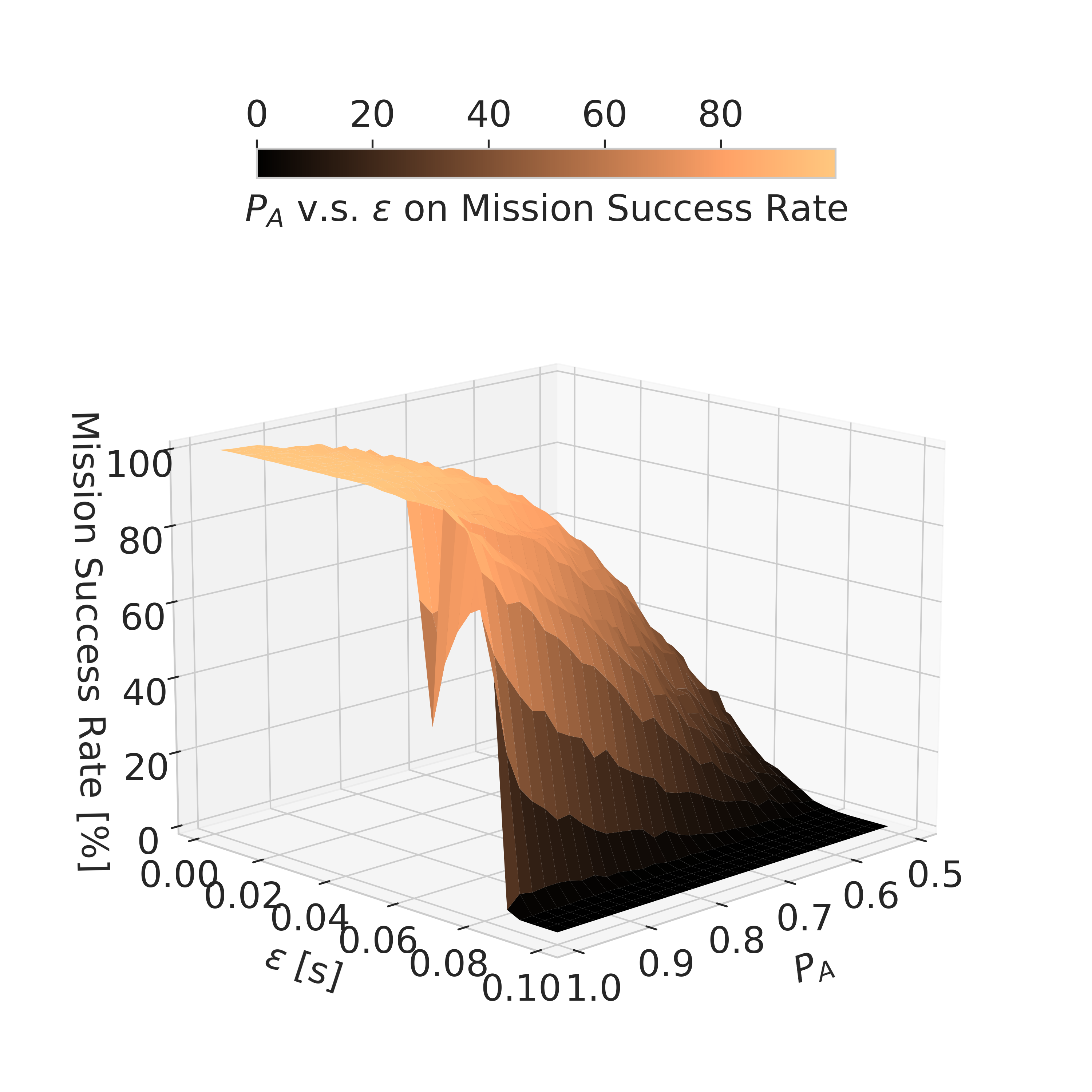}
    \end{subfigure}
    \begin{subfigure}[t]{0.32\textwidth}
        \centering
        \includegraphics[width=\textwidth,height=4cm]{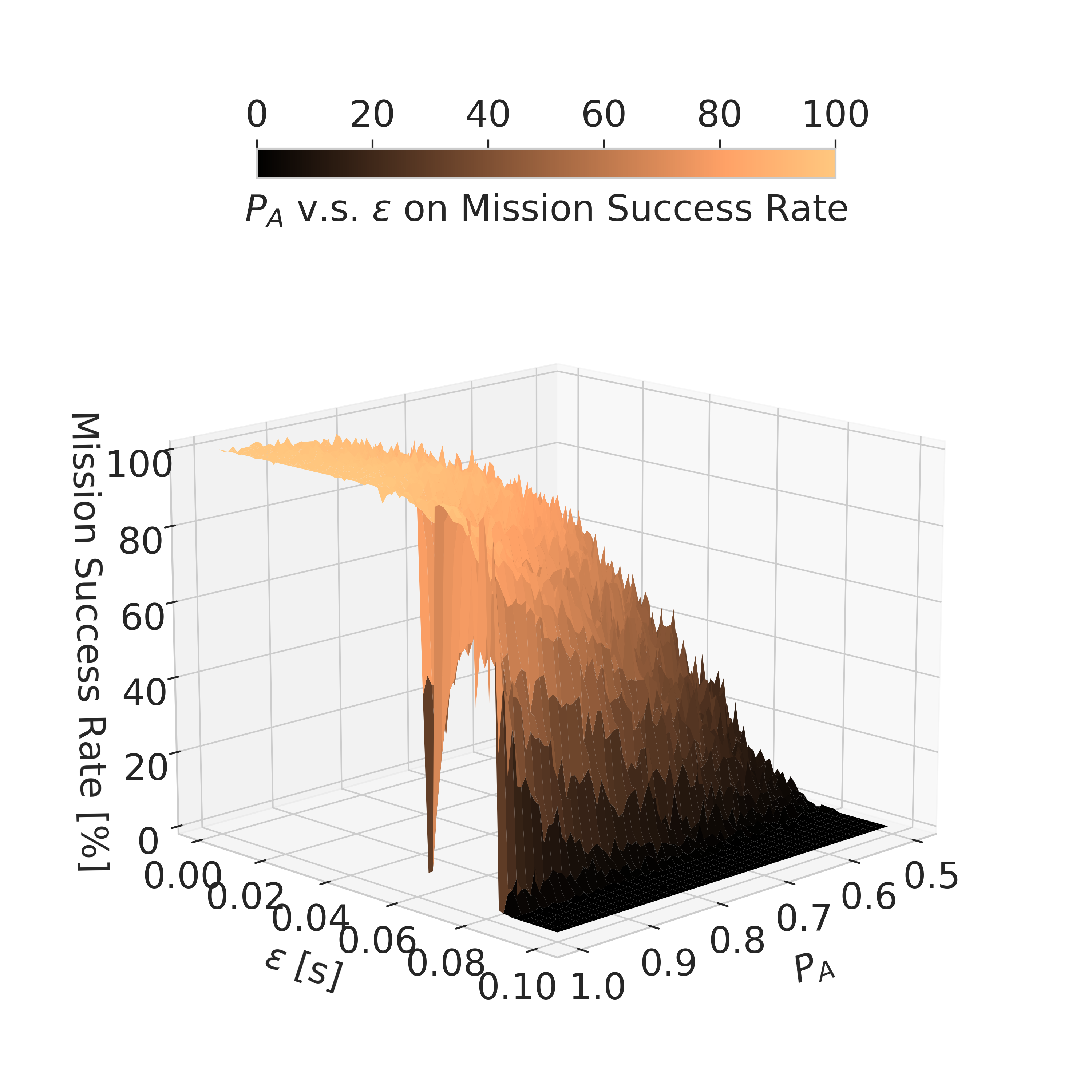}
    \end{subfigure}
    ~
    \begin{subfigure}[t]{0.32\textwidth}
        \centering
        \includegraphics[width=\textwidth,height=3cm]{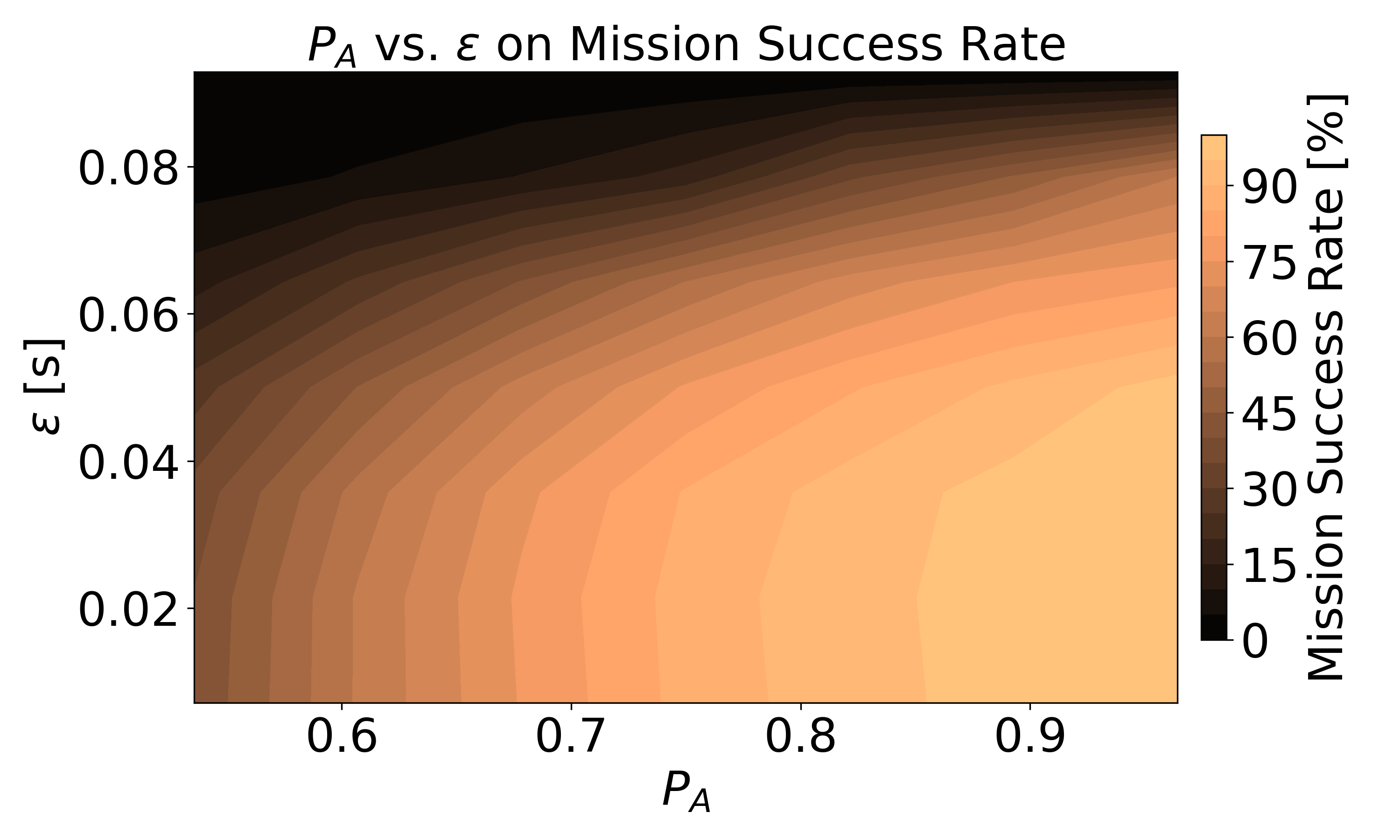}
    \end{subfigure}
    \begin{subfigure}[t]{0.32\textwidth}
        \centering
        \includegraphics[width=\textwidth,height=3cm]{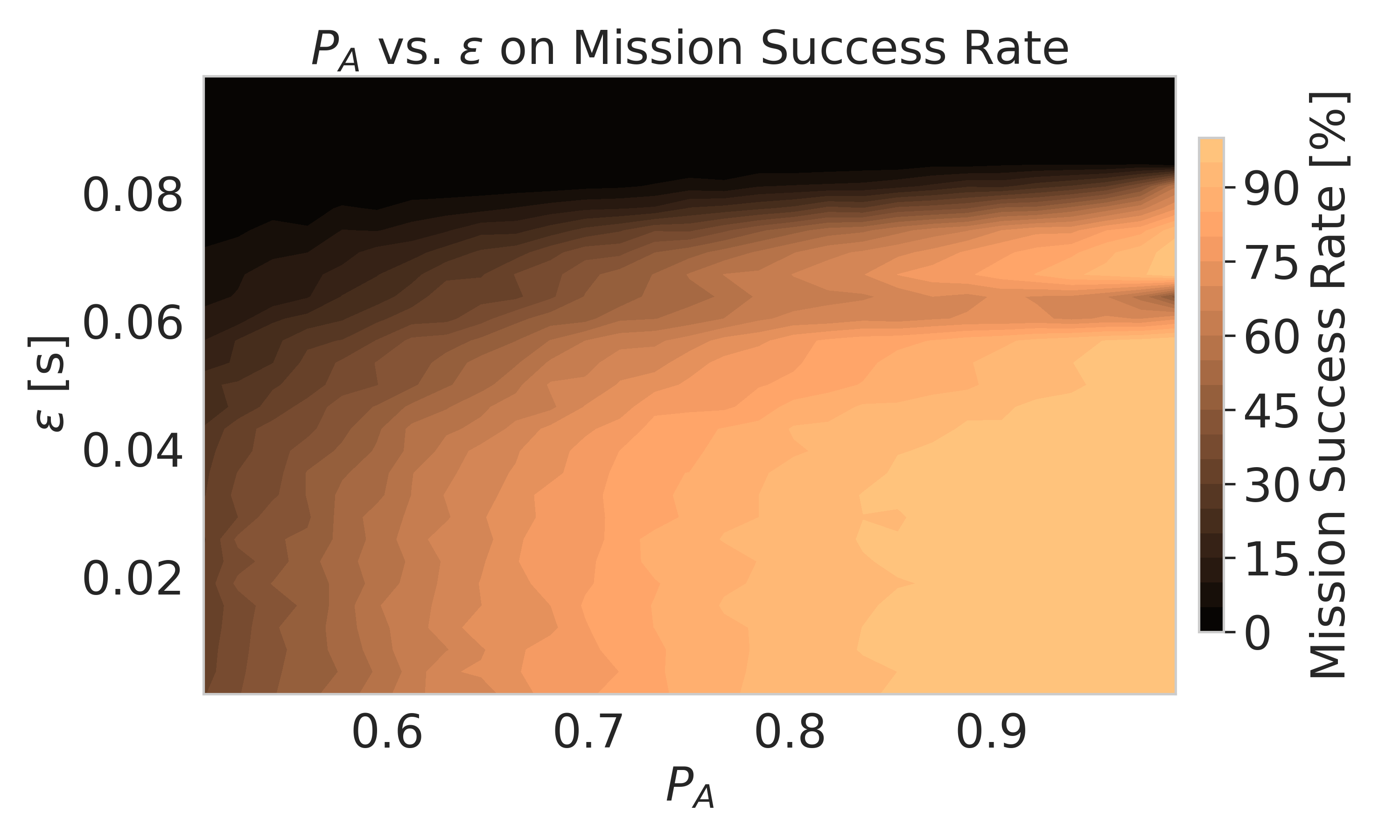}
    \end{subfigure}
    \begin{subfigure}[t]{0.32\textwidth}
        \centering
        \includegraphics[width=\textwidth,height=3cm]{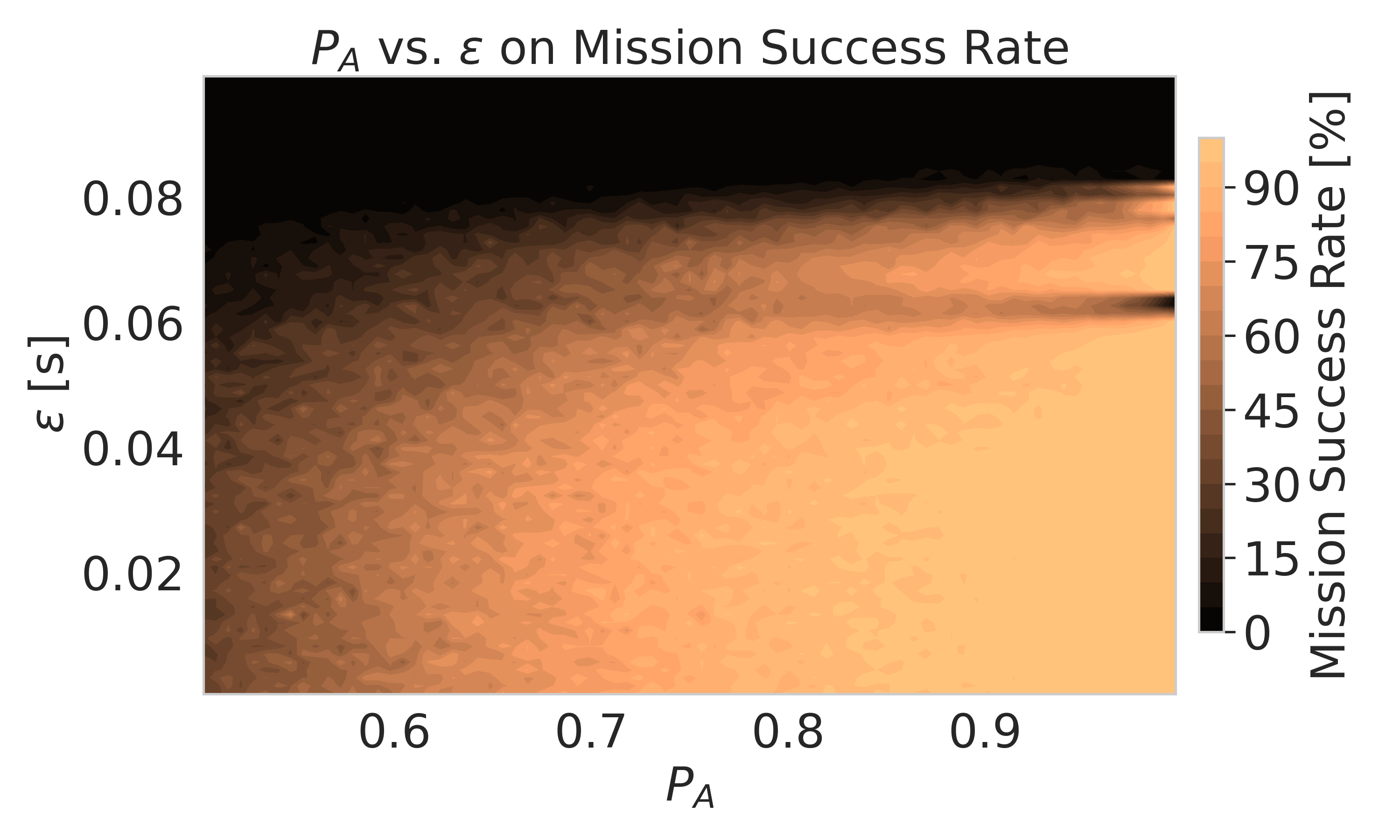}
    \end{subfigure}
    \caption{Scenario 1: mission success surface envelop (top row) and mission success surface contour (bottom row). }
    \label{fig: wp1-success-rates}
\end{figure}

\begin{figure}[H]
    \centering
    \begin{subfigure}[t]{0.32\textwidth}
        \centering
        \includegraphics[width=\textwidth,height=4cm]{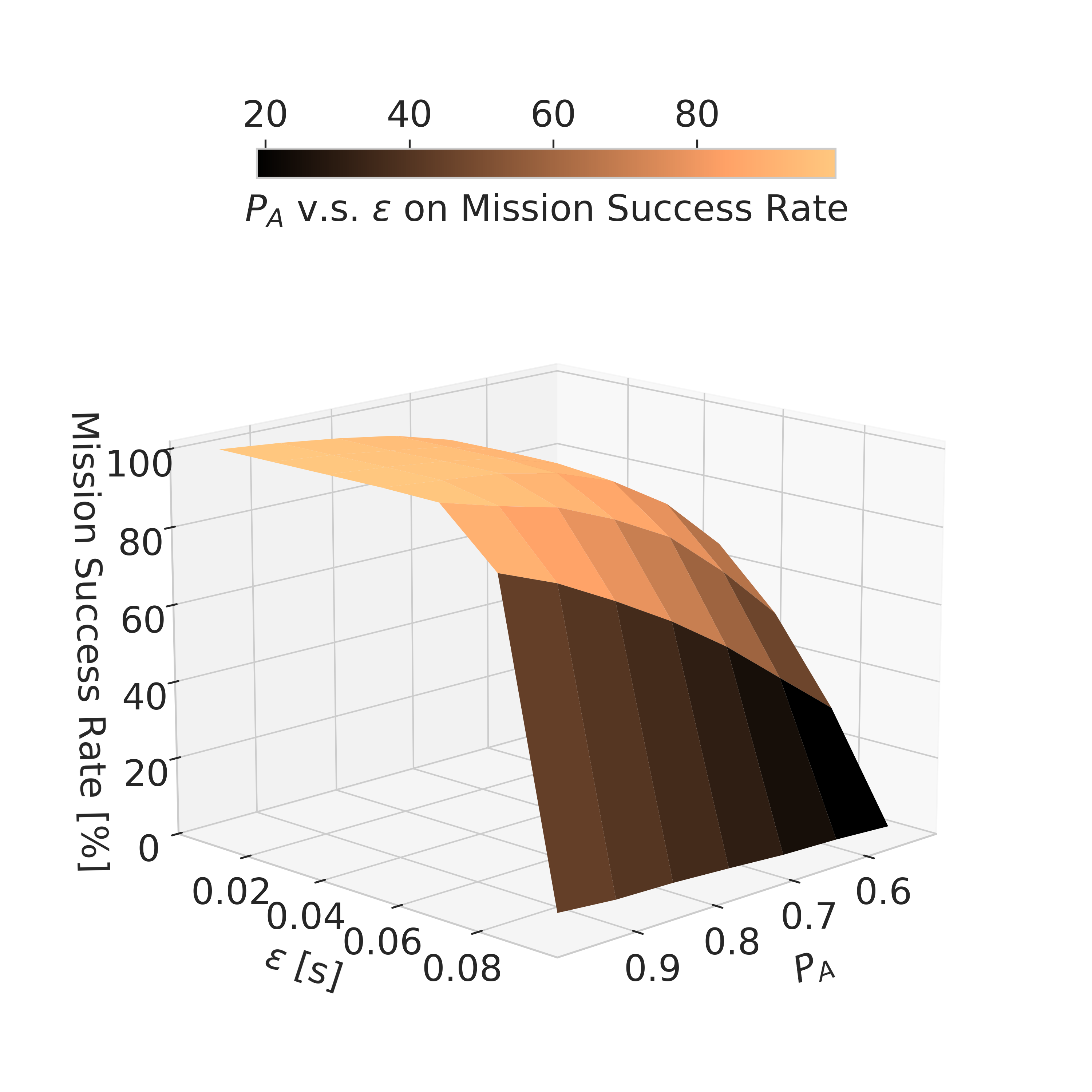}
    \end{subfigure}
    \begin{subfigure}[t]{0.32\textwidth}
        \centering
        \includegraphics[width=\textwidth,height=4cm]{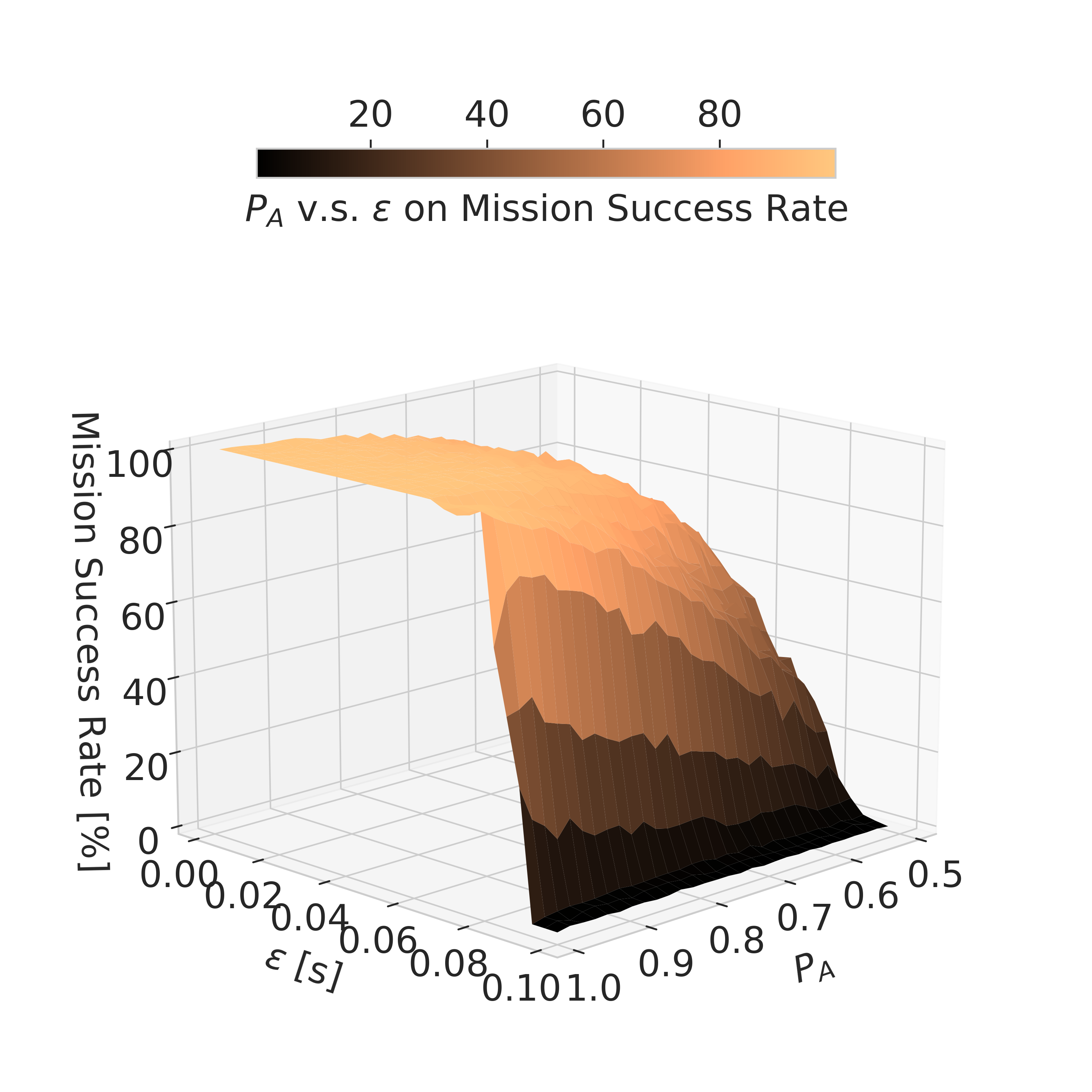}
    \end{subfigure}
    \begin{subfigure}[t]{0.32\textwidth}
        \centering
        \includegraphics[width=\textwidth,height=4cm]{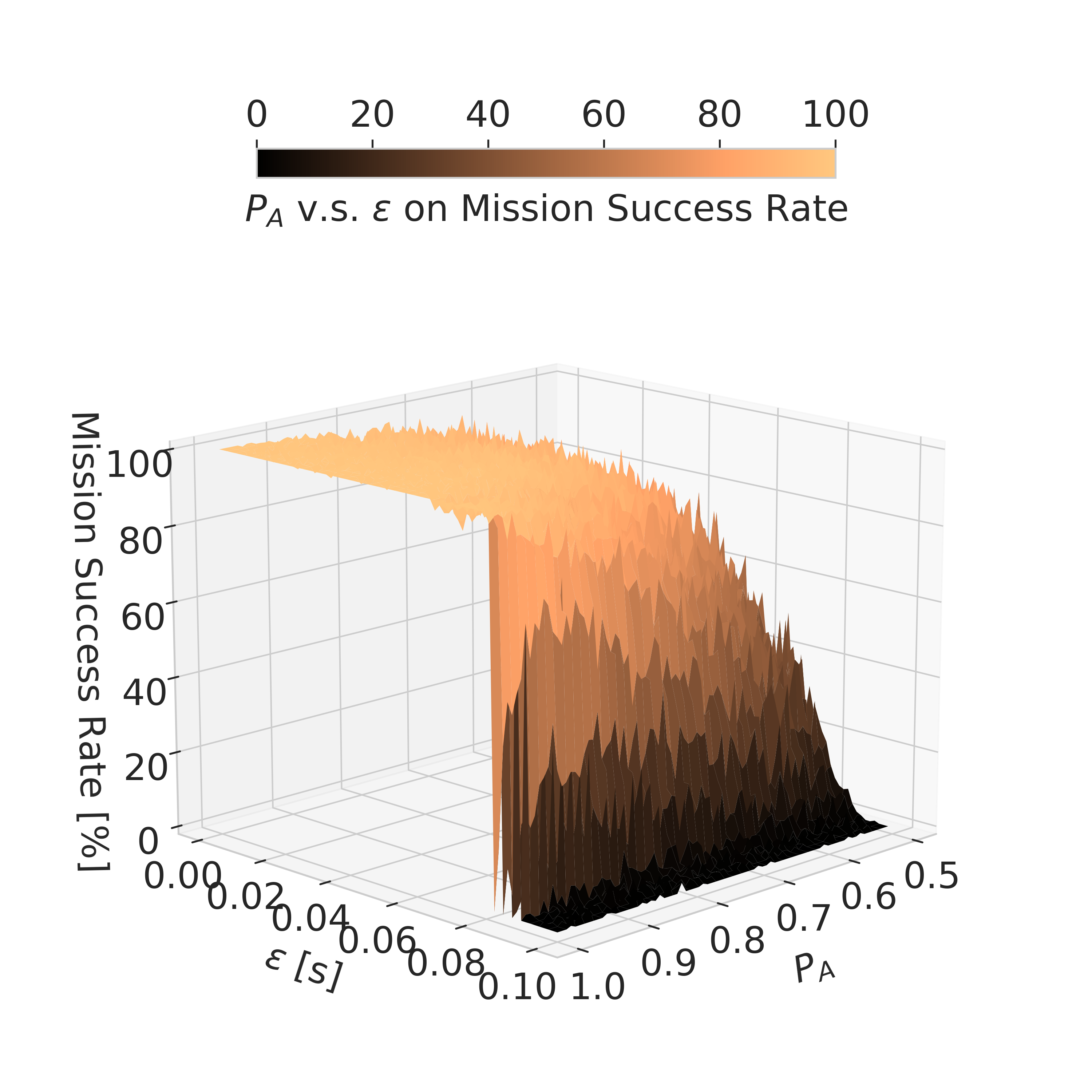}
    \end{subfigure}
    ~
    \begin{subfigure}[t]{0.32\textwidth}
        \centering
        \includegraphics[width=\textwidth,height=3cm]{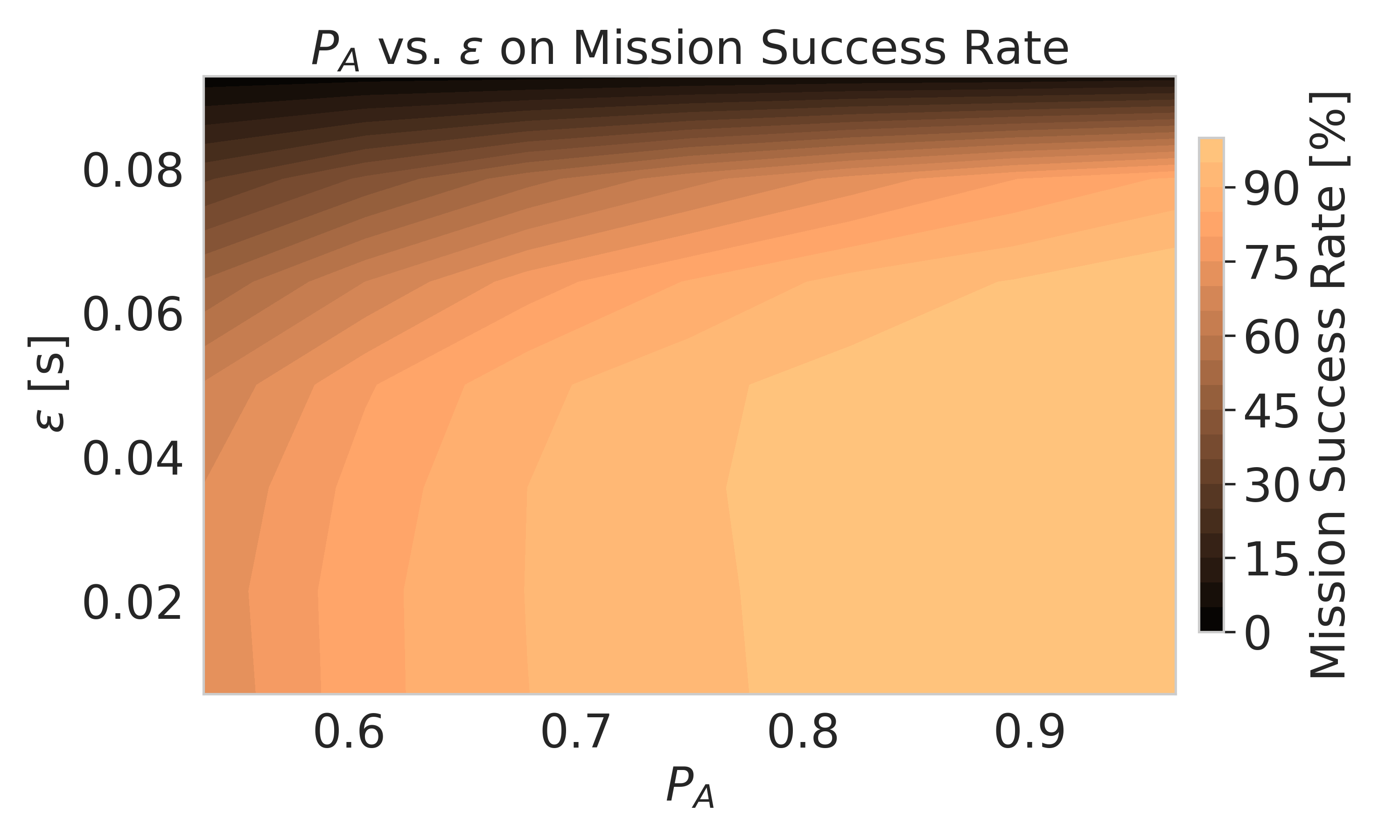}
    \end{subfigure}
    \begin{subfigure}[t]{0.32\textwidth}
        \centering
        \includegraphics[width=\textwidth,height=3cm]{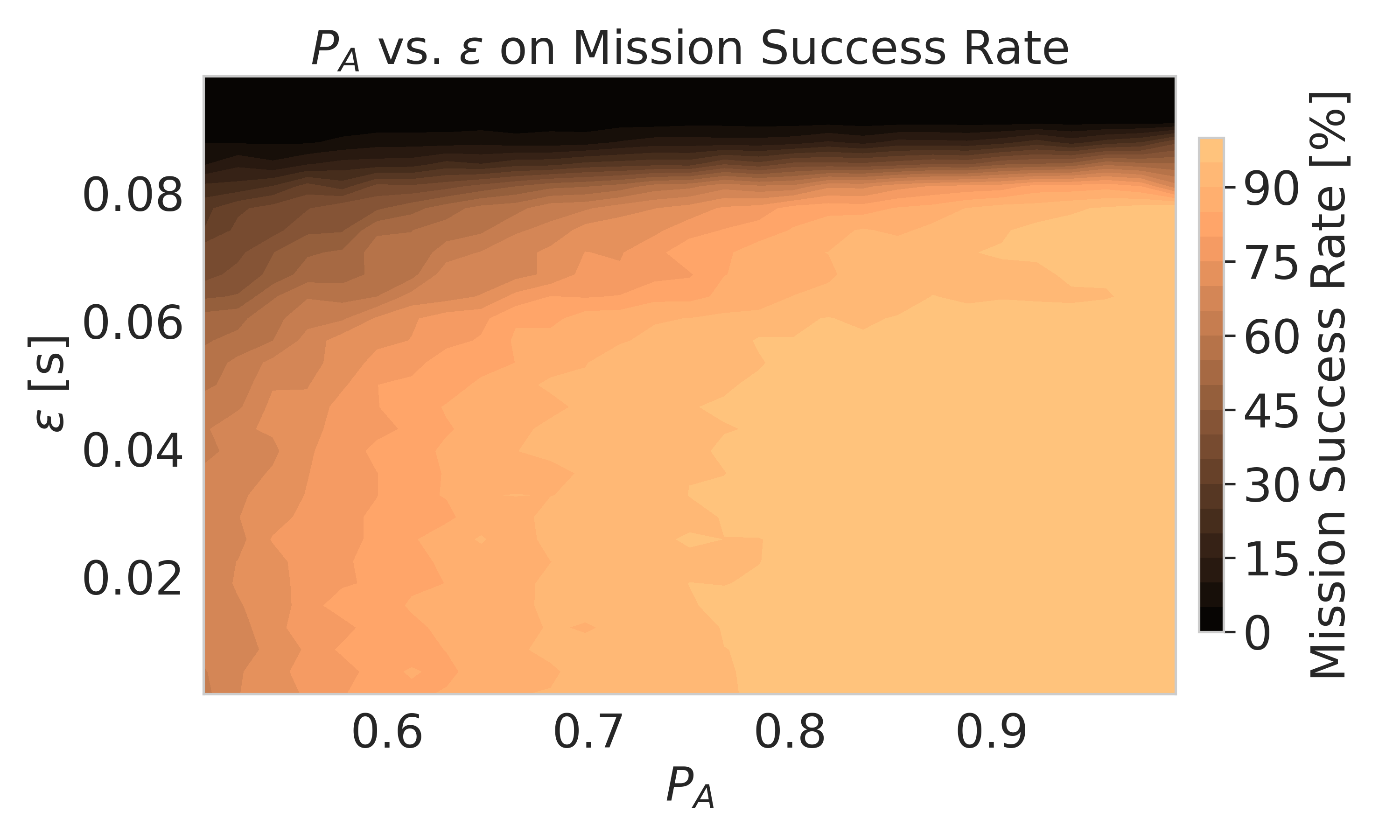}
    \end{subfigure}
    \begin{subfigure}[t]{0.32\textwidth}
        \centering
        \includegraphics[width=\textwidth,height=3cm]{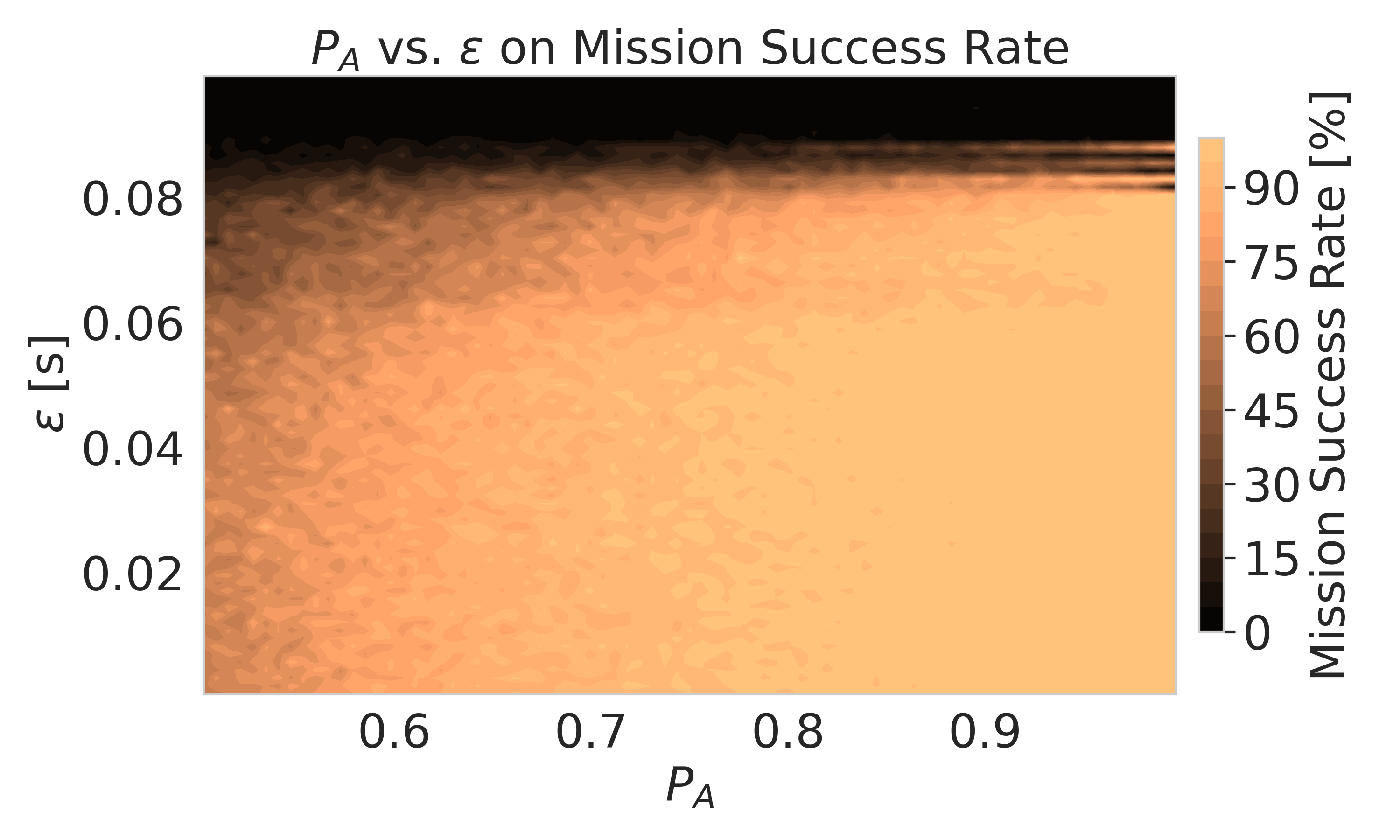}
    \end{subfigure}
    \caption{Scenario 2: mission success surface (top row) and mission success surface contour (bottom row). }
    \label{fig: wp2-success-rates}
\end{figure}

\Cref{fig: wp1-success-rates} and \Cref{fig: wp2-success-rates} present more results on both scenarios. The top row is the mission success envelop in 2D space (i.e., surface), and the bottom row is for the 3D space (i.e., contour). The resolution (e.g., the number of histogram bins on the x-axis and y-axis) increases from left column to right column. Complex scenarios clearly requires lower latency and higher availability to achieve the same level of mission success, but the performance deterioration behavior remains. The phase margin measure of stability and robustness, attainable via Nyquist and Bode analysis, provides an insight into the deterioration of performance with increasing latency. The closed-loop system stability and performance degrades when the control system is pushed to its margins \citep{brinker1996stability, pachter1996literal, hess2018design}. More theoretical analysis on verifying the Nyquist stability criterion under increased latency is highly desired, and is listed as a topic of future study. 

Additionally, the higher resolution mission success envelope of scenario 1 reveals the behavior where certain latency values $\varepsilon$ can fail the mission, even with very high availability (i.e., $P_A = 1$). This pattern is not significant in the simpler scenario. We name it \textit{latency blockage}. Slightly decreasing or increasing latency alleviates the issue. More analysis on \textit{latency blockage} is given in \Cref{subsec: discussion}. 

\subsubsection{Mission Completion Time}
\begin{figure}[H]
    \centering
    \begin{subfigure}[t]{0.32\textwidth}
        \centering
        \includegraphics[width=\textwidth,height=4cm]{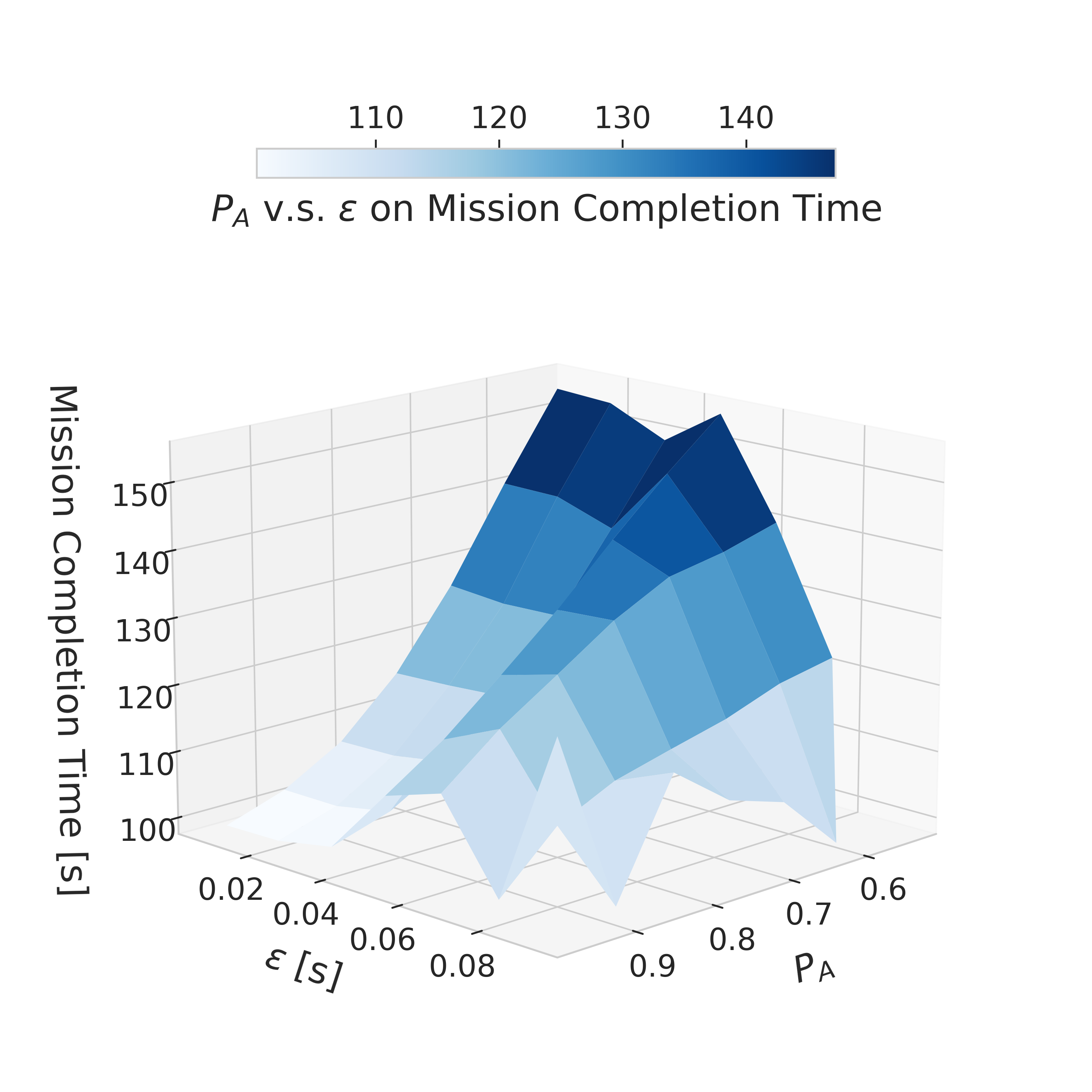}
    \end{subfigure}
    \begin{subfigure}[t]{0.32\textwidth}
        \centering
        \includegraphics[width=\textwidth,height=4cm]{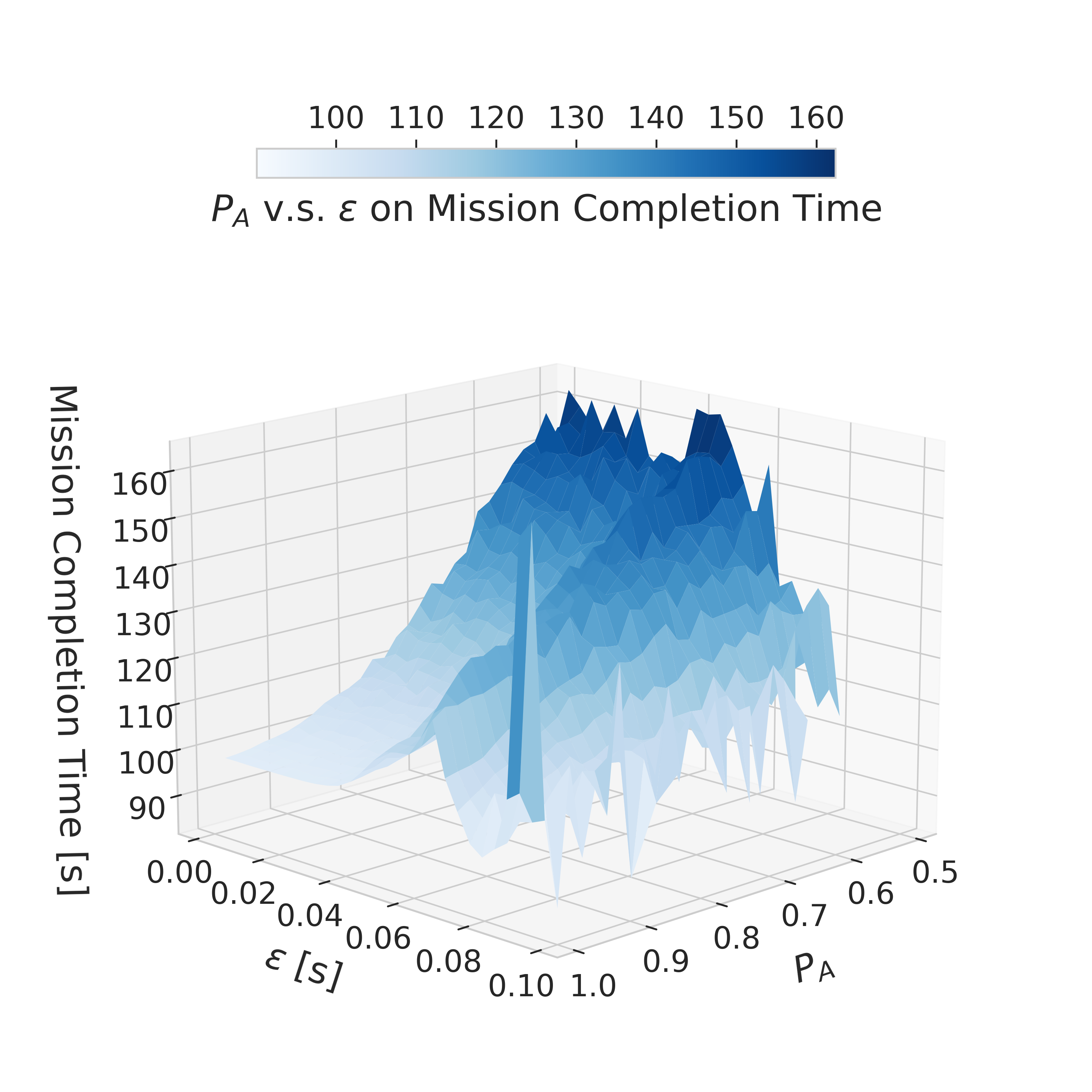}
    \end{subfigure}
    \begin{subfigure}[t]{0.32\textwidth}
        \centering
        \includegraphics[width=\textwidth,height=4cm]{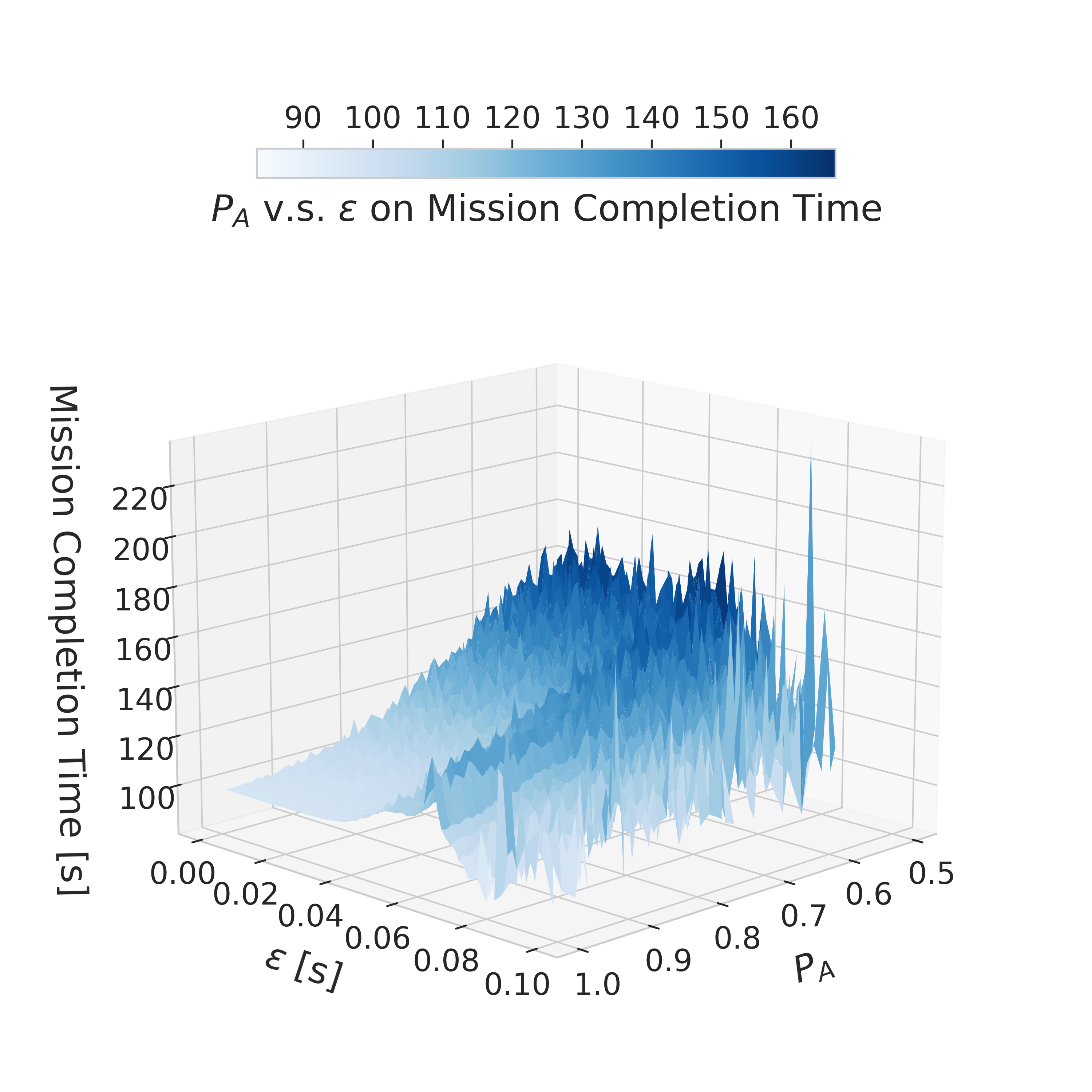}
    \end{subfigure}
    ~
    \begin{subfigure}[t]{0.32\textwidth}
        \centering
        \includegraphics[width=\textwidth,height=3cm]{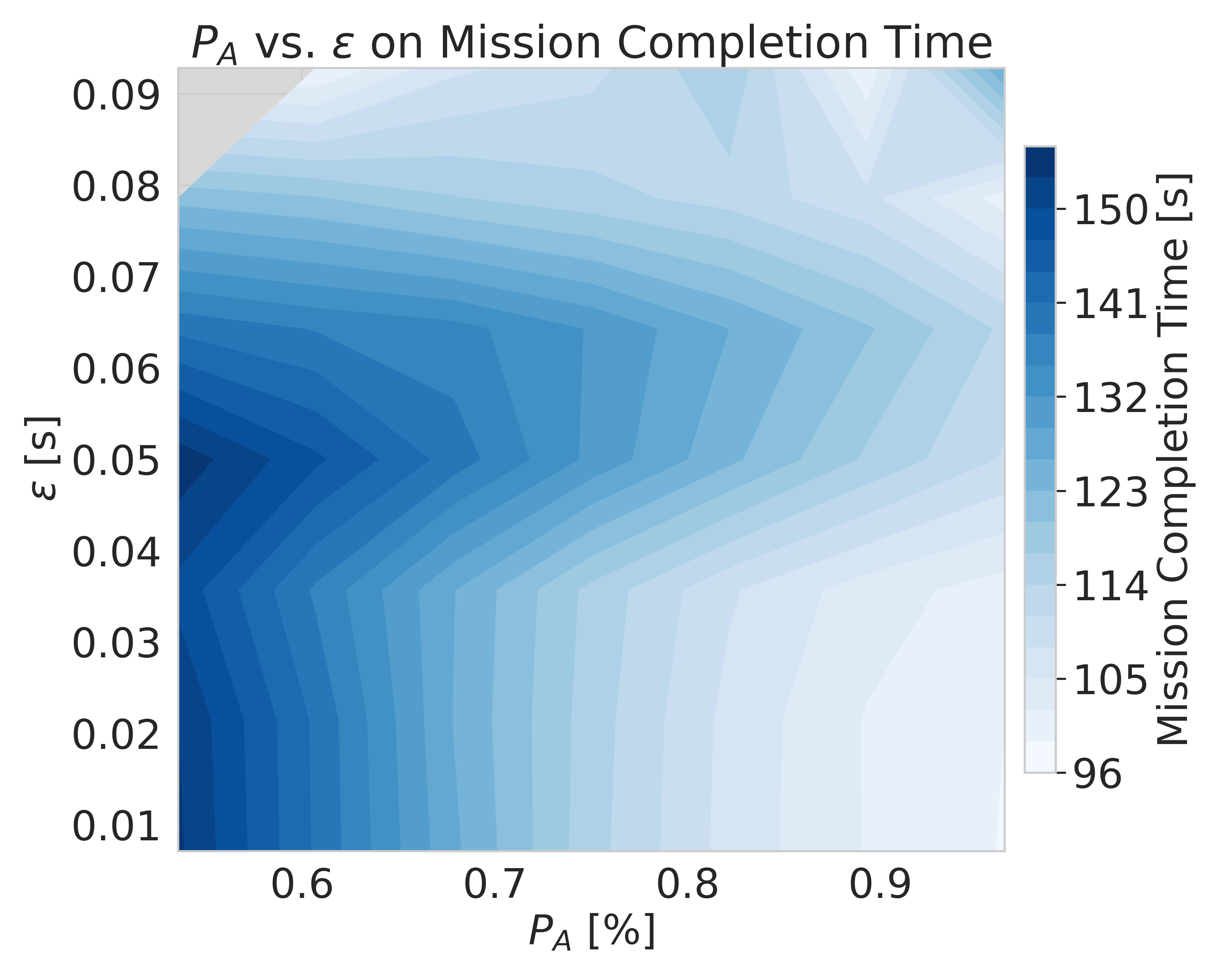}
    \end{subfigure}
    \begin{subfigure}[t]{0.32\textwidth}
        \centering
        \includegraphics[width=\textwidth,height=3cm]{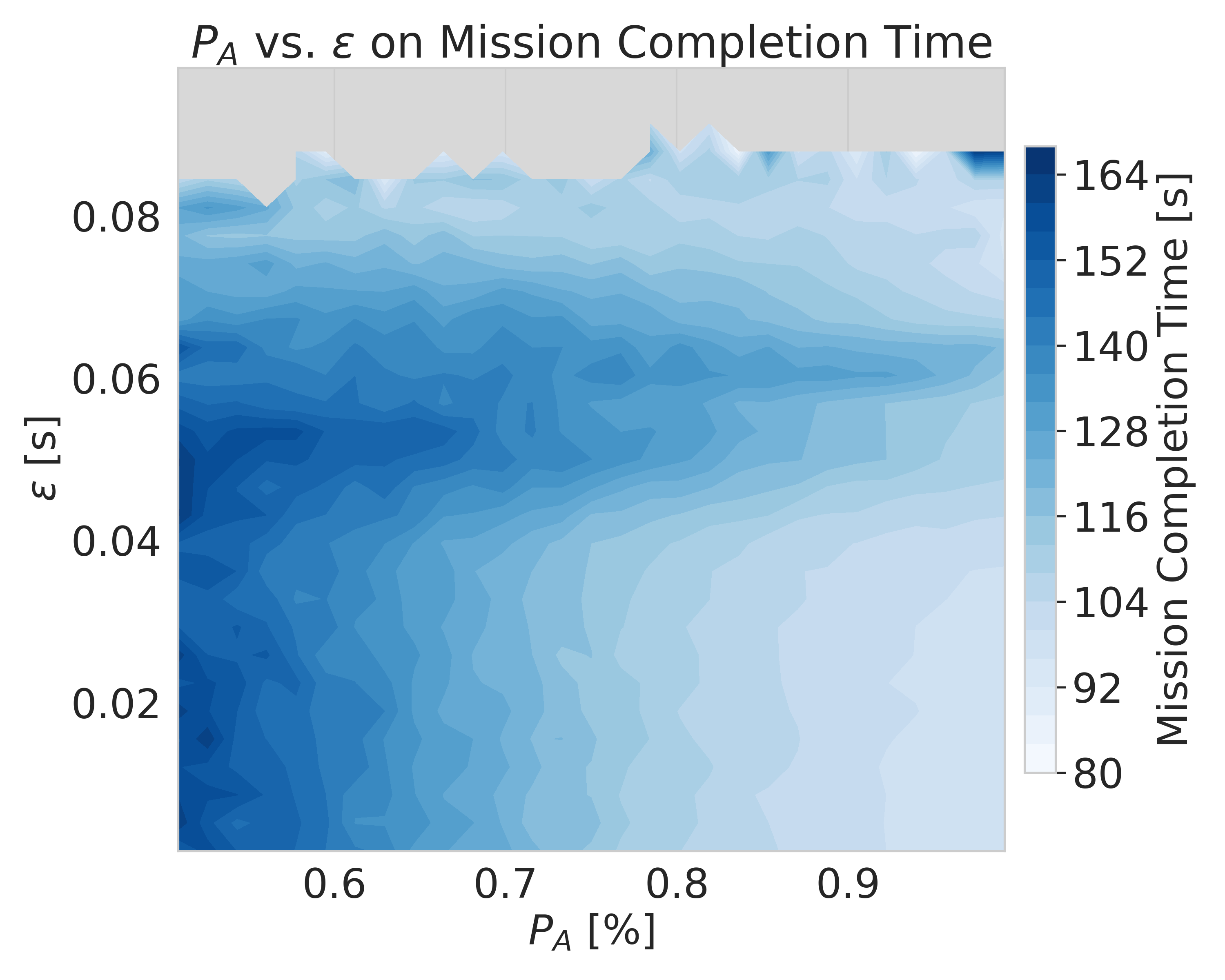}
    \end{subfigure}
    \begin{subfigure}[t]{0.32\textwidth}
        \centering
        \includegraphics[width=\textwidth,height=3cm]{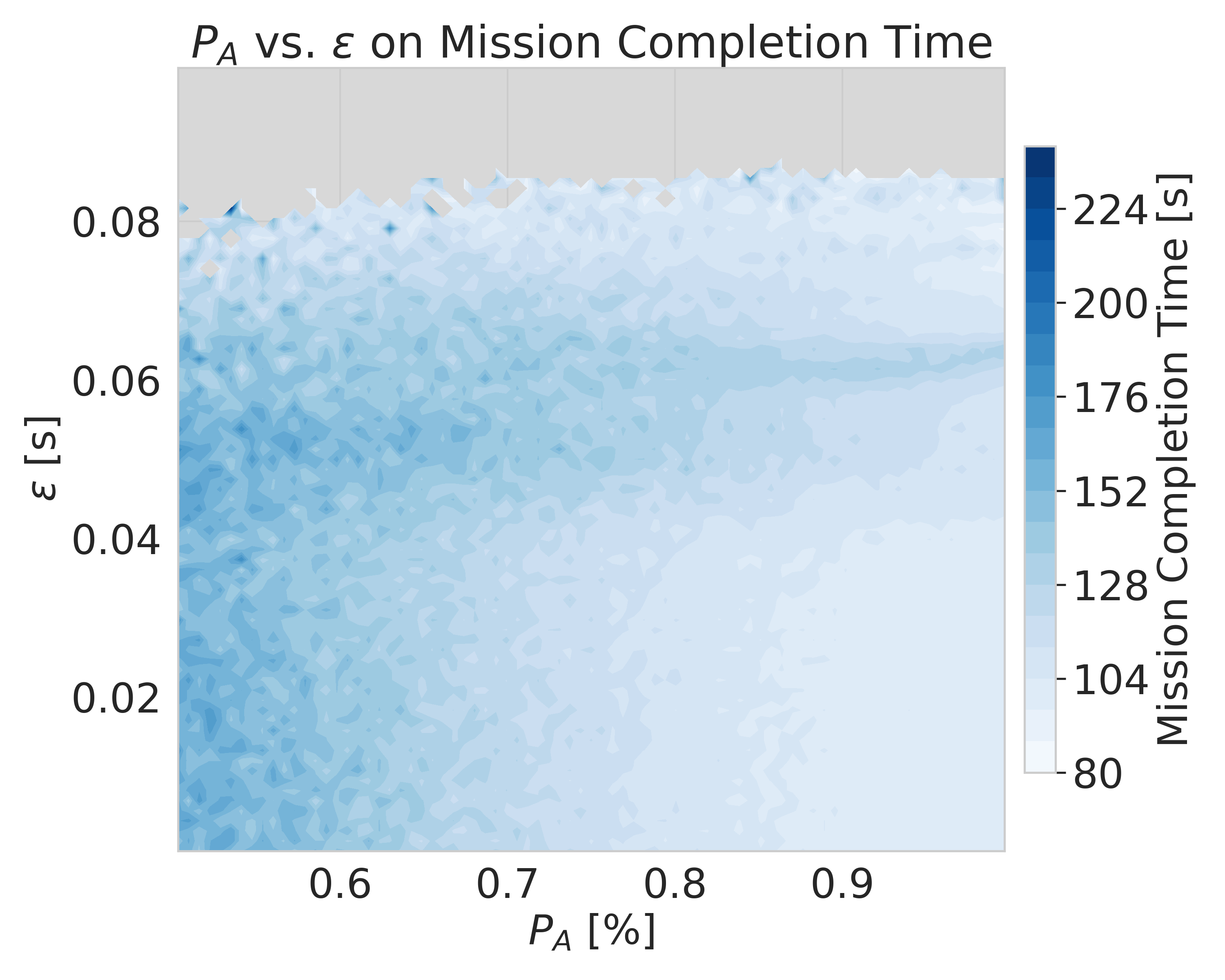}
    \end{subfigure}
    \caption{Scenario 1: mission completion time surface (top row) and mission completion time contour (bottom row). }
    \label{fig: wp1-time-rates}
\end{figure}

\begin{figure}[H]
    \centering
    \begin{subfigure}[t]{0.32\textwidth}
        \centering
        \includegraphics[width=\textwidth,height=4cm]{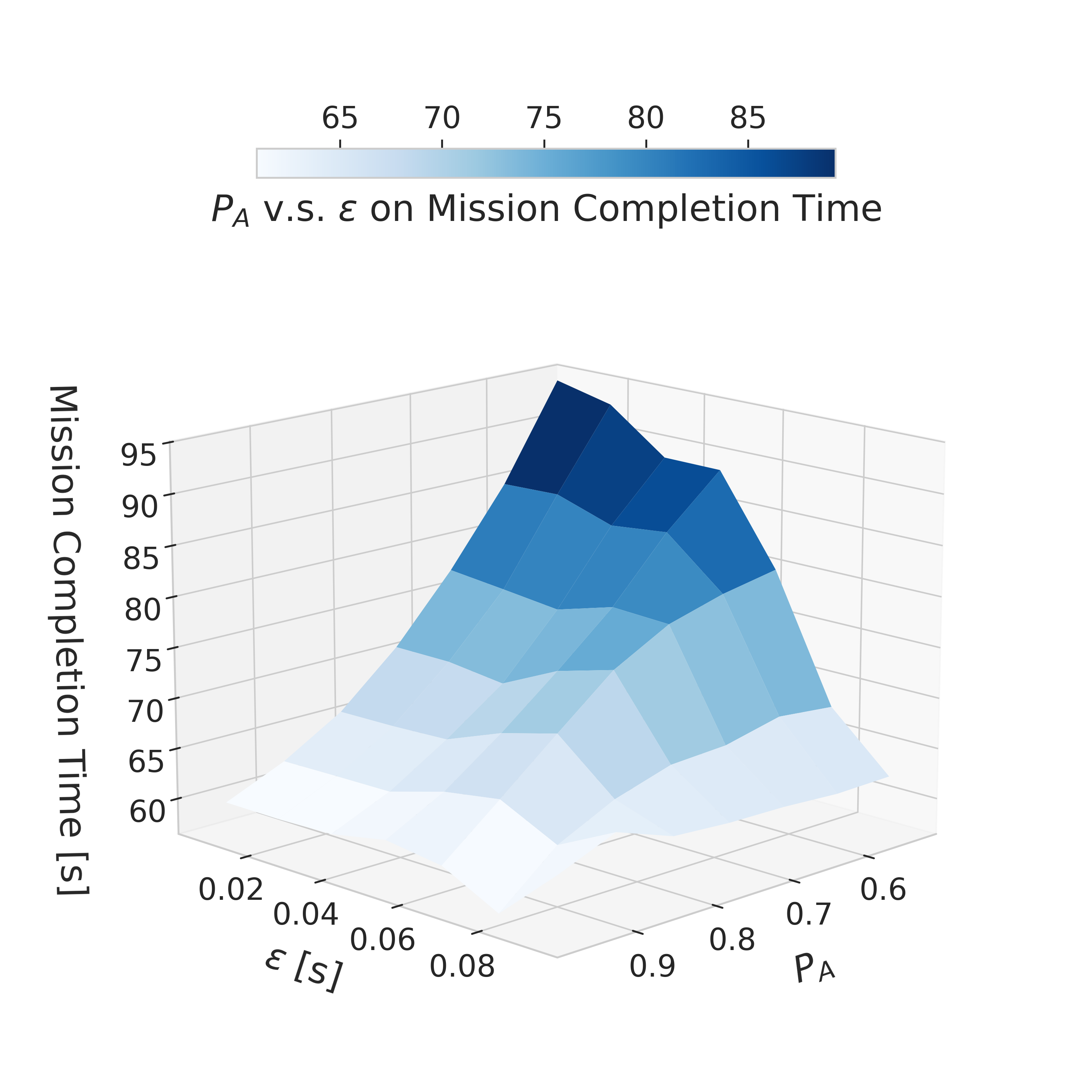}
    \end{subfigure}
    \begin{subfigure}[t]{0.32\textwidth}
        \centering
        \includegraphics[width=\textwidth,height=4cm]{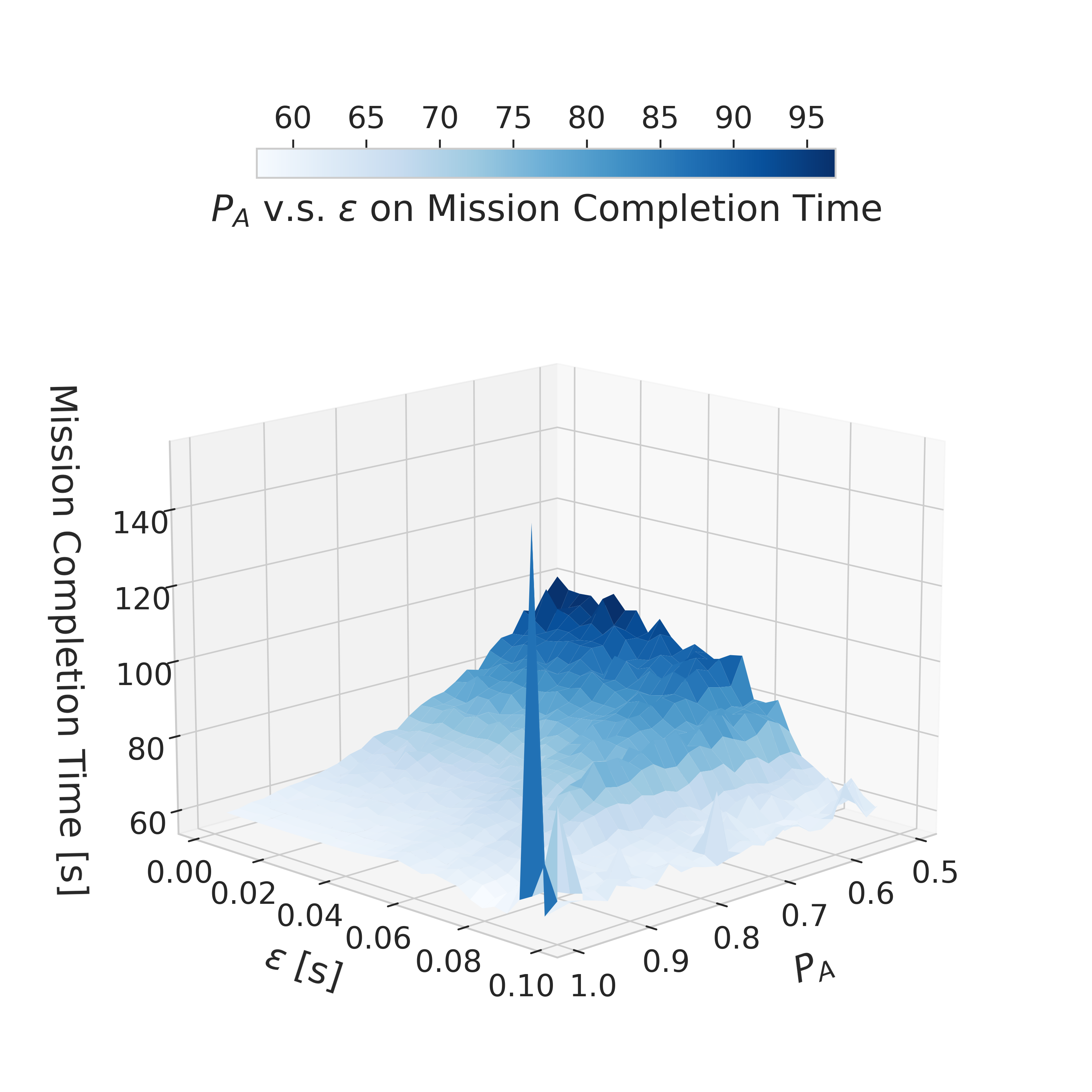}
    \end{subfigure}
    \begin{subfigure}[t]{0.32\textwidth}
        \centering
        \includegraphics[width=\textwidth,height=4cm]{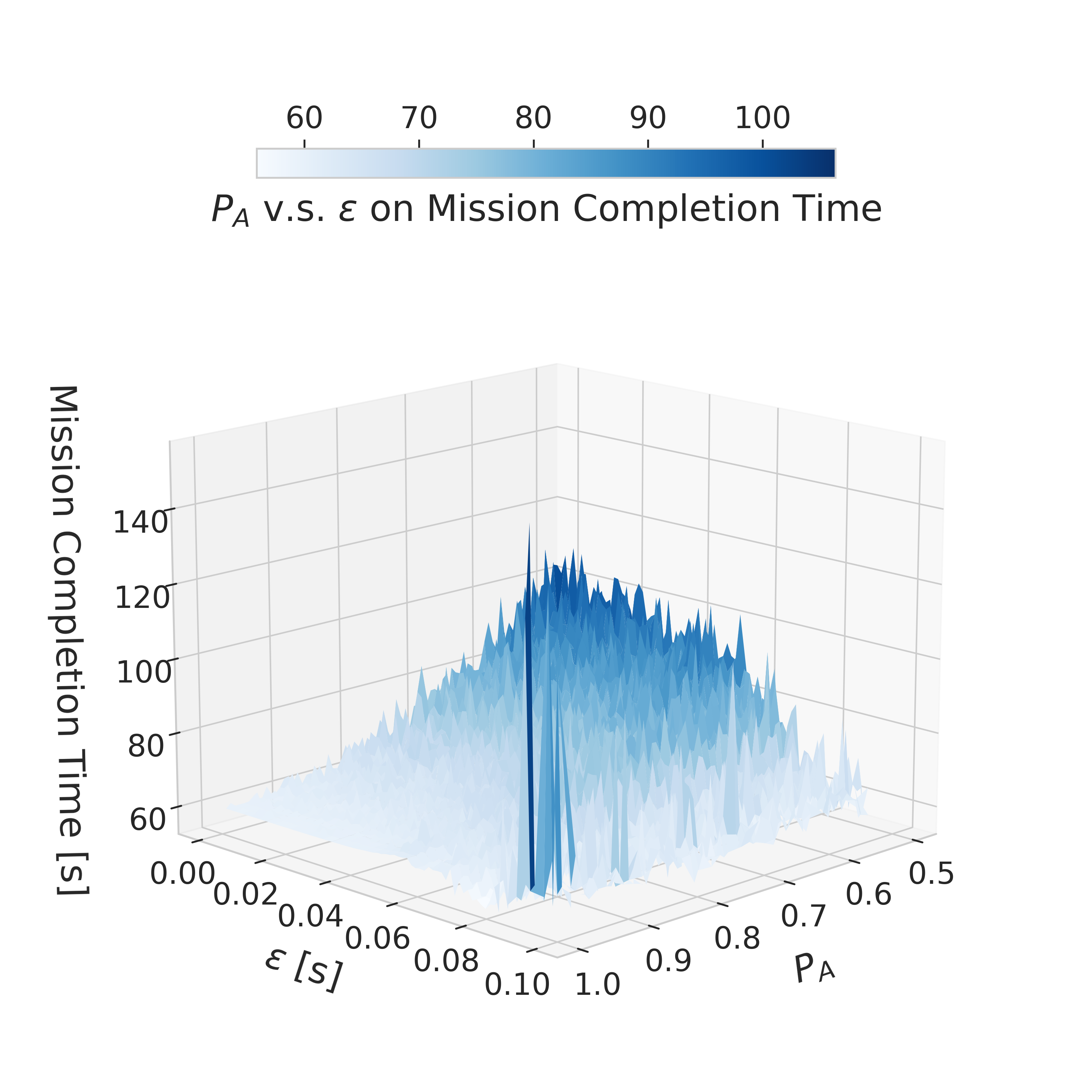}
    \end{subfigure}
    ~
    \begin{subfigure}[t]{0.32\textwidth}
        \centering
        \includegraphics[width=\textwidth,height=3cm]{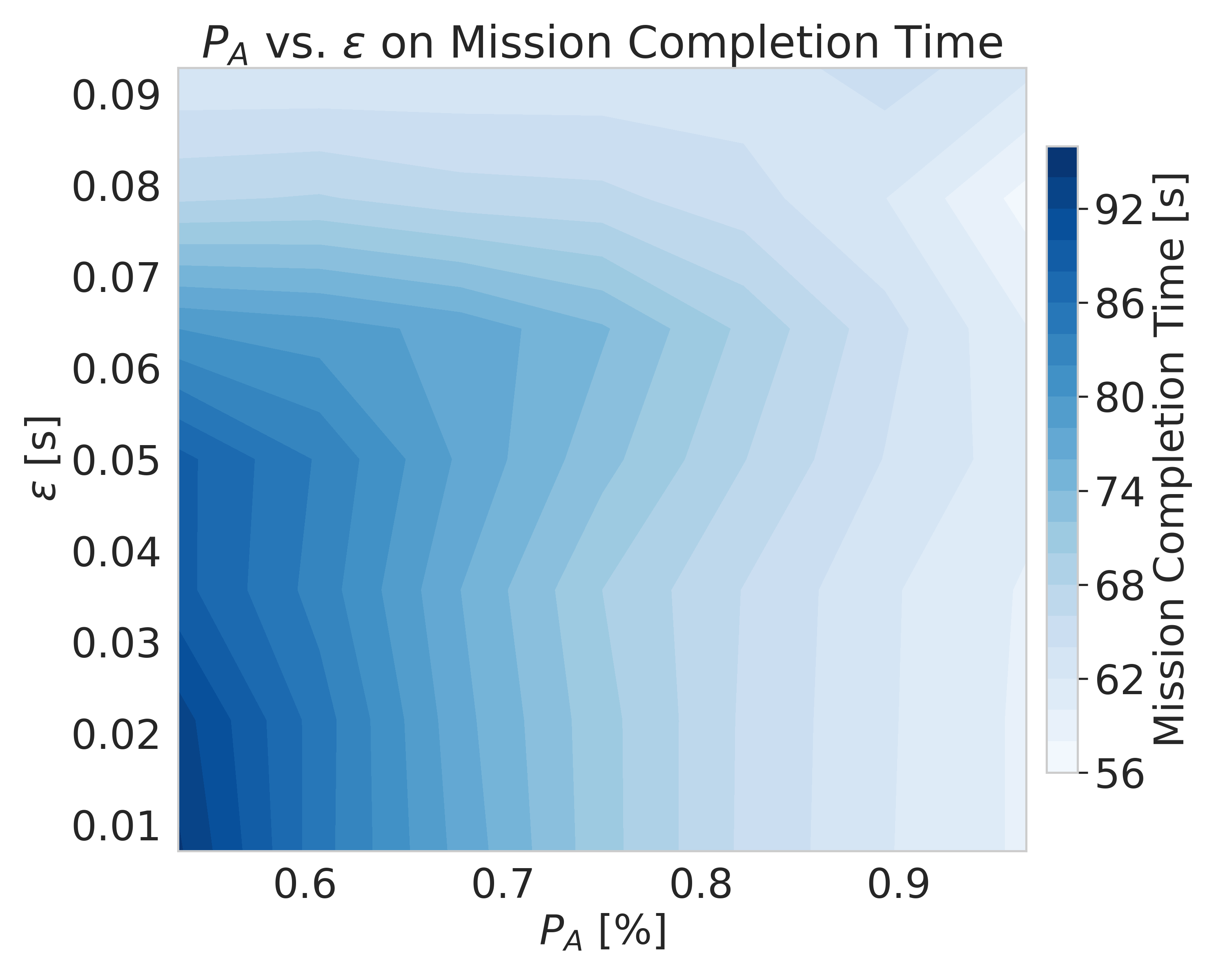}
    \end{subfigure}
    \begin{subfigure}[t]{0.32\textwidth}
        \centering
        \includegraphics[width=\textwidth,height=3cm]{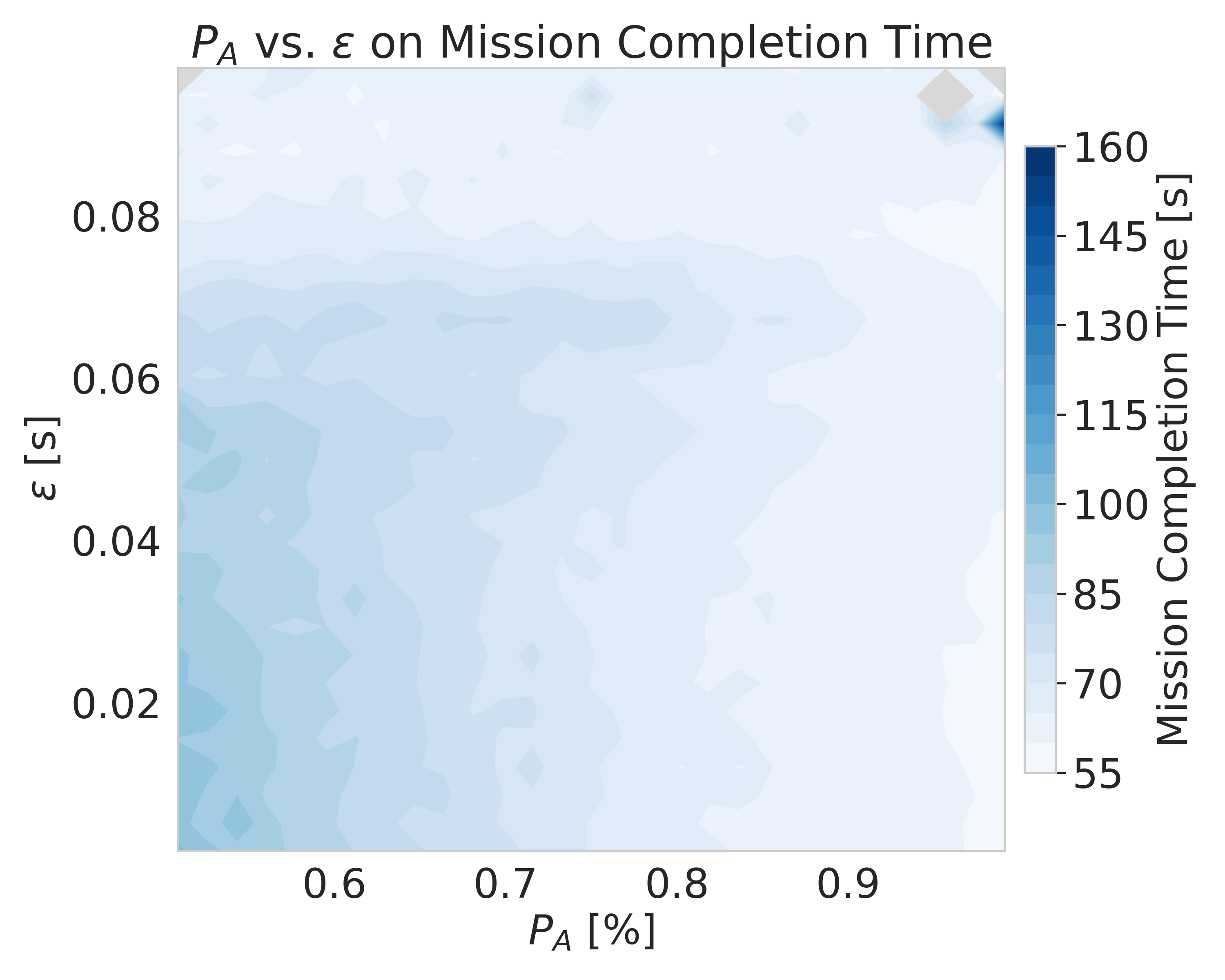}
    \end{subfigure}
    \begin{subfigure}[t]{0.32\textwidth}
        \centering
        \includegraphics[width=\textwidth,height=3cm]{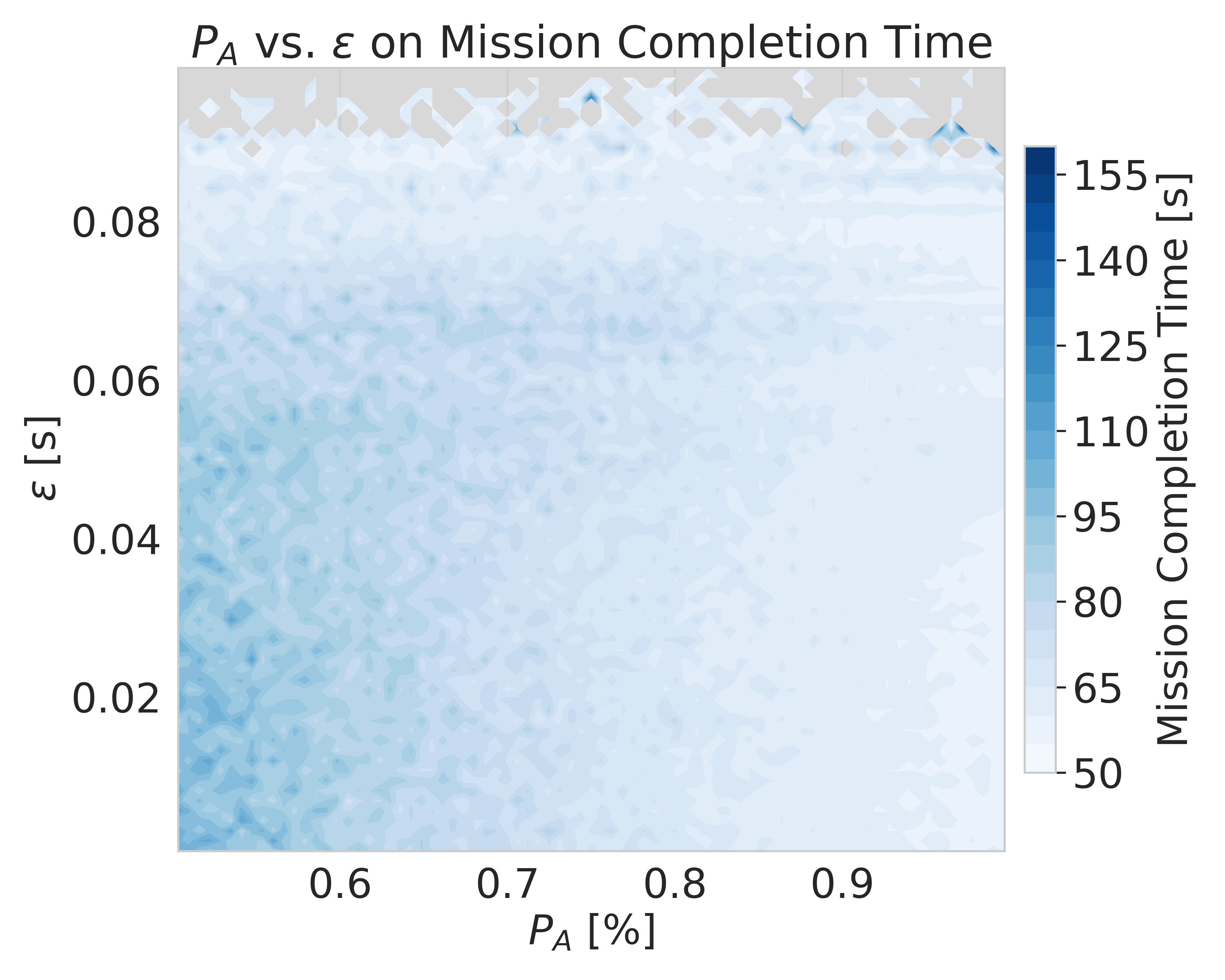}
    \end{subfigure}
    \caption{Scenario 2: mission completion time surface (top row) and mission completion time contour (bottom row). }
    \label{fig: wp2-time-rates}
\end{figure}

Mission completion time is another metric to quantify the RPAS performance, especially in success simulations. Similar to the mission success rate, we present the mission completion surfaces and contours in \Cref{fig: wp1-time-rates} and \Cref{fig: wp2-time-rates}. In general, the mission completion time is lower for better communication situations, and the simpler scenario requires less time to complete the mission. In the simpler scenario (i.e., scenario 2), the mission completion time increases monotonically as latency rises and availability falls. However, the behavior in scenario 1 is more complex. While $P_A$ exerts a strictly monotonic influence on the mission completion time, $\varepsilon$ is again subject to \textit{latency blockage}. 

\subsection{Analysis of Communicability \label{subsec: p-comm}}
As introduced in \Cref{subsec: communicability}, communicability, as a combined metric to represent RCP key metrics, considers both $\varepsilon$ and $P_A$ for a given message transaction time $\tau_{msg}$. $\tau_{msg}$ is computed based on the message size (in bits) and the transmission bitrate (in bits per second) as,
\begin{equation}
    \tau_{msg} = \frac{S_{bits}}{R_{bitrate}}
\end{equation}
\noindent where \(\tau_{msg}\) is the message transmission time (seconds), \(S_{bits}\) is the size of the message in bits, and \(R_{bitrate}\) is the transmission bitrate in bits per second (bps). In this simulation, seven double float control inputs are continuous transmitted between the RPA and RPS, corresponding to a total of 56 bytes, or,
\begin{equation}
S_{bits} = 56 \text{ bytes} \times 8 \frac{\text{bits}}{\text{byte}} = 448 \text{ bits}
\end{equation}

The bitrate is based on the communication infrastructure. We use the Aircraft Communications Addressing and Reporting System (ACARS) in our case study. ACARS is one of the earliest widely adopted digital data link systems in aviation. It enables the secure exchange of short, text-based encrypted messages between aircraft and ground-based operations (i.e., flight plan updates, weather forecasts) \citep{roy2001secure, risley2001experimental}. Traditional VHF ACARS data transmissions use a low data rate around 2.4 kbps \citep{tooley2017aircraft}. That is,
\begin{equation}
R_{bitrate} = 2.4 \times 10^3 \text{ bits/s}
\end{equation}
\noindent and the corresponding message transaction time is,
\begin{equation}
\tau_{msg} = \frac{448 \text{ bits}}{2400 \text{ bits/s}} \approx 0.1867 \text{s}
\end{equation}

\begin{figure}[H]
    \centering
    \begin{subfigure}[t]{0.95\textwidth}
        \centering
        \includegraphics[width=\textwidth,height=6cm]{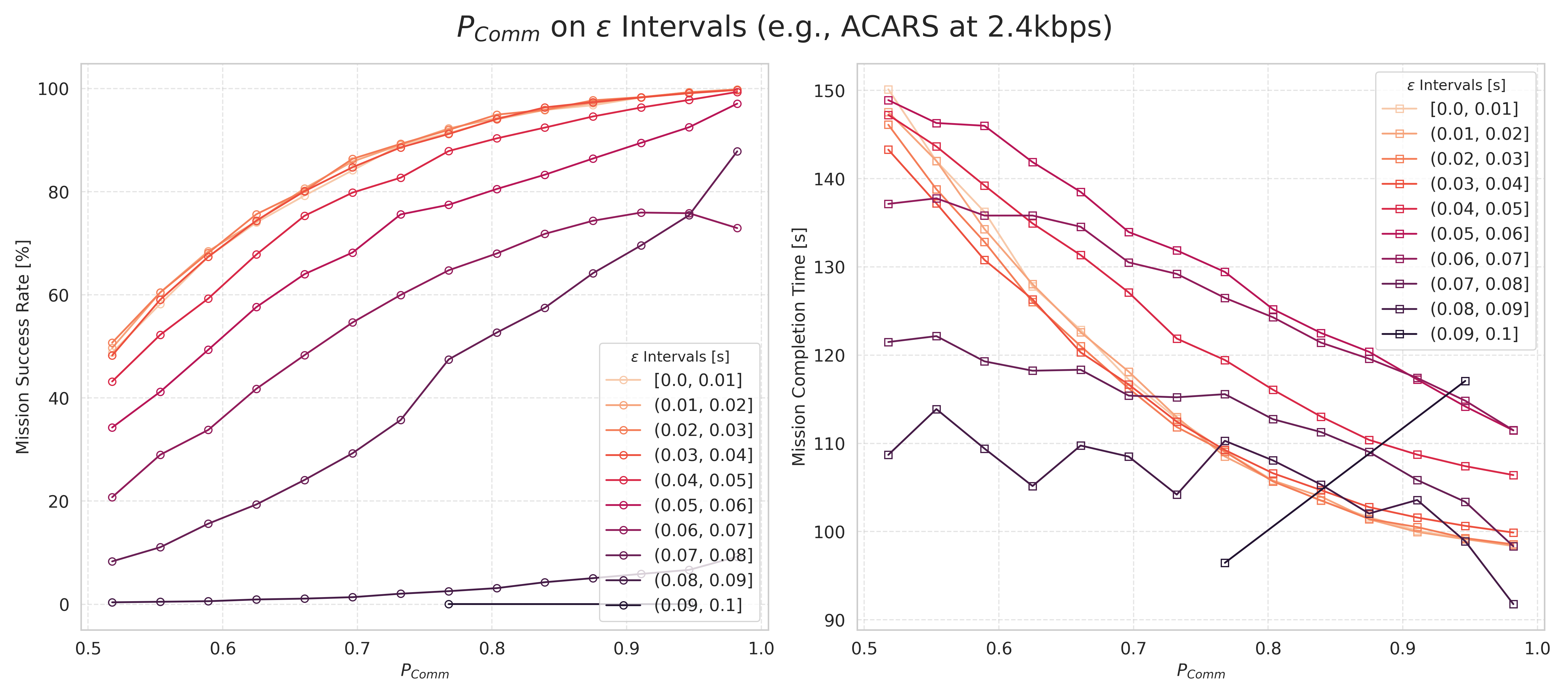}
        \caption{Scenario 1.}
        \label{fig: p-comm-1}
    \end{subfigure}
    ~
    \begin{subfigure}[t]{0.95\textwidth}
        \centering
        \includegraphics[width=\textwidth,height=6cm]{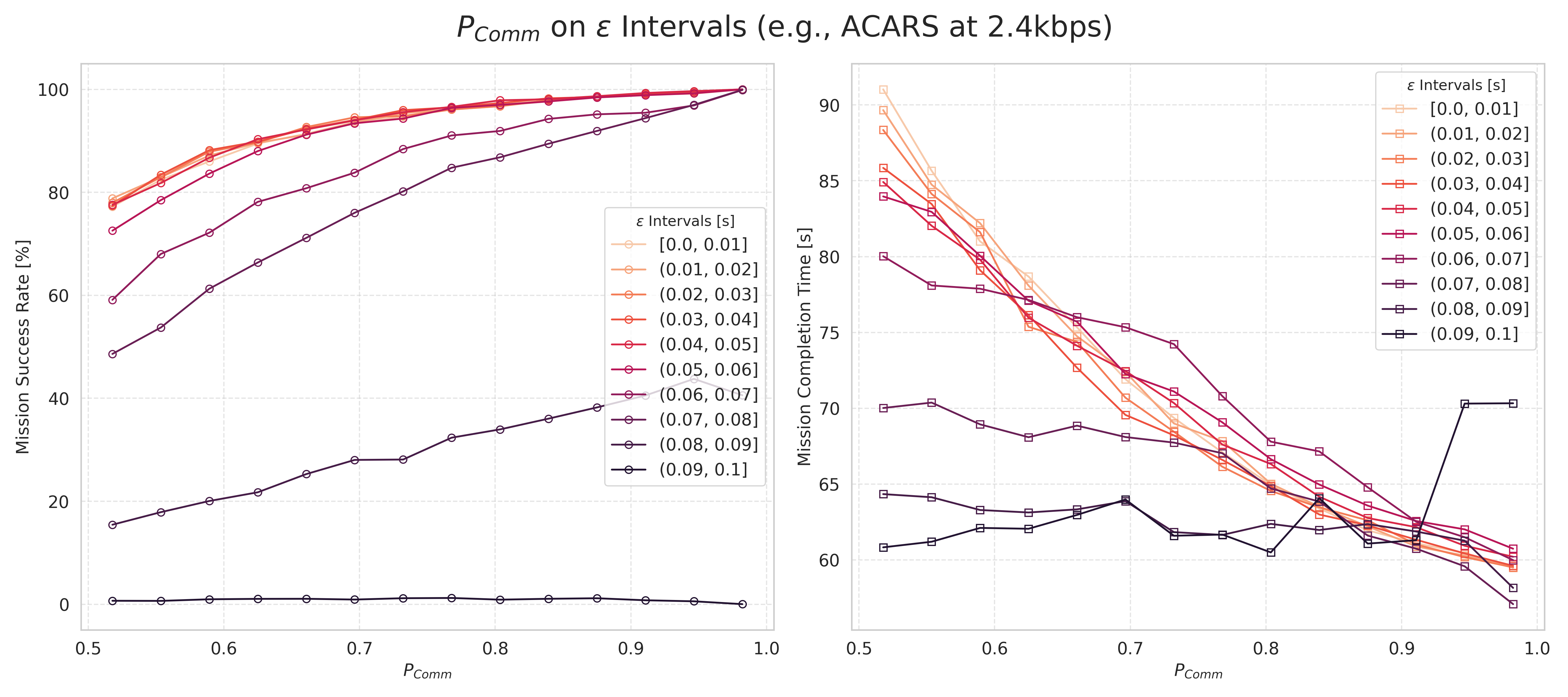}
        \caption{Scenario 2.}
        \label{fig: p-comm-2}
    \end{subfigure}
    \caption{Communicability analysis for two waypoint following scenarios. The bitrate is set as 2.4 kps for VHF ACARS.}
    \label{fig: p-comm-two}
\end{figure}

$P_{\text{comm}}$ is then calculated with \Cref{eq: P_comm}. We further draw the mission success rate and mission completion time versus $P_{\text{comm}}$ for $\varepsilon$ intervals for both scenarios, as in \Cref{fig: p-comm-two}. Communicability analysis facilitates better understanding of the maximum tolerable latency value for a given data transmission system. For instance, it is obvious that in \Cref{fig: p-comm-2}, latency of $0.08$s still leads to high mission success rate for high communicability in simpler scenarios. \Cref{fig: p-comm-1} also shows the impact of \textit{latency blockage} where a higher latency (i.e., $\varepsilon \in (0.07, 0.08]$) can achieve higher mission success rate than lower latency (i.e., $\varepsilon \in (0.06, 0.07]$), even with high communicability (i.e., $P_{\text{Comm}}=1$). Other aviation communication system (i.e.,Ka/Ku-band/L-band SATCOM, Aeronautical Mobile Airport Communication System (AeroMACS)) with different bitrate (i.e.,  AeroMACS maxed at 9.2 Mbps) can be evaluated in a similar way.

\subsection{Investigation on Latency Blockage \label{subsec: discussion}}
\begin{figure}[H]
    \centering
    \begin{subfigure}[t]{0.45\textwidth}
        \centering
        \includegraphics[width=\textwidth,height=6cm]{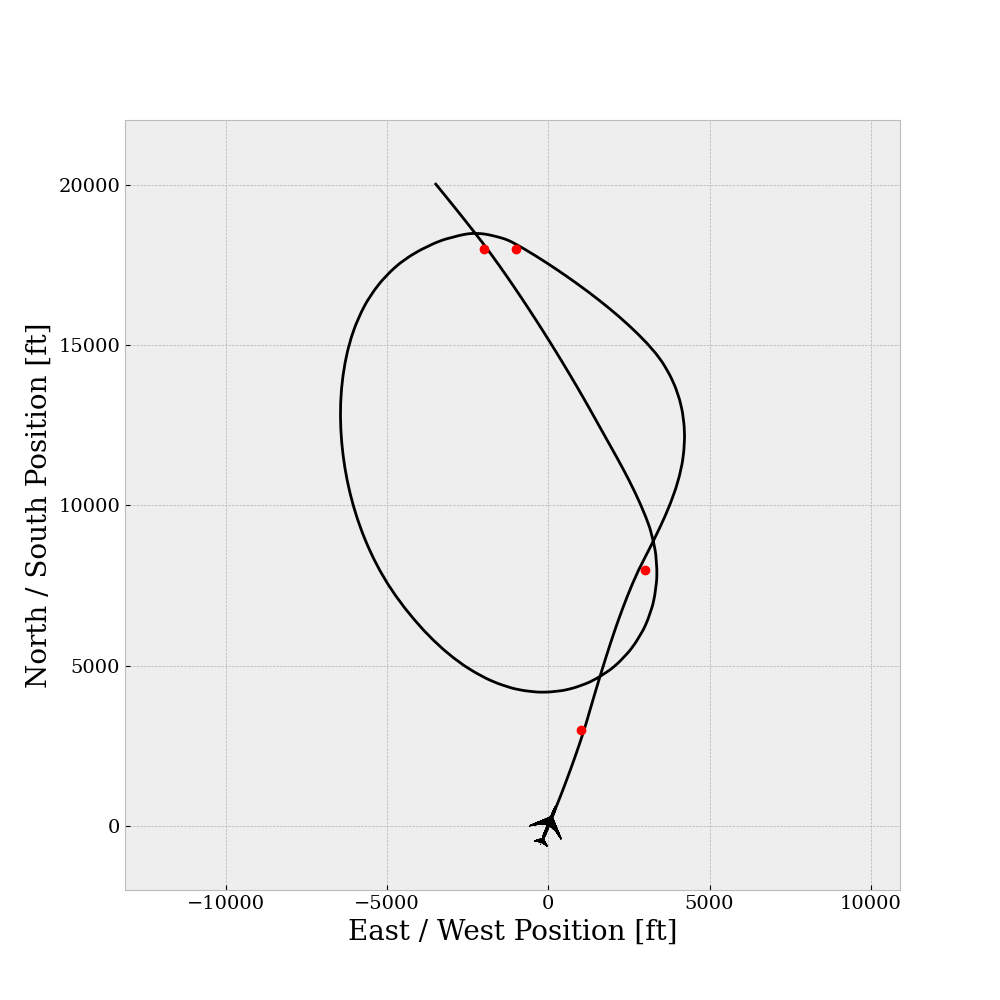}
        \caption{$P_A=1, \varepsilon=0.062$.}
        \label{fig: trajectory-wp1-0.062}
    \end{subfigure}
    \begin{subfigure}[t]{0.45\textwidth}
        \centering
        \includegraphics[width=\textwidth,height=6cm]{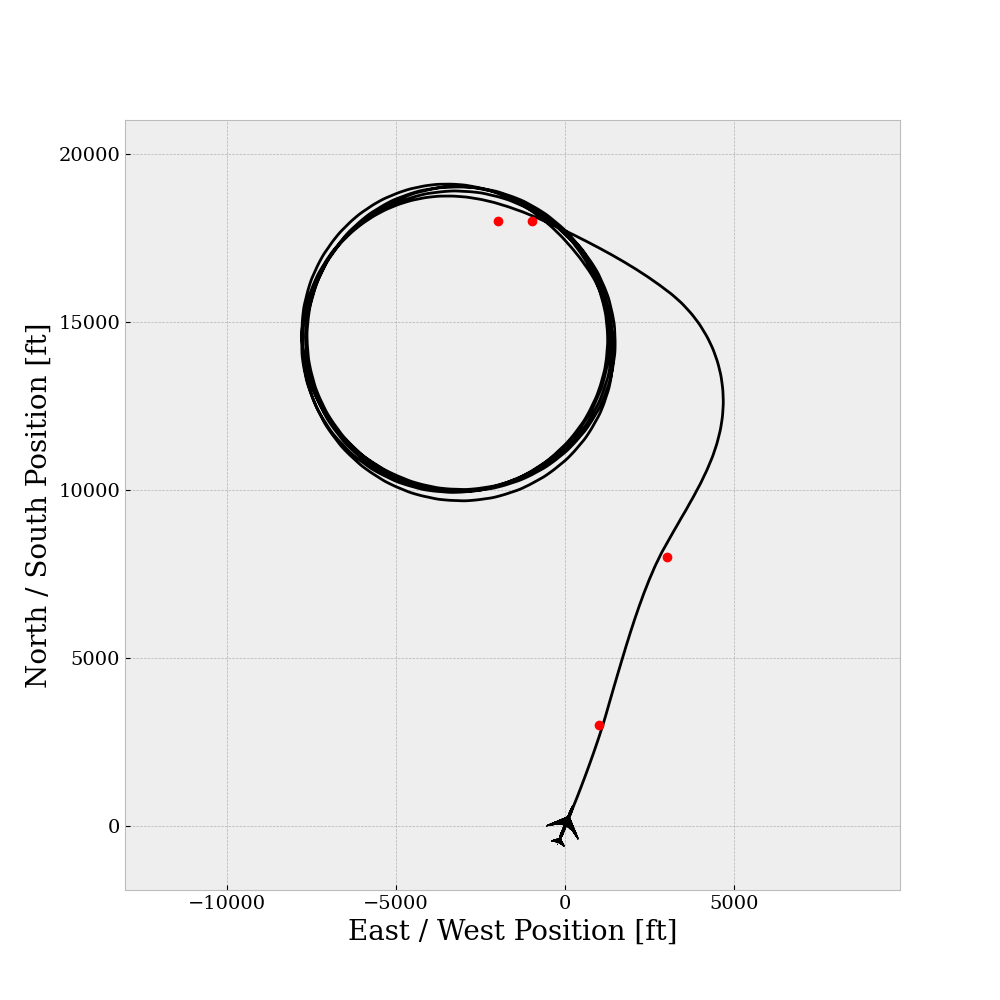}
        \caption{$P_A=1, \varepsilon=0.063$.}
        \label{fig: trajectory-wp1-0.063}
    \end{subfigure}
    ~
    \begin{subfigure}[t]{0.45\textwidth}
        \centering
        \includegraphics[width=\textwidth,height=6cm]{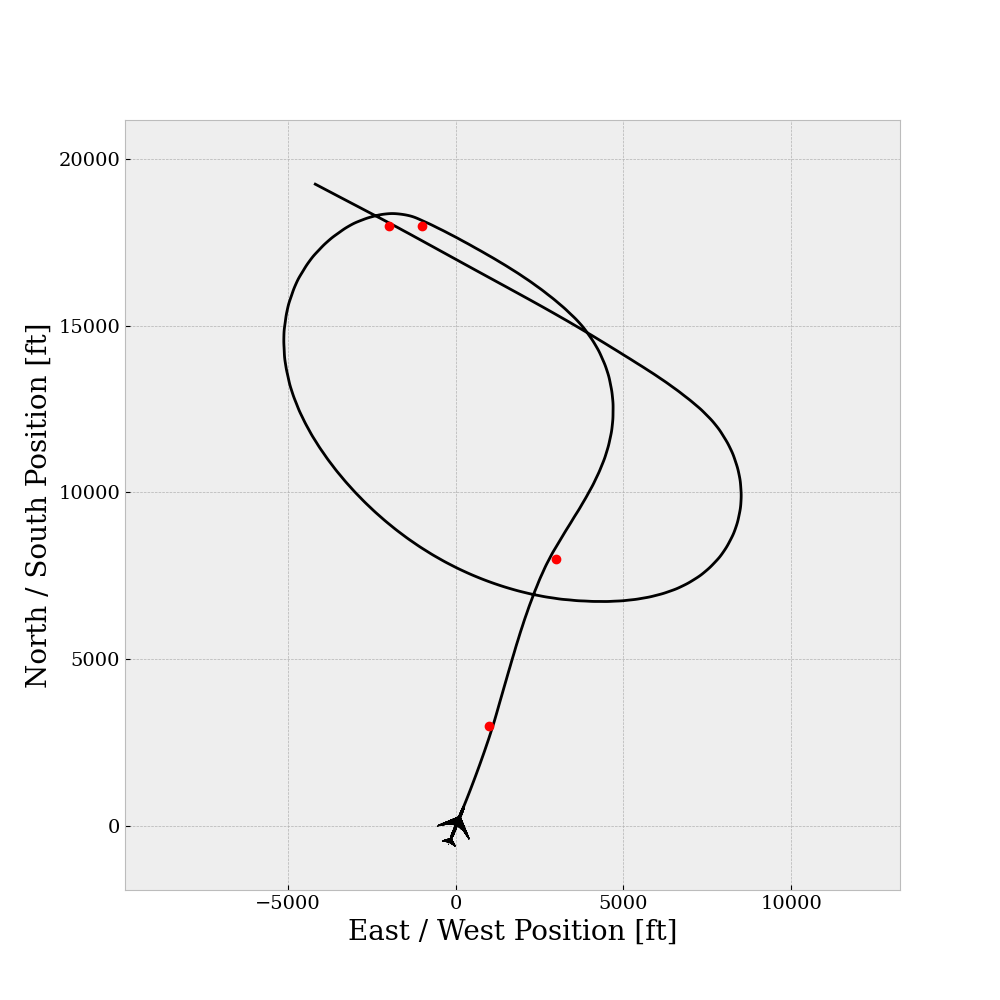}
        \caption{$P_A=1, \varepsilon=0.064$.}
        \label{fig: trajectory-wp1-0.064}
    \end{subfigure}
    \begin{subfigure}[t]{0.45\textwidth}
        \centering
        \includegraphics[width=\textwidth,height=6cm]{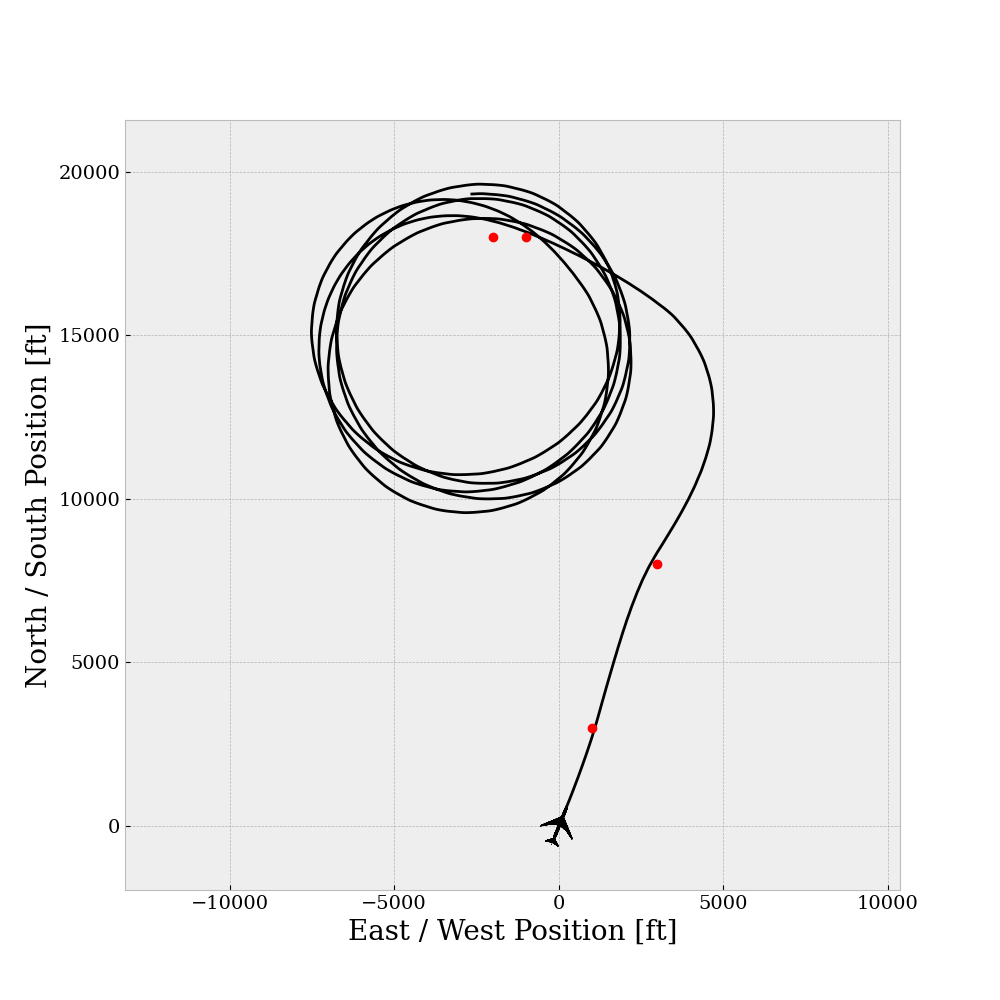}
        \caption{$P_A=1, \varepsilon=0.065$.}
        \label{fig: trajectory-wp1-0.065}
    \end{subfigure}
    ~
    \begin{subfigure}[t]{0.45\textwidth}
        \centering
        \includegraphics[width=\textwidth,height=6cm]{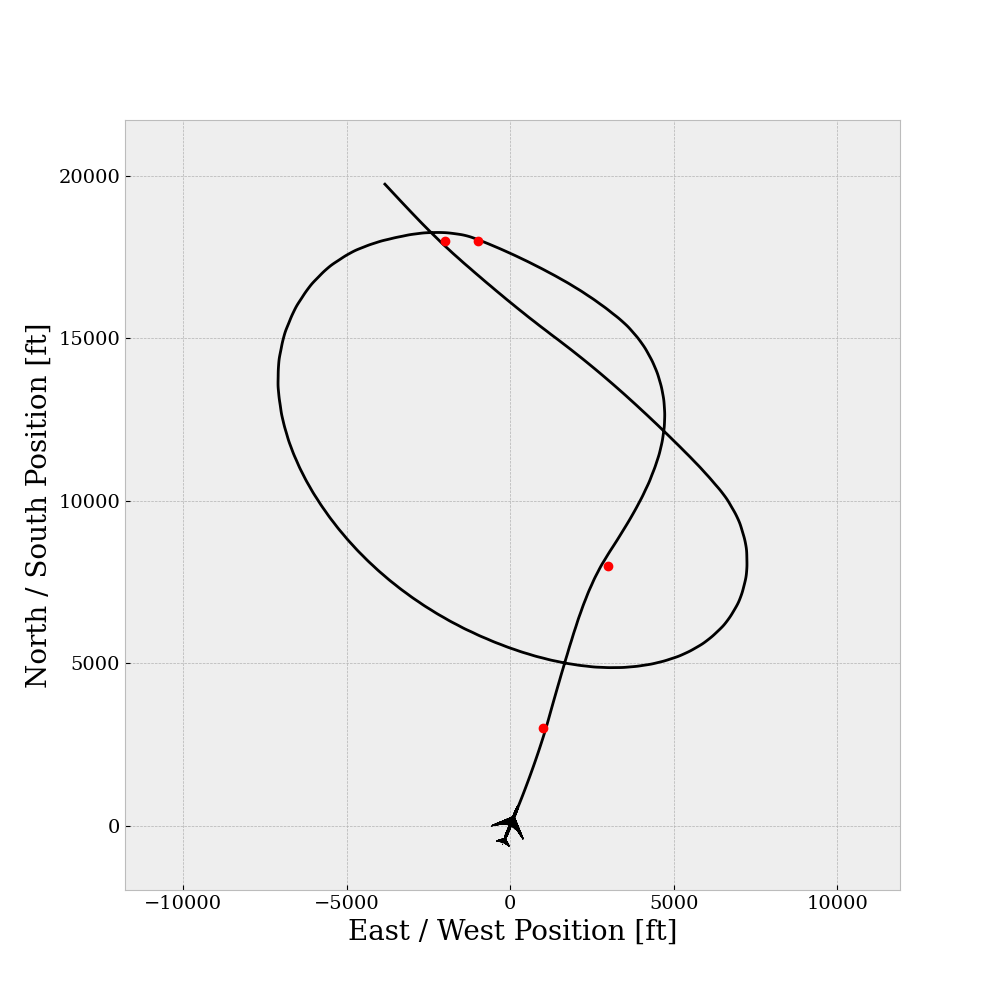}
        \caption{$P_A=1, \varepsilon=0.066$.}
        \label{fig: trajectory-wp1-0.066}
    \end{subfigure}
    \begin{subfigure}[t]{0.45\textwidth}
        \centering
        \includegraphics[width=\textwidth,height=6cm]{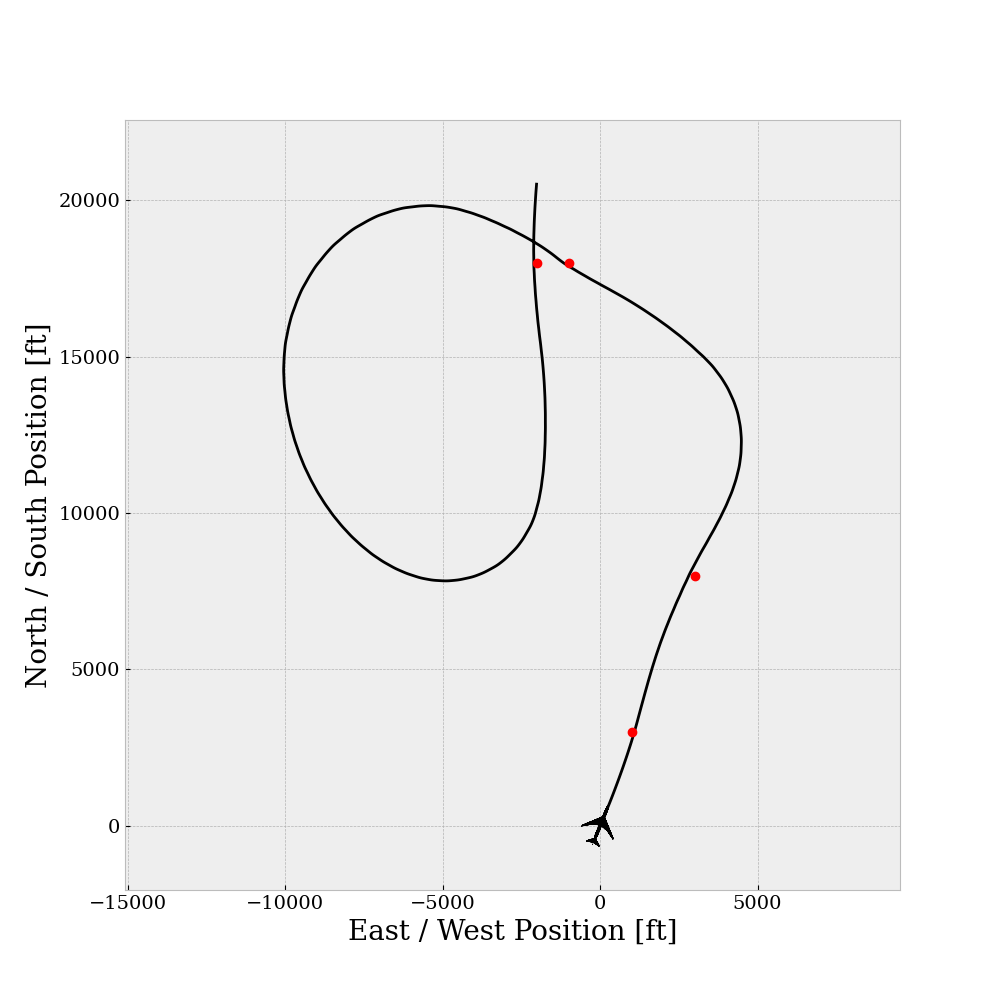}
        \caption{$P_A=1, \varepsilon=0.067$.}
        \label{fig: trajectory-wp1-0.067}
    \end{subfigure}
    \caption{Flight trajectory in the vertical plane for scenario 1 under different single way latency values $\varepsilon$, showing the communication \textit{latency blockage} for certain $\varepsilon$ values for the given task. }
    \label{fig: trajectory-s1-latency-blockage}
\end{figure}

We look further to the \textit{latency blockage} issue highlighted in scenario 1 in this section. \Cref{fig: trajectory-s1-latency-blockage} provides flight trajectories when $P_A=1$, and $\varepsilon$ ranges from $0.0062$s to $0.0067$s. In these simulations, no parameter-induced randomness existed due to $P_A=1$. The simulations in \Cref{fig: trajectory-wp1-0.062} and \Cref{fig: trajectory-wp1-0.064} can complete the task while \Cref{fig: trajectory-wp1-0.063} cannot. The insight is that, in \Cref{fig: trajectory-wp1-0.063}, the flight control is \textit{tricked} into thinking it can reach the last waypoint by conducting minimum radius turns, but if $\varepsilon=0.062$ or $\varepsilon=0.064$ the flight control gives up and moves further away to reach the last waypoint. Similar behavior exist for scenarios \Cref{fig: trajectory-wp1-0.065}. This particular behavior is believed to be very scenario and control system dependent due to a latency induced resonance in the minimum radius turn, however, the breakdown of monotonicity with latency may be a general phenomenon that appears in other RPAS missions, scenarios, and control systems.

\section{Conclusions\label{sec: conclusion}}

This study examines how latency and communication availability influence the flight performance of a Remotely Piloted Aircraft System (RPAS), modeled on flight dynamics, during a static waypoint-following task. We begin by reviewing relevant aviation communication standards, including ICAO’s Required Communication Performance (RCP) and JARUS’s Required Link Performance (RLP), and then derive their corresponding mathematical formulations. In addition, we propose a new metric, communicability, that collectively represents availability, continuity, and transaction time.

To evaluate flight performance in the context of these communication standards, we define both a complex and a simple scenario, conducting extensive Monte Carlo simulations with randomness from latency and availabilities. These simulations employ key performance measures, such as mission success rate and mission completion time, assessed against varying latency and availability conditions. Finally, we generate a communication reliability surface envelope from the simulation results, offering valuable insights into how communication parameters affect overall RPAS flight performance. We further provide the approach to utilize the communicability concept to help understand the maximum tolerable latency value of assuring mission success and evaluate the flight performance in general. Additionally, we discover the interesting behavior of \textit{latency blockage} to trick the flight control systems into circling, in scenarios with no signal loss at all. 

\subsection{Limitations \label{subsec: limitations}}
We have several limitations. The experiment was conducted on the simplified open-sourced simulation code of F-16 dynamics, with many assumptions (i.e., without leading edge flaps, approximated aerodynamics coefficients from subsonic flight). A better simulator with broader effective flight envelop makes our study closer to real world scenarios. Moreover, the simulator assume a decoupled flight control on the horizontal and longitudinal planes with two LQRs. Better nonlinear flight control such as NDI exist for F-16 and has been used in the literature. Lastly, we only consider static waypoint reaching task, where a large portion of RPAS usage involves interactions between multiple aircraft in highly dynamic environments. 

\subsection{Future Studies \label{subsec: future}}
The future studies for this work are manifolds,

\begin{itemize}
    \item Further investigations into more effective optimal control designs are desired. In the current case studies, we consider only the LQR around a single trim point, resulting in suboptimal performance when the state vector operates far from the design setpoint. Gain scheduling at runtime, or control approaches accounting for nonlinear dynamics (e.g., nonlinear dynamics inversion), represent promising directions for future research.
    \item The current work focuses on static waypoint-following tasks. However, a significant portion of interests lies in flight performance within highly dynamic environments, typically involving multiple moving agents (e.g., collision avoidance, separation violations, tracking). Performance in such dynamic scenarios can be substantially more complex than those examined here. In system-of-systems contexts, system importance measures could help identify the most vulnerable subsystems and guide design improvements for resilience \citep{uday_system_2019}.
    \item As mentioned in \Cref{subsubsec: msr}, Nyquist stability analysis is highly desired to understand and verify phase and gain margins. The open-loop transfer function and the frequency response will need to be derived, especially for Multi-Input Multi-Output (MIMO) systems. 
    \item The experimentation process requires a substantial number of Monte Carlo simulations to produce a high-quality mission success envelope, making real-time estimates of mission completion probabilities impractical. Parallel computing serves as a potential solution \citep{esselink1995parallel}. Additionally, NVIDIA CUDA enables the agile deployment of thousands of parallel threads, significantly reducing Monte Carlo simulation times at large scales to enable onboard real-time computation.
\end{itemize}

\section*{Acknowledgment}
This work was supported by the National Aeronautics and Space Administration (NASA) University Leadership Initiative (ULI) program under project “Autonomous Aerial Cargo Operations at Scale”, via grant No. 80NSSC21M071 to the University of Texas at Austin. The support is greatfully appreciated. 

\bibliography{ref}

\end{document}